%
%
%

\ifx\mnmacrosloaded\undefined 
%
%
%
%

\catcode `\@=11 

\def\@version{1.6}
\def\@verdate{18th September 1995}

%
%


\newif\ifprod@font

\ifx\@typeface\undefined
  \def\@typeface{Comp. Modern}\prod@fontfalse
\else
  \prod@fonttrue 
\fi

\def\newfam{\alloc@8\fam\chardef\sixt@@n} 

\ifprod@font
\font\fiverm=mtr10 at 5pt
\font\fivebf=mtbx10 at 5pt
\font\fiveit=mtti10 at 5pt
\font\fivesl=mtsl10 at 5pt
\font\fivett=cmtt8 at 5pt     \hyphenchar\fivett=-1
\font\fivecsc=mtcsc10 at 5pt
\font\fivesf=mtss10 at 5pt
\font\fivei=mtmi10 at 5pt      \skewchar\fivei='177
\font\fivesy=mtsy10 at 5pt     \skewchar\fivesy='60

\font\sixrm=mtr10 at 6pt
\font\sixbf=mtbx10 at 6pt
\font\sixit=mtti10 at 6pt
\font\sixsl=mtsl10 at 6pt
\font\sixtt=cmtt8 at 6pt      \hyphenchar\sixtt=-1
\font\sixcsc=mtcsc10 at 6pt
\font\sixsf=mtss10 at 6pt
\font\sixi=mtmi10 at 6pt       \skewchar\sixi='177
\font\sixsy=mtsy10 at 6pt      \skewchar\sixsy='60

\font\sevenrm=mtr10 at 7pt
\font\sevenbf=mtbx10 at 7pt
\font\sevenit=mtti10 at 7pt
\font\sevensl=mtsl10 at 7pt
\font\seventt=cmtt8 at 7pt     \hyphenchar\seventt=-1
\font\sevencsc=mtcsc10 at 7pt
\font\sevensf=mtss10 at 7pt
\font\seveni=mtmi10 at 7pt      \skewchar\seveni='177
\font\sevensy=mtsy10 at 7pt     \skewchar\sevensy='60

\font\eightrm=mtr10 at 8pt
\font\eightbf=mtbx10 at 8pt
\font\eightit=mtti10 at 8pt
\font\eighti=mtmi10 at 8pt      \skewchar\eighti='177
\font\eightsy=mtsy10 at 8pt     \skewchar\eightsy='60
\font\eightsl=mtsl10 at 8pt
\font\eighttt=cmtt8             \hyphenchar\eighttt=-1
\font\eightcsc=mtcsc10 at 8pt
\font\eightsf=mtss10 at 8pt

\font\ninerm=mtr10 at 9pt
\font\ninebf=mtbx10 at 9pt
\font\nineit=mtti10 at 9pt
\font\ninei=mtmi10 at 9pt      \skewchar\ninei='177
\font\ninesy=mtsy10 at 9pt     \skewchar\ninesy='60
\font\ninesl=mtsl10 at 9pt
\font\ninett=cmtt9             \hyphenchar\ninett=-1
\font\ninecsc=mtcsc10 at 9pt
\font\ninesf=mtss10 at 9pt

\font\tenrm=mtr10
\font\tenbf=mtbx10
\font\tenit=mtti10
\font\teni=mtmi10		\skewchar\teni='177
\font\tensy=mtsy10		\skewchar\tensy='60
\font\tenex=cmex10
\font\tensl=mtsl10
\font\tentt=cmtt10		\hyphenchar\tentt=-1
\font\tencsc=mtcsc10
\font\tensf=mtss10

\font\elevenrm=mtr10 at 11pt
\font\elevenbf=mtbx10 at 11pt
\font\elevenit=mtti10 at 11pt
\font\eleveni=mtmi10 at 11pt      \skewchar\eleveni='177
\font\elevensy=mtsy10 at 11pt     \skewchar\elevensy='60
\font\elevensl=mtsl10 at 11pt
\font\eleventt=cmtt10 at 11pt     \hyphenchar\eleventt=-1
\font\elevencsc=mtcsc10 at 11pt
\font\elevensf=mtss10 at 11pt

\font\twelverm=mtr10 at 12pt
\font\twelvebf=mtbx10 at 12pt
\font\twelveit=mtti10 at 12pt
\font\twelvesl=mtsl10 at 12pt
\font\twelvett=cmtt12             \hyphenchar\twelvett=-1
\font\twelvecsc=mtcsc10 at 12pt
\font\twelvesf=mtss10 at 12pt
\font\twelvei=mtmi10 at 12pt      \skewchar\twelvei='177
\font\twelvesy=mtsy10 at 12pt     \skewchar\twelvesy='60

\font\fourteenrm=mtr10 at 14pt
\font\fourteenbf=mtbx10 at 14pt
\font\fourteenit=mtti10 at 14pt
\font\fourteeni=mtmi10 at 14pt      \skewchar\fourteeni='177
\font\fourteensy=mtsy10 at 14pt     \skewchar\fourteensy='60
\font\fourteensl=mtsl10 at 14pt
\font\fourteentt=cmtt12 at 14pt     \hyphenchar\fourteentt=-1
\font\fourteencsc=mtcsc10 at 14pt
\font\fourteensf=mtss10 at 14pt

\font\seventeenrm=mtr10 at 17pt
\font\seventeenbf=mtbx10 at 17pt
\font\seventeenit=mtti10 at 17pt
\font\seventeeni=mtmi10 at 17pt      \skewchar\seventeeni='177
\font\seventeensy=mtsy10 at 17pt     \skewchar\seventeensy='60
\font\seventeensl=mtsl10 at 17pt
\font\seventeentt=cmtt12 at 17pt     \hyphenchar\seventeentt=-1
\font\seventeencsc=mtcsc10 at 17pt
\font\seventeensf=mtss10 at 17pt
\else
\font\fiverm=cmr5
\font\fivei=cmmi5             \skewchar\fivei='177
\font\fivesy=cmsy5            \skewchar\fivesy='60
\font\fivebf=cmbx5

\font\sixrm=cmr6
\font\sixi=cmmi6             \skewchar\sixi='177
\font\sixsy=cmsy6            \skewchar\sixsy='60
\font\sixbf=cmbx6

\font\sevenrm=cmr7
\font\sevenit=cmti7
\font\seveni=cmmi7             \skewchar\seveni='177
\font\sevensy=cmsy7            \skewchar\sevensy='60
\font\sevenbf=cmbx7

\font\eightrm=cmr8
\font\eightbf=cmbx8
\font\eightit=cmti8
\font\eighti=cmmi8			\skewchar\eighti='177
\font\eightsy=cmsy8			\skewchar\eightsy='60
\font\eightsl=cmsl8
\font\eighttt=cmtt8			\hyphenchar\eighttt=-1
\font\eightcsc=cmcsc10 at 8pt
\font\eightsf=cmss8

\font\ninerm=cmr9
\font\ninebf=cmbx9
\font\nineit=cmti9
\font\ninei=cmmi9			\skewchar\ninei='177
\font\ninesy=cmsy9			\skewchar\ninesy='60
\font\ninesl=cmsl9
\font\ninett=cmtt9			\hyphenchar\ninett=-1
\font\ninecsc=cmcsc10 at 9pt
\font\ninesf=cmss9

\font\tenrm=cmr10
\font\tenbf=cmbx10
\font\tenit=cmti10
\font\teni=cmmi10		\skewchar\teni='177
\font\tensy=cmsy10		\skewchar\tensy='60
\font\tenex=cmex10
\font\tensl=cmsl10
\font\tentt=cmtt10		\hyphenchar\tentt=-1
\font\tencsc=cmcsc10
\font\tensf=cmss10

\font\elevenrm=cmr10 scaled \magstephalf
\font\elevenbf=cmbx10 scaled \magstephalf
\font\elevenit=cmti10 scaled \magstephalf
\font\eleveni=cmmi10 scaled \magstephalf	\skewchar\eleveni='177
\font\elevensy=cmsy10 scaled \magstephalf	\skewchar\elevensy='60
\font\elevensl=cmsl10 scaled \magstephalf
\font\eleventt=cmtt10 scaled \magstephalf	\hyphenchar\eleventt=-1
\font\elevencsc=cmcsc10 scaled \magstephalf
\font\elevensf=cmss10 scaled \magstephalf

\font\twelverm=cmr10 scaled \magstep1
\font\twelvebf=cmbx10 scaled \magstep1
\font\twelvei=cmmi10 scaled \magstep1      \skewchar\twelvei='177
\font\twelvesy=cmsy10 scaled \magstep1     \skewchar\twelvesy='60

\font\fourteenrm=cmr10 scaled \magstep2
\font\fourteenbf=cmbx10 scaled \magstep2
\font\fourteenit=cmti10 scaled \magstep2
\font\fourteeni=cmmi10 scaled \magstep2		\skewchar\fourteeni='177
\font\fourteensy=cmsy10 scaled \magstep2	\skewchar\fourteensy='60
\font\fourteensl=cmsl10 scaled \magstep2
\font\fourteentt=cmtt10 scaled \magstep2	\hyphenchar\fourteentt=-1
\font\fourteencsc=cmcsc10 scaled \magstep2
\font\fourteensf=cmss10 scaled \magstep2

\font\seventeenrm=cmr10 scaled \magstep3
\font\seventeenbf=cmbx10 scaled \magstep3
\font\seventeenit=cmti10 scaled \magstep3
\font\seventeeni=cmmi10 scaled \magstep3	\skewchar\seventeeni='177
\font\seventeensy=cmsy10 scaled \magstep3	\skewchar\seventeensy='60
\font\seventeensl=cmsl10 scaled \magstep3
\font\seventeentt=cmtt10 scaled \magstep3	\hyphenchar\seventeentt=-1
\font\seventeencsc=cmcsc10 scaled \magstep3
\font\seventeensf=cmss10 scaled \magstep3
\fi

\def\hexnumber#1{\ifcase#1 0\or1\or2\or3\or4\or5\or6\or7\or8\or9\or
  A\or B\or C\or D\or E\or F\fi}

\def\makestrut{%
  \setbox\strutbox=\hbox{%
    \vrule height.7\baselineskip depth.3\baselineskip width \z@}%
}

\def\baselinestretch{1}
\newskip\tmp@bls

\def\b@ls#1{
  \tmp@bls=#1\relax
  \baselineskip=#1\relax\makestrut
  \normalbaselineskip=\baselinestretch\tmp@bls
  \normalbaselines
}

\def\nostb@ls#1{
  \normalbaselineskip=#1\relax
  \normalbaselines
  \makestrut
}

%

\newfam\scfam  
\newfam\sffam  

\def\mit{\fam\@ne}
\def\cal{\fam\tw@}
\def\em{\ifdim\fontdimen1\font>\z@ \rm\else\it\fi}

\textfont3=\tenex
\scriptfont3=\tenex
\scriptscriptfont3=\tenex

\setbox0=\hbox{\tenex B} \p@renwd=\wd0 

\def\eightpoint{
  \def\rm{\fam0\eightrm}%
  \textfont0=\eightrm \scriptfont0=\sixrm \scriptscriptfont0=\fiverm%
  \textfont1=\eighti  \scriptfont1=\sixi  \scriptscriptfont1=\fivei%
  \textfont2=\eightsy \scriptfont2=\sixsy \scriptscriptfont2=\fivesy%
  \textfont\itfam=\eightit\def\it{\fam\itfam\eightit}%
  \ifprod@font
    \scriptfont\itfam=\sixit
      \scriptscriptfont\itfam=\fiveit
  \else
    \scriptfont\itfam=\eightit
      \scriptscriptfont\itfam=\eightit
  \fi
  \textfont\bffam=\eightbf%
    \scriptfont\bffam=\sixbf%
      \scriptscriptfont\bffam=\fivebf%
  \def\bf{\fam\bffam\eightbf}%
  \textfont\slfam=\eightsl\def\sl{\fam\slfam\eightsl}%
  \ifprod@font
    \scriptfont\slfam=\sixsl
      \scriptscriptfont\slfam=\fivesl
  \else
    \scriptfont\slfam=\eightsl
      \scriptscriptfont\slfam=\eightsl
  \fi
  \textfont\ttfam=\eighttt\def\tt{\fam\ttfam\eighttt}%
  \ifprod@font
    \scriptfont\ttfam=\sixtt
      \scriptscriptfont\ttfam=\fivett
  \else
    \scriptfont\ttfam=\eighttt
      \scriptscriptfont\ttfam=\eighttt
  \fi
  \textfont\scfam=\eightcsc\def\sc{\fam\scfam\eightcsc}%
  \ifprod@font
    \scriptfont\scfam=\sixcsc
      \scriptscriptfont\scfam=\fivecsc
  \else
    \scriptfont\scfam=\eightcsc
      \scriptscriptfont\scfam=\eightcsc
  \fi
  \textfont\sffam=\eightsf\def\sf{\fam\sffam\eightsf}%
  \ifprod@font
    \scriptfont\sffam=\sixsf
      \scriptscriptfont\sffam=\fivesf
  \else
    \scriptfont\sffam=\eightsf
      \scriptscriptfont\sffam=\eightsf
  \fi
  \def\oldstyle{\fam\@ne\eighti}%
  \b@ls{10pt}\rm\@viiipt%
}
\def\@viiipt{}

\def\ninepoint{
  \def\rm{\fam0\ninerm}%
  \textfont0=\ninerm \scriptfont0=\sixrm \scriptscriptfont0=\fiverm%
  \textfont1=\ninei  \scriptfont1=\sixi  \scriptscriptfont1=\fivei%
  \textfont2=\ninesy \scriptfont2=\sixsy \scriptscriptfont2=\fivesy%
  \textfont\itfam=\nineit\def\it{\fam\itfam\nineit}%
  \ifprod@font
    \scriptfont\itfam=\sixit
      \scriptscriptfont\itfam=\fiveit
  \else
    \scriptfont\itfam=\nineit
      \scriptscriptfont\itfam=\nineit
  \fi
  \textfont\bffam=\ninebf%
    \scriptfont\bffam=\sixbf%
      \scriptscriptfont\bffam=\fivebf%
  \def\bf{\fam\bffam\ninebf}%
  \textfont\slfam=\ninesl\def\sl{\fam\slfam\ninesl}%
  \ifprod@font
    \scriptfont\slfam=\sixsl
      \scriptscriptfont\slfam=\fivesl
  \else
    \scriptfont\slfam=\ninesl
      \scriptscriptfont\slfam=\ninesl
  \fi
  \textfont\ttfam=\ninett\def\tt{\fam\ttfam\ninett}%
  \ifprod@font
    \scriptfont\ttfam=\sixtt
      \scriptscriptfont\ttfam=\fivett
  \else
    \scriptfont\ttfam=\ninett
      \scriptscriptfont\ttfam=\ninett
  \fi
  \textfont\scfam=\ninecsc\def\sc{\fam\scfam\ninecsc}%
  \ifprod@font
    \scriptfont\scfam=\sixcsc
      \scriptscriptfont\scfam=\fivecsc
  \else
    \scriptfont\scfam=\ninecsc
      \scriptscriptfont\scfam=\ninecsc
  \fi
  \textfont\sffam=\ninesf\def\sf{\fam\sffam\ninesf}%
  \ifprod@font
    \scriptfont\sffam=\sixsf
      \scriptscriptfont\sffam=\fivesf
  \else
    \scriptfont\sffam=\ninesf
      \scriptscriptfont\sffam=\ninesf
  \fi
  \def\oldstyle{\fam\@ne\ninei}%
  \b@ls{\TextLeading plus \Feathering}\rm\@ixpt%
}
\def\@ixpt{}

\def\tenpoint{
  \def\rm{\fam0\tenrm}%
  \textfont0=\tenrm \scriptfont0=\sevenrm \scriptscriptfont0=\fiverm%
  \textfont1=\teni  \scriptfont1=\seveni  \scriptscriptfont1=\fivei%
  \textfont2=\tensy \scriptfont2=\sevensy \scriptscriptfont2=\fivesy%
  \textfont\itfam=\tenit\def\it{\fam\itfam\tenit}%
  \ifprod@font
    \scriptfont\itfam=\sevenit
      \scriptscriptfont\itfam=\fiveit
  \else
    \scriptfont\itfam=\tenit
      \scriptscriptfont\itfam=\tenit
  \fi
  \textfont\bffam=\tenbf%
    \scriptfont\bffam=\sevenbf%
      \scriptscriptfont\bffam=\fivebf%
  \def\bf{\fam\bffam\tenbf}%
  \textfont\slfam=\tensl\def\sl{\fam\slfam\tensl}%
  \ifprod@font
    \scriptfont\slfam=\sevensl
      \scriptscriptfont\slfam=\fivesl
  \else
    \scriptfont\slfam=\tensl
      \scriptscriptfont\slfam=\tensl
  \fi
  \textfont\ttfam=\tentt\def\tt{\fam\ttfam\tentt}%
  \ifprod@font
    \scriptfont\ttfam=\seventt
      \scriptscriptfont\ttfam=\fivett
  \else
    \scriptfont\ttfam=\tentt
      \scriptscriptfont\ttfam=\tentt
  \fi
  \textfont\scfam=\tencsc\def\sc{\fam\scfam\tencsc}%
  \ifprod@font
    \scriptfont\scfam=\sevencsc
      \scriptscriptfont\scfam=\fivecsc
  \else
    \scriptfont\scfam=\tencsc
      \scriptscriptfont\scfam=\tencsc
  \fi
  \textfont\sffam=\tensf\def\sf{\fam\sffam\tensf}%
  \ifprod@font
    \scriptfont\sffam=\sevensf
      \scriptscriptfont\sffam=\fivesf
  \else
    \scriptfont\sffam=\tensf
      \scriptscriptfont\sffam=\tensf
  \fi
  \def\oldstyle{\fam\@ne\teni}%
  \b@ls{11pt}\rm\@xpt%
}
\def\@xpt{}

\def\elevenpoint{
  \def\rm{\fam0\elevenrm}%
  \textfont0=\elevenrm \scriptfont0=\eightrm \scriptscriptfont0=\sixrm%
  \textfont1=\eleveni  \scriptfont1=\eighti  \scriptscriptfont1=\sixi%
  \textfont2=\elevensy \scriptfont2=\eightsy \scriptscriptfont2=\sixsy%
  \textfont\itfam=\elevenit\def\it{\fam\itfam\elevenit}%
  \ifprod@font
    \scriptfont\itfam=\eightit
      \scriptscriptfont\itfam=\sixit
  \else
    \scriptfont\itfam=\elevenit
      \scriptscriptfont\itfam=\elevenit
  \fi
  \textfont\bffam=\elevenbf%
    \scriptfont\bffam=\eightbf%
      \scriptscriptfont\bffam=\sixbf%
  \def\bf{\fam\bffam\elevenbf}%
  \textfont\slfam=\elevensl\def\sl{\fam\slfam\elevensl}%
  \ifprod@font
    \scriptfont\slfam=\eightsl
      \scriptscriptfont\slfam=\sixsl
  \else
    \scriptfont\slfam=\elevensl
      \scriptscriptfont\slfam=\elevensl
  \fi
  \textfont\ttfam=\eleventt\def\tt{\fam\ttfam\eleventt}%
  \ifprod@font
    \scriptfont\ttfam=\eighttt
      \scriptscriptfont\ttfam=\sixtt
  \else
    \scriptfont\ttfam=\eleventt
      \scriptscriptfont\ttfam=\eleventt
  \fi
  \textfont\scfam=\elevencsc\def\sc{\fam\scfam\elevencsc}%
  \ifprod@font
    \scriptfont\scfam=\eightcsc
      \scriptscriptfont\scfam=\sixcsc
  \else
    \scriptfont\scfam=\elevencsc
      \scriptscriptfont\scfam=\elevencsc
  \fi
  \textfont\sffam=\elevensf\def\sf{\fam\sffam\elevensf}%
  \ifprod@font
    \scriptfont\sffam=\eightsf
      \scriptscriptfont\sffam=\sixsf
  \else
    \scriptfont\sffam=\elevensf
      \scriptscriptfont\sffam=\elevensf
  \fi
  \def\oldstyle{\fam\@ne\eleveni}%
  \b@ls{13pt}\rm\@xipt%
}
\def\@xipt{}

\def\fourteenpoint{
  \def\rm{\fam0\fourteenrm}%
  \textfont0\fourteenrm  \scriptfont0\tenrm  \scriptscriptfont0\sevenrm%
  \textfont1\fourteeni   \scriptfont1\teni   \scriptscriptfont1\seveni%
  \textfont2\fourteensy  \scriptfont2\tensy  \scriptscriptfont2\sevensy%
  \textfont\itfam=\fourteenit\def\it{\fam\itfam\fourteenit}%
  \ifprod@font
    \scriptfont\itfam=\tenit
      \scriptscriptfont\itfam=\sevenit
  \else
    \scriptfont\itfam=\fourteenit
      \scriptscriptfont\itfam=\fourteenit
  \fi
  \textfont\bffam=\fourteenbf%
    \scriptfont\bffam=\tenbf%
      \scriptscriptfont\bffam=\sevenbf%
  \def\bf{\fam\bffam\fourteenbf}%
  \textfont\slfam=\fourteensl\def\sl{\fam\slfam\fourteensl}%
  \ifprod@font
    \scriptfont\slfam=\tensl
      \scriptscriptfont\slfam=\sevensl
  \else
    \scriptfont\slfam=\fourteensl
      \scriptscriptfont\slfam=\fourteensl
  \fi
  \textfont\ttfam=\fourteentt\def\tt{\fam\ttfam\fourteentt}%
  \ifprod@font
    \scriptfont\ttfam=\tentt
      \scriptscriptfont\ttfam=\seventt
  \else
    \scriptfont\ttfam=\fourteentt
      \scriptscriptfont\ttfam=\fourteentt
  \fi
  \textfont\scfam=\fourteencsc\def\sc{\fam\scfam\fourteencsc}%
  \ifprod@font
    \scriptfont\scfam=\tencsc
      \scriptscriptfont\scfam=\sevencsc
  \else
    \scriptfont\scfam=\fourteencsc
      \scriptscriptfont\scfam=\fourteencsc
  \fi
  \textfont\sffam=\fourteensf\def\sf{\fam\sffam\fourteensf}%
  \ifprod@font
    \scriptfont\sffam=\tensf
      \scriptscriptfont\sffam=\sevensf
  \else
    \scriptfont\sffam=\fourteensf
      \scriptscriptfont\sffam=\fourteensf
  \fi
  \def\oldstyle{\fam\@ne\fourteeni}%
  \b@ls{17pt}\rm\@xivpt%
}
\def\@xivpt{}

\def\seventeenpoint{
  \def\rm{\fam0\seventeenrm}%
  \textfont0\seventeenrm  \scriptfont0\twelverm  \scriptscriptfont0\tenrm%
  \textfont1\seventeeni   \scriptfont1\twelvei   \scriptscriptfont1\teni%
  \textfont2\seventeensy  \scriptfont2\twelvesy  \scriptscriptfont2\tensy%
  \textfont\itfam=\seventeenit\def\it{\fam\itfam\seventeenit}%
  \ifprod@font
    \scriptfont\itfam=\twelveit
      \scriptscriptfont\itfam=\tenit
  \else
    \scriptfont\itfam=\seventeenit
      \scriptscriptfont\itfam=\seventeenit
  \fi
  \textfont\bffam=\seventeenbf%
    \scriptfont\bffam=\twelvebf%
      \scriptscriptfont\bffam=\tenbf%
  \def\bf{\fam\bffam\seventeenbf}%
  \textfont\slfam=\seventeensl\def\sl{\fam\slfam\seventeensl}%
  \ifprod@font
    \scriptfont\slfam=\twelvesl
      \scriptscriptfont\slfam=\tensl
  \else
    \scriptfont\slfam=\seventeensl
      \scriptscriptfont\slfam=\seventeensl
  \fi
  \textfont\ttfam=\seventeentt\def\tt{\fam\ttfam\seventeentt}%
  \ifprod@font
    \scriptfont\ttfam=\twelvett
      \scriptscriptfont\ttfam=\tentt
  \else
    \scriptfont\ttfam=\seventeentt
      \scriptscriptfont\ttfam=\seventeentt
  \fi
  \textfont\scfam=\seventeencsc\def\sc{\fam\scfam\seventeencsc}%
  \ifprod@font
    \scriptfont\scfam=\twelvecsc
      \scriptscriptfont\scfam=\tencsc
  \else
    \scriptfont\scfam=\seventeencsc
      \scriptscriptfont\scfam=\seventeencsc
  \fi
  \textfont\sffam=\seventeensf\def\sf{\fam\sffam\seventeensf}%
  \ifprod@font
    \scriptfont\sffam=\twelvesf
      \scriptscriptfont\sffam=\tensf
  \else
    \scriptfont\sffam=\seventeensf
      \scriptscriptfont\sffam=\seventeensf
  \fi
  \def\oldstyle{\fam\@ne\seventeeni}%
  \b@ls{20pt}\rm\@xviipt%
}
\def\@xviipt{}

\lineskip=1pt      \normallineskip=\lineskip
\lineskiplimit=\z@ \normallineskiplimit=\lineskiplimit


\def\loadboldmathnames{%
  \def\balpha{{\bmath{\alpha}}}%
  \def\bbeta{{\bmath{\beta}}}%
  \def\bgamma{{\bmath{\gamma}}}%
  \def\bdelta{{\bmath{\delta}}}%
  \def\bepsilon{{\bmath{\epsilon}}}%
  \def\bzeta{{\bmath{\zeta}}}%
  \def\boldeta{{\bmath{\eta}}}%
  \def\btheta{{\bmath{\theta}}}%
  \def\biota{{\bmath{\iota}}}%
  \def\bkappa{{\bmath{\kappa}}}%
  \def\blambda{{\bmath{\lambda}}}%
  \def\bmu{{\bmath{\mu}}}%
  \def\bnu{{\bmath{\nu}}}%
  \def\bxi{{\bmath{\xi}}}%
  \def\bpi{{\bmath{\pi}}}%
  \def\brho{{\bmath{\rho}}}%
  \def\bsigma{{\bmath{\sigma}}}%
  \def\btau{{\bmath{\tau}}}%
  \def\bupsilon{{\bmath{\upsilon}}}%
  \def\bphi{{\bmath{\phi}}}%
  \def\bchi{{\bmath{\chi}}}%
  \def\bpsi{{\bmath{\psi}}}%
  \def\bomega{{\bmath{\omega}}}%
  \def\bvarepsilon{{\bmath{\varepsilon}}}%
  \def\bvartheta{{\bmath{\vartheta}}}%
  \def\bvarpi{{\bmath{\varpi}}}%
  \def\bvarrho{{\bmath{\varrho}}}%
  \def\bvarsigma{{\bmath{\varsigma}}}%
  \def\bvarphi{{\bmath{\varphi}}}%
  \def\baleph{{\bmath{\aleph}}}%
  \def\bimath{{\bmath{\imath}}}%
  \def\bjmath{{\bmath{\jmath}}}%
  \def\bell{{\bmath{\ell}}}%
  \def\bwp{{\bmath{\wp}}}%
  \def\bRe{{\bmath{\Re}}}%
  \def\bIm{{\bmath{\Im}}}%
  \def\bpartial{{\bmath{\partial}}}%
  \def\binfty{{\bmath{\infty}}}%
  \def\bprime{{\bmath{\prime}}}%
  \def\bemptyset{{\bmath{\emptyset}}}%
  \def\bnabla{{\bmath{\nabla}}}%
  \def\btop{{\bmath{\top}}}%
  \def\bbot{{\bmath{\bot}}}%
  \def\btriangle{{\bmath{\triangle}}}%
  \def\bforall{{\bmath{\forall}}}%
  \def\bexists{{\bmath{\exists}}}%
  \def\bneg{{\bmath{\neg}}}%
  \def\bflat{{\bmath{\flat}}}%
  \def\bnatural{{\bmath{\natural}}}%
  \def\bsharp{{\bmath{\sharp}}}%
  \def\bclubsuit{{\bmath{\clubsuit}}}%
  \def\bdiamondsuit{{\bmath{\diamondsuit}}}%
  \def\bheartsuit{{\bmath{\heartsuit}}}%
  \def\bspadesuit{{\bmath{\spadesuit}}}%
  \def\bsmallint{{\bmath{\smallint}}}%
  \def\btriangleleft{{\bmath{\triangleleft}}}%
  \def\btriangleright{{\bmath{\triangleright}}}%
  \def\bbigtriangleup{{\bmath{\bigtriangleup}}}%
  \def\bbigtriangledown{{\bmath{\bigtriangledown}}}%
  \def\bwedge{{\bmath{\wedge}}}%
  \def\bvee{{\bmath{\vee}}}%
  \def\bcap{{\bmath{\cap}}}%
  \def\bcup{{\bmath{\cup}}}%
  \def\bddagger{{\bmath{\ddagger}}}%
  \def\bdagger{{\bmath{\dagger}}}%
  \def\bsqcap{{\bmath{\sqcap}}}%
  \def\bsqcup{{\bmath{\sqcup}}}%
  \def\buplus{{\bmath{\uplus}}}%
  \def\bamalg{{\bmath{\amalg}}}%
  \def\bdiamond{{\bmath{\diamond}}}%
  \def\bbullet{{\bmath{\bullet}}}%
  \def\bwr{{\bmath{\wr}}}%
  \def\bdiv{{\bmath{\div}}}%
  \def\bodot{{\bmath{\odot}}}%
  \def\boslash{{\bmath{\oslash}}}%
  \def\botimes{{\bmath{\otimes}}}%
  \def\bominus{{\bmath{\ominus}}}%
  \def\boplus{{\bmath{\oplus}}}%
  \def\bmp{{\bmath{\mp}}}%
  \def\bpm{{\bmath{\pm}}}%
  \def\bcirc{{\bmath{\circ}}}%
  \def\bbigcirc{{\bmath{\bigcirc}}}%
  \def\bsetminus{{\bmath{\setminus}}}%
  \def\bcdot{{\bmath{\cdot}}}%
  \def\bast{{\bmath{\ast}}}%
  \def\btimes{{\bmath{\times}}}%
  \def\bstar{{\bmath{\star}}}%
  \def\bpropto{{\bmath{\propto}}}%
  \def\bsqsubseteq{{\bmath{\sqsubseteq}}}%
  \def\bsqsupseteq{{\bmath{\sqsupseteq}}}%
  \def\bparallel{{\bmath{\parallel}}}%
  \def\bmid{{\bmath{\mid}}}%
  \def\bdashv{{\bmath{\dashv}}}%
  \def\bvdash{{\bmath{\vdash}}}%
  \def\bnearrow{{\bmath{\nearrow}}}%
  \def\bsearrow{{\bmath{\searrow}}}%
  \def\bnwarrow{{\bmath{\nwarrow}}}%
  \def\bswarrow{{\bmath{\swarrow}}}%
  \def\bLeftrightarrow{{\bmath{\Leftrightarrow}}}%
  \def\bLeftarrow{{\bmath{\Leftarrow}}}%
  \def\bRightarrow{{\bmath{\Rightarrow}}}%
  \def\bleq{{\bmath{\leq}}}%
  \def\bgeq{{\bmath{\geq}}}%
  \def\bsucc{{\bmath{\succ}}}%
  \def\bprec{{\bmath{\prec}}}%
  \def\bapprox{{\bmath{\approx}}}%
  \def\bsucceq{{\bmath{\succeq}}}%
  \def\bpreceq{{\bmath{\preceq}}}%
  \def\bsupset{{\bmath{\supset}}}%
  \def\bsubset{{\bmath{\subset}}}%
  \def\bsupseteq{{\bmath{\supseteq}}}%
  \def\bsubseteq{{\bmath{\subseteq}}}%
  \def\bin{{\bmath{\in}}}%
  \def\bni{{\bmath{\ni}}}%
  \def\bgg{{\bmath{\gg}}}%
  \def\bll{{\bmath{\ll}}}%
  \def\bnot{{\bmath{\not}}}%
  \def\bleftrightarrow{{\bmath{\leftrightarrow}}}%
  \def\bleftarrow{{\bmath{\leftarrow}}}%
  \def\brightarrow{{\bmath{\rightarrow}}}%
  \def\bmapstochar{{\bmath{\mapstochar}}}%
  \def\bsim{{\bmath{\sim}}}%
  \def\bsimeq{{\bmath{\simeq}}}%
  \def\bperp{{\bmath{\perp}}}%
  \def\bequiv{{\bmath{\equiv}}}%
  \def\basymp{{\bmath{\asymp}}}%
  \def\bsmile{{\bmath{\smile}}}%
  \def\bfrown{{\bmath{\frown}}}%
  \def\bleftharpoonup{{\bmath{\leftharpoonup}}}%
  \def\bleftharpoondown{{\bmath{\leftharpoondown}}}%
  \def\brightharpoonup{{\bmath{\rightharpoonup}}}%
  \def\brightharpoondown{{\bmath{\rightharpoondown}}}%
  \def\blhook{{\bmath{\lhook}}}%
  \def\brhook{{\bmath{\rhook}}}%
  \def\bldotp{{\bmath{\ldotp}}}%
  \def\bcdotp{{\bmath{\cdotp}}}%
}

\def\,{\relax\ifmmode \mskip\thinmuskip\else \thinspace\fi}
\let\protect=\relax

\long\def\@ifundefined#1#2#3{\expandafter\ifx\csname
  #1\endcsname\relax#2\else#3\fi}




\newtoks\math@groups \math@groups={}
\def\addtom@thgroup#1#2{#1\expandafter{\the#1#2}} 



\def\addtosizeh@ok#1#2#3#4{%
  \expandafter\def\csname @#1pt\endcsname{%
    \def\s@ze{#2}\def\ss@ze{#3}\def\sss@ze{#4}\the\math@groups%
  }%
}



\let\resetsizehook=\addtosizeh@ok


\ifprod@font
  \addtosizeh@ok{viii} {8} {6}  {5}
  \addtosizeh@ok{ix}   {9} {6}  {5}
  \addtosizeh@ok{x}    {10}{7}  {5}
  \addtosizeh@ok{xi}   {11}{8}  {6}
  \addtosizeh@ok{xiv}  {14}{10} {7}
  \addtosizeh@ok{xvii} {17}{12}{10}
\else
  \addtosizeh@ok{viii} {8}     {6}     {5}
  \addtosizeh@ok{ix}   {9}     {6}     {5}
  \addtosizeh@ok{x}    {10}    {7}     {5}
  \addtosizeh@ok{xi}   {10.95} {8}     {6}
  \addtosizeh@ok{xiv}  {14.4}  {10}    {7}
  \addtosizeh@ok{xvii} {17.28} {12}    {10}
\fi

\def\get@font#1#2#3{%
  \edef\fonts@ze{\romannumeral#3}
  \edef\fontn@me{\fonts@ze#1}
  \@ifundefined{\fontn@me}%
    {
     \global\expandafter\font\csname \fontn@me\endcsname=#2 at #3pt}%
    {}%
}

\def\ass@tfont#1#2{%
  \xdef\fam@name{\csname #1\endcsname}%
  \xdef\font@name{\csname #2\endcsname}%
  \let\textfont@name\font@name
  \textfont\fam@name\textfont@name
}

\def\ass@sfont#1#2{%
  \xdef\fam@name{\csname #1\endcsname}%
  \xdef\font@name{\csname #2\endcsname}%
  \let\textfont@name\font@name
  \scriptfont\fam@name\textfont@name
}

\def\ass@ssfont#1#2{%
  \xdef\fam@name{\csname #1\endcsname}%
  \xdef\font@name{\csname #2\endcsname}%
  \let\textfont@name\font@name
  \scriptscriptfont\fam@name\textfont@name
}


\def\NewSymbolFont#1#2{%
  \expandafter\ifx\csname sym#1fam\endcsname\relax 
    \expandafter\newfam\csname sym#1fam\endcsname
    \expandafter\edef\csname sym#1fam\endcsname{\the\allocationnumber}%
    \addtom@thgroup\math@groups{%
      \get@font{#1}{#2}{\s@ze}%
      \ass@tfont{sym#1fam}{\fontn@me}%
      \get@font{#1}{#2}{\ss@ze}%
      \ass@sfont{sym#1fam}{\fontn@me}%
      \get@font{#1}{#2}{\sss@ze}%
      \ass@ssfont{sym#1fam}{\fontn@me}%
    }%
  \else
    \errmessage{Family `#1' already defined}%
  \fi
}


\def\NewMathSymbol#1#2#3#4{%
  \edef\f@mly{\expandafter\hexnumber{\csname sym#3fam\endcsname}}%
  \mathchardef#1="#2\f@mly#4\relax
}


\newif\ifd@f

\def\NewMathDelimiter#1#2#3#4#5#6{%
  \d@ftrue
  \expandafter\ifx\csname sym#3fam\endcsname\relax
    \d@ffalse \errmessage{Family `#3' is not defined}%
  \fi
  \expandafter\ifx\csname sym#5fam\endcsname\relax
    \d@ffalse \errmessage{Family `#5' is not defined}%
  \fi
  \ifd@f
    \edef\f@mly{\expandafter\hexnumber{\csname sym#3fam\endcsname}}%
    \edef\f@mlytw@{\expandafter\hexnumber{\csname sym#5fam\endcsname}}%
    \xdef#1{\delimiter"#2\f@mly #4\f@mlytw@ #6\relax}%
  \fi
}


\def\setboxz@h{\setbox\z@\hbox}
\def\wdz@{\wd\z@}
\def\boxz@{\box\z@}
\def\setbox@ne{\setbox\@ne}
\def\wd@ne{\wd\@ne}

\def\math@atom#1#2{%
   \binrel@{#1}\binrel@@{#2}}
\def\binrel@#1{\setboxz@h{\thinmuskip0mu
  \medmuskip\m@ne mu\thickmuskip\@ne mu$#1\m@th$}%
 \setbox@ne\hbox{\thinmuskip0mu\medmuskip\m@ne mu\thickmuskip
  \@ne mu${}#1{}\m@th$}%
 \setbox\tw@\hbox{\hskip\wd@ne\hskip-\wdz@}}
\def\binrel@@#1{\ifdim\wd2<\z@\mathbin{#1}\else\ifdim\wd\tw@>\z@
 \mathrel{#1}\else{#1}\fi\fi}

\def\m@thit{1}

\def\set@skchar#1{\global\expandafter\skewchar
  \csname\fontn@me\endcsname=#1\relax}

\def\NewMathAlphabet#1#2#3{%
  \def\tst{#3}%
  \ifx\tst\empty\else 
    \expandafter\gdef\csname #1@sc\endcsname{}
  \fi
  \expandafter\def\csname #1\endcsname{
    \protect\csname @#1\endcsname}%
  \expandafter\def\csname @#1\endcsname##1{
    {%
    \begingroup
      \get@font{#1}{#2}{\s@ze}%
      \@ifundefined{#1@sc}{}{\set@skchar{#3}}%
      \ass@tfont{m@thit}{\fontn@me}%
      \get@font{#1}{#2}{\ss@ze}%
      \@ifundefined{#1@sc}{}{\set@skchar{#3}}%
      \ass@sfont{m@thit}{\fontn@me}%
      \get@font{#1}{#2}{\sss@ze}%
      \@ifundefined{#1@sc}{}{\set@skchar{#3}}%
      \ass@ssfont{m@thit}{\fontn@me}%
      \math@atom{##1}{%
      \mathchoice%
        {\hbox{$\m@th\displaystyle##1$}}%
        {\hbox{$\m@th\textstyle##1$}}%
        {\hbox{$\m@th\scriptstyle##1$}}%
        {\hbox{$\m@th\scriptscriptstyle##1$}}}%
    \endgroup
    }%
  }%
}


\newif\iffirstta  \firsttatrue

\def\set@hchar#1{\global\expandafter\hyphenchar
  \csname\fontn@me\endcsname=#1\relax}

\def\NewTextAlphabet#1#2#3{%
  \iffirstta
    \global\firsttafalse
    \newfam\scratchfam
    \edef\scrt@fam{\the\allocationnumber}%
  \fi
  \def\tst{#3}%
  \ifx\tst\empty\else 
    \expandafter\gdef\csname #1@hc\endcsname{}
  \fi
  \expandafter\def\csname #1\endcsname{
    \protect\csname t@#1\endcsname}%
  \long\expandafter\def\csname t@#1\endcsname##1{
    \ifmmode
      \typeout{Warning: do not use \expandafter\string\csname #1\endcsname
        \space in math mode}\fi%
    {%
      \get@font{#1}{#2}{\s@ze}\let\t@xtfnt=\fontn@me\relax
      \@ifundefined{#1@hc}{}{\set@hchar{#3}}%
      \ass@tfont{scrt@fam}{\fontn@me}%
      \get@font{#1}{#2}{\ss@ze}%
      \@ifundefined{#1@hc}{}{\set@hchar{#3}}%
      \ass@sfont{scrt@fam}{\fontn@me}%
      \get@font{#1}{#2}{\sss@ze}%
      \@ifundefined{#1@hc}{}{\set@hchar{#3}}%
      \ass@ssfont{scrt@fam}{\fontn@me}%
      \fam\scratchfam\csname\t@xtfnt\endcsname
    ##1%
    }%
  }%
  \expandafter\def\csname #1shape
    \endcsname{\protect\csname @#1shape\endcsname}%
  \expandafter\def\csname @#1shape\endcsname{
    \ifmmode
      \typeout{Warning: do not use \expandafter\string\csname
        #1shape\endcsname \space in math mode}\fi
      \get@font{#1}{#2}{\s@ze}\let\t@xtfnt=\fontn@me\relax
      \@ifundefined{#1@hc}{}{\set@hchar{#3}}%
      \ass@tfont{scrt@fam}{\fontn@me}%
      \get@font{#1}{#2}{\ss@ze}%
      \@ifundefined{#1@hc}{}{\set@hchar{#3}}%
      \ass@sfont{scrt@fam}{\fontn@me}%
      \get@font{#1}{#2}{\sss@ze}%
      \@ifundefined{#1@hc}{}{\set@hchar{#3}}%
      \ass@ssfont{scrt@fam}{\fontn@me}%
      \fam\scratchfam\csname\t@xtfnt\endcsname
  }%
}


\ifprod@font
  \def\math@itfnt{mtmib10}
  \def\math@syfnt{mtbsy10}
\else
  \def\math@itfnt{cmmib10}
  \def\math@syfnt{cmbsy10}
\fi

\def\m@thsy{2}

\def\bmath{\protect\@bmath}
\def\@bmath#1{%
  {%
  \begingroup
    \get@font{mthit}{\math@itfnt}{\s@ze}\set@skchar{'177}%
    \ass@tfont{m@thit}{\fontn@me}%
    \get@font{mthit}{\math@itfnt}{\ss@ze}\set@skchar{'177}%
    \ass@sfont{m@thit}{\fontn@me}%
    \get@font{mthit}{\math@itfnt}{\sss@ze}\set@skchar{'177}%
    \ass@ssfont{m@thit}{\fontn@me}%
    \get@font{mthsy}{\math@syfnt}{\s@ze}\set@skchar{'60}%
    \ass@tfont{m@thsy}{\fontn@me}%
    \get@font{mthsy}{\math@syfnt}{\ss@ze}\set@skchar{'60}%
    \ass@sfont{m@thsy}{\fontn@me}%
    \get@font{mthsy}{\math@syfnt}{\sss@ze}\set@skchar{'60}%
    \ass@ssfont{m@thsy}{\fontn@me}%
    \math@atom{#1}{%
    \mathchoice%
      {\hbox{$\m@th\displaystyle#1$}}%
      {\hbox{$\m@th\textstyle#1$}}%
      {\hbox{$\m@th\scriptstyle#1$}}%
      {\hbox{$\m@th\scriptscriptstyle#1$}}}%
  \endgroup
  }%
}



\def\diameter{{\ifmmode\mathchoice
{\ooalign{\hfil\hbox{$\displaystyle/$}\hfil\crcr
{\hbox{$\displaystyle\mathchar"20D$}}}}
{\ooalign{\hfil\hbox{$\textstyle/$}\hfil\crcr
{\hbox{$\textstyle\mathchar"20D$}}}}
{\ooalign{\hfil\hbox{$\scriptstyle/$}\hfil\crcr
{\hbox{$\scriptstyle\mathchar"20D$}}}}
{\ooalign{\hfil\hbox{$\scriptscriptstyle/$}\hfil\crcr
{\hbox{$\scriptscriptstyle\mathchar"20D$}}}}
\else{\ooalign{\hfil/\hfil\crcr\mathhexbox20D}}%
\fi}}

\def\sq{\ifmmode\squareforqed\else{\unskip\nobreak\hfil
\penalty50\hskip1em\null\nobreak\hfil\squareforqed
\parfillskip=0pt\finalhyphendemerits=0\endgraf}\fi}
\def\squareforqed{\hbox{\rlap{$\sqcap$}$\sqcup$}}


\def\bbbc{{\mathchoice {\setbox0=\hbox{$\displaystyle\rm C$}\hbox{\hbox
to0pt{\kern0.4\wd0\vrule height0.9\ht0\hss}\box0}}
{\setbox0=\hbox{$\textstyle\rm C$}\hbox{\hbox
to0pt{\kern0.4\wd0\vrule height0.9\ht0\hss}\box0}}
{\setbox0=\hbox{$\scriptstyle\rm C$}\hbox{\hbox
to0pt{\kern0.4\wd0\vrule height0.9\ht0\hss}\box0}}
{\setbox0=\hbox{$\scriptscriptstyle\rm C$}\hbox{\hbox
to0pt{\kern0.4\wd0\vrule height0.9\ht0\hss}\box0}}}}
\def\bbbq{{\mathchoice {\setbox0=\hbox{$\displaystyle\rm
Q$}\hbox{\raise
0.15\ht0\hbox to0pt{\kern0.4\wd0\vrule height0.8\ht0\hss}\box0}}
{\setbox0=\hbox{$\textstyle\rm Q$}\hbox{\raise
0.15\ht0\hbox to0pt{\kern0.4\wd0\vrule height0.8\ht0\hss}\box0}}
{\setbox0=\hbox{$\scriptstyle\rm Q$}\hbox{\raise
0.15\ht0\hbox to0pt{\kern0.4\wd0\vrule height0.7\ht0\hss}\box0}}
{\setbox0=\hbox{$\scriptscriptstyle\rm Q$}\hbox{\raise
0.15\ht0\hbox to0pt{\kern0.4\wd0\vrule height0.7\ht0\hss}\box0}}}}
\def\bbbt{{\mathchoice {\setbox0=\hbox{$\displaystyle\rm
T$}\hbox{\hbox to0pt{\kern0.3\wd0\vrule height0.9\ht0\hss}\box0}}
{\setbox0=\hbox{$\textstyle\rm T$}\hbox{\hbox
to0pt{\kern0.3\wd0\vrule height0.9\ht0\hss}\box0}}
{\setbox0=\hbox{$\scriptstyle\rm T$}\hbox{\hbox
to0pt{\kern0.3\wd0\vrule height0.9\ht0\hss}\box0}}
{\setbox0=\hbox{$\scriptscriptstyle\rm T$}\hbox{\hbox
to0pt{\kern0.3\wd0\vrule height0.9\ht0\hss}\box0}}}}
\def\bbbs{{\mathchoice
{\setbox0=\hbox{$\displaystyle     \rm S$}\hbox{\raise0.5\ht0\hbox
to0pt{\kern0.35\wd0\vrule height0.45\ht0\hss}\hbox
to0pt{\kern0.55\wd0\vrule height0.5\ht0\hss}\box0}}
{\setbox0=\hbox{$\textstyle        \rm S$}\hbox{\raise0.5\ht0\hbox
to0pt{\kern0.35\wd0\vrule height0.45\ht0\hss}\hbox
to0pt{\kern0.55\wd0\vrule height0.5\ht0\hss}\box0}}
{\setbox0=\hbox{$\scriptstyle      \rm S$}\hbox{\raise0.5\ht0\hbox
to0pt{\kern0.35\wd0\vrule height0.45\ht0\hss}\raise0.05\ht0\hbox
to0pt{\kern0.5\wd0\vrule height0.45\ht0\hss}\box0}}
{\setbox0=\hbox{$\scriptscriptstyle\rm S$}\hbox{\raise0.5\ht0\hbox
to0pt{\kern0.4\wd0\vrule height0.45\ht0\hss}\raise0.05\ht0\hbox
to0pt{\kern0.55\wd0\vrule height0.45\ht0\hss}\box0}}}}
\def\bbbz{{\mathchoice {\hbox{$\sf\textstyle Z\kern-0.4em Z$}}
{\hbox{$\sf\textstyle Z\kern-0.4em Z$}}
{\hbox{$\sf\scriptstyle Z\kern-0.3em Z$}}
{\hbox{$\sf\scriptscriptstyle Z\kern-0.2em Z$}}}}


\def\Nulle{0} 
\def\Afe{1}   
\def\Hae{2}   
\def\Hbe{3}   
\def\Hce{4}   
\def\Hde{5}   


\newcount\LastMac       \LastMac=\Nulle

\newskip\half      \half=5.5pt plus 1.5pt minus 2.25pt
\newskip\one       \one=11pt plus 3pt minus 5.5pt
\newskip\onehalf   \onehalf=16.5pt plus 5.5pt minus 8.25pt
\newskip\two       \two=22pt plus 5.5pt minus 11pt

\def\Half{\addvspace{\half}}
\def\One{\addvspace{\one}}
\def\OneHalf{\addvspace{\onehalf}}
\def\Two{\addvspace{\two}}

\def\Raggedright{
  \rightskip=\z@ plus \hsize\relax
}

\def\Fullout{
  \rightskip=\z@\relax
}

\def\Hang#1#2{
  \hangindent=#1%
  \hangafter=#2\relax
}


\newif\ifsp@page
\def\pagestyle#1{\csname ps@#1\endcsname}
\def\thispagestyle#1{\global\sp@pagetrue\gdef\sp@type{#1}}

\def\ps@titlepage{%
  \def\@oddhead{\eightpoint\noindent \the\CatchLine
    \ifprod@font\else\qquad Printed\ \today\qquad
      (MN plain \TeX\ macros\ v\@version)\fi \hfil}%
  \let\@evenhead=\@oddhead
  \def\@oddfoot{\eightpoint\copyright\ \@pubyear\ RAS\hfil}%
  \def\@evenfoot{\hfil\eightpoint\noindent\copyright\ \@pubyear\ RAS}%
}

\def\ps@headings{%
  \def\@oddhead{\elevenpoint\it\noindent
    \hfill\the\RightHeader\hskip1.5em\rm\folio}%
  \def\@evenhead{\elevenpoint\noindent
    \folio\hskip1.5em\it\the\LeftHeader\hfill}%
  \def\@oddfoot{\eightpoint\noindent\copyright\ \@pubyear\ RAS,
    MNRAS {\bf \@volume}, \@pagerange\hfil}%
  \def\@evenfoot{\hfil\eightpoint\copyright\ \@pubyear\ RAS,
    MNRAS {\bf \@volume}, \@pagerange}%
}

\def\ps@plate{%
  \def\@oddhead{\eightpoint\noindent\plt@cap\hfil}%
  \def\@evenhead{\eightpoint\noindent\plt@cap\hfil}%
  \def\@oddfoot{\eightpoint\noindent\copyright\ \@pubyear\ RAS,
    MNRAS {\bf \@volume}, \@pagerange\hfil}%
  \def\@evenfoot{\hfil\eightpoint\copyright\ \@pubyear\ RAS,
    MNRAS {\bf \@volume}, \@pagerange}%
}



\def\title#1{
  \bgroup
    \vbox to 8pt{\vss}%
    \seventeenpoint
    \Raggedright
    \noindent \strut{\bf #1}\par
  \egroup
}

\def\author#1{
  \bgroup
    \ifnum\LastMac=\Afe \OneHalf\else \vskip 21pt\fi
    \fourteenpoint
    \Raggedright
    \noindent \strut #1\par
    \vskip 3pt%
  \egroup
}

\def\affiliation#1{
  \bgroup
    \vskip -4pt%
    \eightpoint
    \Raggedright
    \noindent \strut {\it #1}\par
  \egroup
  \LastMac=\Afe\relax
}

\def\acceptedline#1{
  \bgroup
    \Two
    \eightpoint
    \Raggedright
    \noindent \strut #1\par
  \egroup
}

\long\def\abstract#1{%
  \bgroup
    \vskip 20pt%
    \leftskip 11pc\rightskip\z@
    \noindent{\ninebf ABSTRACT}\par
    \tenpoint
    \Fullout
    \noindent #1\par
  \egroup
}

\long\def\keywords#1{
  \bgroup
    \Half
    \leftskip 11pc\rightskip\z@
    \tenpoint
    \Fullout
    \noindent\hbox{\bf Key words:}\ #1\par
  \egroup
}


\def\maketitle{%
  \EndOpening
  \ifsinglecol \else \MakePage\fi
}


\def\pageoffset#1#2{\hoffset=#1\relax\voffset=#2\relax}


\def\@nameuse#1{\csname #1\endcsname}
\def\arabic#1{\@arabic{\@nameuse{#1}}}
\def\alph#1{\@alph{\@nameuse{#1}}}
\def\Alph#1{\@Alph{\@nameuse{#1}}}
\def\@arabic#1{\number #1}
\def\@Alph#1{\ifcase#1\or A\or B\or C\or D\else\@Ialph{#1}\fi}
\def\@Ialph#1{\ifcase#1\or \or \or \or \or E\or F\or G\or H\or I\or J\or
   K\or L\or M\or N\or O\or P\or Q\or R\or S\or T\or U\or V\or W\or X\or
   Y\or Z\else\errmessage{Counter out of range}\fi}
\def\@alph#1{\ifcase#1\or a\or b\or c\or d\else\@ialph{#1}\fi}
\def\@ialph#1{\ifcase#1\or \or \or \or \or e\or f\or g\or h\or i\or j\or
   k\or l\or m\or n\or o\or p\or q\or r\or s\or t\or u\or v\or w\or x\or y\or
   z\else\errmessage{Counter out of range}\fi}


\newcount\Eqnno
\newcount\SubEqnno

\def\theeq{\arabic{Eqnno}}
\def\thesubeq{\alph{SubEqnno}}

\def\stepeq{\relax
  \global\SubEqnno \z@
  \global\advance\Eqnno \@ne\relax
  {\rm (\theeq)}%
}

\def\startsubeq{\relax
  \global\SubEqnno \z@
  \global\advance\Eqnno \@ne\relax
  \stepsubeq
}

\def\stepsubeq{\relax
  \global\advance\SubEqnno \@ne\relax
  {\rm (\theeq\thesubeq)}%
}


\newcount\Sec        
\newcount\SecSec
\newcount\SecSecSec

\def\thesection{\arabic{Sec}}
\def\thesubsection{\thesection.\arabic{SecSec}}
\def\thesubsubsection{\thesubsection.\arabic{SecSecSec}}

\Sec=\z@

\def\:{\let\@sptoken= } \:  
\def\:{\@xifnch} \expandafter\def\: {\futurelet\@tempc\@ifnch}

\def\@ifnextchar#1#2#3{%
  \let\@tempMACe #1%
  \def\@tempMACa{#2}%
  \def\@tempMACb{#3}%
  \futurelet \@tempMACc\@ifnch%
}

\def\@ifnch{%
\ifx \@tempMACc \@sptoken%
  \let\@tempMACd\@xifnch%
\else%
  \ifx \@tempMACc \@tempMACe%
    \let\@tempMACd\@tempMACa%
  \else%
    \let\@tempMACd\@tempMACb%
  \fi%
\fi%
\@tempMACd%
}

\def\@ifstar#1#2{\@ifnextchar *{\def\@tempMACa*{#1}\@tempMACa}{#2}}

\newskip\@tempskipb

\def\addvspace#1{%
  \ifvmode\else \endgraf\fi%
  \ifdim\lastskip=\z@%
    \vskip #1\relax%
  \else%
    \@tempskipb#1\relax\@xaddvskip%
  \fi%
}

\def\@xaddvskip{%
  \ifdim\lastskip<\@tempskipb%
    \vskip-\lastskip%
    \vskip\@tempskipb\relax%
  \else%
    \ifdim\@tempskipb<\z@%
      \ifdim\lastskip<\z@ \else%
        \advance\@tempskipb\lastskip%
        \vskip-\lastskip\vskip\@tempskipb%
      \fi%
    \fi%
  \fi%
}

\newskip\@tmpSKIP

\def\addpen#1{%
  \ifvmode
    \if@nobreak
    \else
      \ifdim\lastskip=\z@
        \penalty#1\relax
      \else
        \@tmpSKIP=\lastskip
        \vskip -\lastskip
        \penalty#1\vskip\@tmpSKIP
      \fi
    \fi
  \fi
}

\newcount\@clubpen   \@clubpen=\clubpenalty
\newif\if@nobreak    \@nobreakfalse

\def\@noafterindent{%
  \global\@nobreaktrue
  \everypar{\if@nobreak
              \global\@nobreakfalse
              \clubpenalty \@M
              {\setbox\z@\lastbox}%
              \LastMac=\Nulle\relax%
            \else
              \clubpenalty \@clubpen
              \everypar{}%
            \fi}%
}

\newcount\gds@cbrk   \gds@cbrk=-300

\def\@nohdbrk{\interlinepenalty \@M\relax}

\let\@par=\par
\def\@restorepar{\def\par{\@par}}

\newif\if@endpe   \@endpefalse
 
\def\@doendpe{\@endpetrue \@nobreakfalse \LastMac=\Nulle\relax%
     \def\par{\@restorepar\everypar{}\par\@endpefalse}%
              \everypar{\setbox\z@\lastbox\everypar{}\@endpefalse}%
}

\def\section{\@ifstar{\@ssection}{\@section}}

\def\@section#1{
  \if@nobreak
    \everypar{}%
    \ifnum\LastMac=\Hae \addvspace{\half}\fi
  \else
    \addpen{\gds@cbrk}%
    \addvspace{\two}%
  \fi
  \bgroup
    \ninepoint\bf
    \Raggedright
    \global\advance\Sec \@ne
    \ifappendix
      \global\Eqnno=\z@ \global\SubEqnno=\z@\relax
      \def\ch@ck{#1}%
      \ifx\ch@ck\empty \def\c@lon{}\else\def\c@lon{:}\fi
      \noindent\@nohdbrk APPENDIX\ \thesection\c@lon\hskip 0.5em%
        \uppercase{#1}\par
    \else
      \noindent\@nohdbrk\thesection\hskip 1pc \uppercase{#1}\par
    \fi
    \global\SecSec=\z@
  \egroup
  \nobreak
  \vskip\half
  \nobreak
  \@noafterindent
  \LastMac=\Hae\relax
}

\def\@ssection#1{
  \if@nobreak
    \everypar{}%
    \ifnum\LastMac=\Hae \addvspace{\half}\fi
  \else
    \addpen{\gds@cbrk}%
    \addvspace{\two}%
  \fi
  \bgroup
    \ninepoint\bf
    \Raggedright
    \noindent\@nohdbrk\uppercase{#1}\par
  \egroup
  \nobreak
  \vskip\half
  \nobreak
  \@noafterindent
  \LastMac=\Hae\relax
}

\def\subsection{\@ifstar{\@ssubsection}{\@subsection}}

\def\@subsection#1{
  \if@nobreak
    \everypar{}%
    \ifnum\LastMac=\Hae \addvspace{1pt plus 1pt minus .5pt}\fi
  \else
    \addpen{\gds@cbrk}%
    \addvspace{\onehalf}%
  \fi
  \bgroup
    \ninepoint\bf
    \Raggedright
    \global\advance\SecSec \@ne
    \noindent\@nohdbrk\thesubsection \hskip 1pc\relax #1\par
    \global\SecSecSec=\z@
  \egroup
  \nobreak
  \vskip\half
  \nobreak
  \@noafterindent
  \LastMac=\Hbe\relax
}

\def\@ssubsection#1{
  \if@nobreak
    \everypar{}%
    \ifnum\LastMac=\Hae \addvspace{1pt plus 1pt minus .5pt}\fi
  \else
    \addpen{\gds@cbrk}%
    \addvspace{\onehalf}%
  \fi
  \bgroup
    \ninepoint\bf
    \Raggedright
    \noindent\@nohdbrk #1\par
  \egroup
  \nobreak
  \vskip\half
  \nobreak
  \@noafterindent
  \LastMac=\Hbe\relax
}

\def\subsubsection{\@ifstar{\@ssubsubsection}{\@subsubsection}}

\def\@subsubsection#1{
  \if@nobreak
    \everypar{}%
    \ifnum\LastMac=\Hbe \addvspace{1pt plus 1pt minus .5pt}\fi
  \else
    \addpen{\gds@cbrk}%
    \addvspace{\onehalf}%
  \fi
  \bgroup
    \ninepoint\it
    \Raggedright
    \global\advance\SecSecSec \@ne
    \noindent\@nohdbrk\thesubsubsection \hskip 1pc\relax #1\par
  \egroup
  \nobreak
  \vskip\half
  \nobreak
  \@noafterindent
  \LastMac=\Hce\relax
}

\def\@ssubsubsection#1{
  \if@nobreak
    \everypar{}%
    \ifnum\LastMac=\Hbe \addvspace{1pt plus 1pt minus .5pt}\fi
  \else
    \addpen{\gds@cbrk}%
    \addvspace{\onehalf}%
  \fi
  \bgroup
    \ninepoint\it
    \Raggedright
    \noindent\@nohdbrk #1\par
  \egroup
  \nobreak
  \vskip\half
  \nobreak
  \@noafterindent
  \LastMac=\Hce\relax
}

\def\paragraph#1{
  \if@nobreak
    \everypar{}%
  \else
    \addpen{\gds@cbrk}%
    \addvspace{\one}%
  \fi%
  \bgroup%
    \ninepoint\it
    \noindent #1\ \nobreak%
  \egroup
  \LastMac=\Hde\relax
  \ignorespaces
}


\newif\ifappendix

\def\appendix{%
  \global\appendixtrue
  \def\thesection{\Alph{Sec}}%
  \def\thesubsection{\thesection\arabic{SecSec}}%
  \def\theeq{\thesection\arabic{Eqnno}}%
  \Sec=\z@ \SecSec=\z@ \SecSecSec=\z@ \Eqnno=\z@ \SubEqnno=\z@\relax
}




\def\beginlist{%
  \par\if@nobreak \else\addvspace{\half}\fi%
  \bgroup%
    \ninepoint
    \let\item=\list@item%
}

\def\list@item{%
  \par\noindent\hskip 1em\relax%
  \ignorespaces%
}

\def\endlist{\par\egroup\addvspace{\half}\@doendpe}


\def\beginrefs{%
  \par
  \bgroup
    \eightpoint
    \Fullout
    \let\bibitem=\bib@item
}

\def\bib@item{%
  \par\parindent=1.5em\Hang{1.5em}{1}%
  \everypar={\Hang{1.5em}{1}\ignorespaces}%
  \noindent\ignorespaces
}

\def\endrefs{\par\egroup\@doendpe}


\newtoks\CatchLine

\def\@journal{Mon.\ Not.\ R.\ Astron.\ Soc.\ }  
\def\@pubyear{1994}        
\def\@pagerange{000--000}  
\def\@volume{000}          
\def\@microfiche{}         %

\def\pubyear#1{\gdef\@pubyear{#1}\@makecatchline}
\def\pagerange#1{\gdef\@pagerange{#1}\@makecatchline}
\def\volume#1{\gdef\@volume{#1}\@makecatchline}
\def\microfiche#1{\gdef\@microfiche{and Microfiche\ #1}\@makecatchline}

\def\@makecatchline{%
  \global\CatchLine{%
    {\rm \@journal {\bf \@volume},\ \@pagerange\ (\@pubyear)\ \@microfiche}}%
}

\@makecatchline 

\newtoks\LeftHeader
\def\shortauthor#1{
  \global\LeftHeader{#1}%
}

\newtoks\RightHeader
\def\shorttitle#1{
  \global\RightHeader{#1}%
}

\def\PageHead{
  \begingroup
    \ifsp@page
      \csname ps@\sp@type\endcsname
    \fi
    \ifodd\pageno
      \let\the@head=\@oddhead
    \else
      \let\the@head=\@evenhead
    \fi
    \vbox to \z@{\vskip-22.5\p@%
      \hbox to \PageWidth{\vbox to8.5\p@{}%
        \the@head
      }%
    \vss}%
  \endgroup
  \nointerlineskip
}

\gdef\PageFoot{%
  \nointerlineskip%
  \begingroup
  \ifsp@page
    \csname ps@\sp@type\endcsname
    \global\sp@pagefalse
  \fi
  \vbox to 22pt{\vfil%
    \hbox to \PageWidth{%
      \eightpoint\strut\noindent
      \ifodd\pageno
        \@oddfoot
      \else
        \@evenfoot
      \fi
    }%
  }%
  \endgroup
}

\def\today{%
  \number\day\space
  \ifcase\month\or January\or February\or March\or April\or May\or June\or
    July\or August\or September\or October\or November\or December\fi
  \space\number\year%
}

\def\authorcomment#1{%
  \gdef\PageFoot{%
    \nointerlineskip%
    \vbox to 20pt{\vfil%
      \hbox to \PageWidth{\elevenpoint\noindent \hfil #1 \hfil}}%
  }%
}


\newif\ifplate@page
\newbox\plt@box

\def\beginplatepage{%
  \let\plate=\plate@head
  \let\caption=\fig@caption
  \global\setbox\plt@box=\vbox\bgroup
  \TEMPDIMEN=\PageWidth 
  \hsize=\PageWidth\relax
}

\def\endplatepage{\par\egroup\global\plate@pagetrue}
\def\plate@head#1{\gdef\plt@cap{#1}}


\def\letters{%
  \gdef\folio{\ifnum\pageno<\z@ L\romannumeral-\pageno
    \else L\number\pageno \fi}%
}


\newdimen\mathindent

\global\mathindent=\z@
\global\everydisplay{\global\@dspwd=\displaywidth\displaysetup}


\def\@displaylines#1{
  {}$\displ@y\hbox{\vbox{\halign{$\@lign\hfil\displaystyle##\hfil$\crcr
  #1\crcr}}}${}%
}

\def\@eqalign#1{\null\vcenter{\openup\jot\m@th
  \ialign{\strut\hfil$\displaystyle{##}$&$\displaystyle{{}##}$\hfil
      \crcr#1\crcr}}%
}

\def\@eqalignno#1{
  \global\advance\@dspwd by -\mathindent%
  {}$\displ@y\hbox{\vbox{\halign to\@dspwd%
  {\hfil$\@lign\displaystyle{##}$\tabskip\z@skip
  &$\@lign\displaystyle{{}##}$\hfil\tabskip\centering
  &\llap{$\@lign##$}\tabskip\z@skip\crcr
  #1\crcr}}}${}%
}


\global\let\displaylines=\@displaylines
\global\let\eqalign=\@eqalign
\global\let\eqalignno=\@eqalignno
\global\let\leqalignno=\@eqalignno

\newdimen\@dspwd   \@dspwd=\z@
\newif\if@eqno
\newif\if@leqno
\newtoks\@eqn
\newtoks\@eq

\def\displaysetup#1$${\displaytest#1\eqno\eqno\displaytest}

\def\displaytest#1\eqno#2\eqno#3\displaytest{%
 \if!#3!\ldisplaytest#1\leqno\leqno\ldisplaytest
 \else\@eqnotrue\@leqnofalse\@eqn={#2}\@eq={#1}\fi
 \generaldisplay$$}

\def\ldisplaytest#1\leqno#2\leqno#3\ldisplaytest{%
\@eq={#1}%
 \if!#3!\@eqnofalse\else\@eqnotrue\@leqnotrue
  \@eqn={#2}\fi}

\def\generaldisplay{%
  \if@eqno
    \if@leqno
      \hbox to \displaywidth{\noindent
        \rlap{$\displaystyle\the\@eqn$}%
        \hskip\mathindent$\displaystyle\the\@eq$\hfil}%
    \else
      \hbox to \displaywidth{\noindent
        \hskip\mathindent
        $\displaystyle\the\@eq$\hfil$\displaystyle\the\@eqn$}%
    \fi
  \else
    \hbox to \displaywidth{\noindent
      \hskip\mathindent$\displaystyle\the\@eq$\hfil}%
  \fi
}


\def\@notice{%
  \par\Two%
  \noindent{\b@ls{11pt}\ninerm This paper has been produced using the
    Royal Astronomical Society/Blackwell Science \TeX\ macros.\par}%
}

\outer\def\bye{\@notice\par\vfill\supereject\end}


\def\start@mess{%
  Monthly notices of the RAS journal style (\@typeface)\space
    v\@version,\space \@verdate.%
}

\everyjob{\Warn{\start@mess}}



\newif\if@debug \@debugfalse  

\def\Print#1{\if@debug\immediate\write16{#1}\else \fi}
\def\Warn#1{\immediate\write16{#1}}
\def\wlog#1{}

\newcount\Iteration 

\def\Single{0} \def\Double{1}                 
\def\Figure{0} \def\Table{1}                  

\def\InStack{0}  
\def\InZoneA{1}
\def\InZoneB{2}
\def\InZoneC{3}

\newcount\TEMPCOUNT 
\newdimen\TEMPDIMEN 
\newbox\TEMPBOX     
\newbox\VOIDBOX     

\newcount\LengthOfStack 
\newcount\MaxItems      
\newcount\StackPointer
\newcount\Point         
\newcount\NextFigure    
\newcount\NextTable     
\newcount\NextItem      

\newcount\StatusStack   
\newcount\NumStack      
\newcount\TypeStack     
\newcount\SpanStack     
\newcount\BoxStack      

\newcount\ItemSTATUS    
\newcount\ItemNUMBER    
\newcount\ItemTYPE      
\newcount\ItemSPAN      
\newbox\ItemBOX         
\newdimen\ItemSIZE      

\newdimen\PageHeight    
\newdimen\TextLeading   
\newdimen\Feathering    
\newcount\LinesPerPage  
\newdimen\ColumnWidth   
\newdimen\ColumnGap     
\newdimen\PageWidth     
\newdimen\BodgeHeight   
\newcount\Leading       

\newdimen\ZoneBSize  
\newdimen\TextSize   
\newbox\ZoneABOX     
\newbox\ZoneBBOX     
\newbox\ZoneCBOX     

\newif\ifFirstSingleItem
\newif\ifFirstZoneA
\newif\ifMakePageInComplete
\newif\ifMoreFigures \MoreFiguresfalse 
\newif\ifMoreTables  \MoreTablesfalse  

\newif\ifFigInZoneB 
\newif\ifFigInZoneC 
\newif\ifTabInZoneB 
\newif\ifTabInZoneC

\newif\ifZoneAFullPage

\newbox\MidBOX    
\newbox\LeftBOX
\newbox\RightBOX
\newbox\PageBOX   

\newif\ifLeftCOL  
\LeftCOLtrue

\newdimen\ZoneBAdjust

\newcount\ItemFits
\def\Yes{1}
\def\No{2}


\MaxItems=15
\NextFigure=\z@        
\NextTable=\@ne

\BodgeHeight=6pt
\TextLeading=11pt    
\Leading=11
\Feathering=\z@      
\LinesPerPage=61     
\topskip=\TextLeading
\ColumnWidth=20pc    
\ColumnGap=2pc       

\newskip\ItemSepamount  
\ItemSepamount=\TextLeading plus \TextLeading minus 4pt

\parskip=\z@ plus .1pt
\parindent=18pt
\widowpenalty=\z@
\clubpenalty=10000
\tolerance=1500
\hbadness=1500
\abovedisplayskip=6pt plus 2pt minus 1pt
\belowdisplayskip=6pt plus 2pt minus 1pt
\abovedisplayshortskip=6pt plus 2pt minus 1pt
\belowdisplayshortskip=6pt plus 2pt minus 1pt

\frenchspacing

\ninepoint 

\PageHeight=682pt
\PageWidth=2\ColumnWidth
\advance\PageWidth by \ColumnGap

\pagestyle{headings}




\newcount\DUMMY \StatusStack=\allocationnumber
\newcount\DUMMY \newcount\DUMMY \newcount\DUMMY 
\newcount\DUMMY \newcount\DUMMY \newcount\DUMMY 
\newcount\DUMMY \newcount\DUMMY \newcount\DUMMY
\newcount\DUMMY \newcount\DUMMY \newcount\DUMMY 
\newcount\DUMMY \newcount\DUMMY \newcount\DUMMY

\newcount\DUMMY \NumStack=\allocationnumber
\newcount\DUMMY \newcount\DUMMY \newcount\DUMMY 
\newcount\DUMMY \newcount\DUMMY \newcount\DUMMY 
\newcount\DUMMY \newcount\DUMMY \newcount\DUMMY 
\newcount\DUMMY \newcount\DUMMY \newcount\DUMMY 
\newcount\DUMMY \newcount\DUMMY \newcount\DUMMY

\newcount\DUMMY \TypeStack=\allocationnumber
\newcount\DUMMY \newcount\DUMMY \newcount\DUMMY 
\newcount\DUMMY \newcount\DUMMY \newcount\DUMMY 
\newcount\DUMMY \newcount\DUMMY \newcount\DUMMY 
\newcount\DUMMY \newcount\DUMMY \newcount\DUMMY 
\newcount\DUMMY \newcount\DUMMY \newcount\DUMMY

\newcount\DUMMY \SpanStack=\allocationnumber
\newcount\DUMMY \newcount\DUMMY \newcount\DUMMY 
\newcount\DUMMY \newcount\DUMMY \newcount\DUMMY 
\newcount\DUMMY \newcount\DUMMY \newcount\DUMMY 
\newcount\DUMMY \newcount\DUMMY \newcount\DUMMY 
\newcount\DUMMY \newcount\DUMMY \newcount\DUMMY

\newbox\DUMMY   \BoxStack=\allocationnumber
\newbox\DUMMY   \newbox\DUMMY \newbox\DUMMY 
\newbox\DUMMY   \newbox\DUMMY \newbox\DUMMY 
\newbox\DUMMY   \newbox\DUMMY \newbox\DUMMY 
\newbox\DUMMY   \newbox\DUMMY \newbox\DUMMY 
\newbox\DUMMY   \newbox\DUMMY \newbox\DUMMY

\def\wlog{\immediate\write\m@ne}


\def\GetItemAll#1{%
 \GetItemSTATUS{#1}
 \GetItemNUMBER{#1}
 \GetItemTYPE{#1}
 \GetItemSPAN{#1}
 \GetItemBOX{#1}
}

\def\GetItemSTATUS#1{%
 \Point=\StatusStack
 \advance\Point by #1
 \global\ItemSTATUS=\count\Point
}

\def\GetItemNUMBER#1{%
 \Point=\NumStack
 \advance\Point by #1
 \global\ItemNUMBER=\count\Point
}

\def\GetItemTYPE#1{%
 \Point=\TypeStack
 \advance\Point by #1
 \global\ItemTYPE=\count\Point
}

\def\GetItemSPAN#1{%
 \Point\SpanStack
 \advance\Point by #1
 \global\ItemSPAN=\count\Point
}

\def\GetItemBOX#1{%
 \Point=\BoxStack
 \advance\Point by #1
 \global\setbox\ItemBOX=\vbox{\copy\Point}
 \global\ItemSIZE=\ht\ItemBOX
 \global\advance\ItemSIZE by \dp\ItemBOX
 \TEMPCOUNT=\ItemSIZE
 \divide\TEMPCOUNT by \Leading
 \divide\TEMPCOUNT by 65536
 \advance\TEMPCOUNT \@ne
 \ItemSIZE=\TEMPCOUNT pt
 \global\multiply\ItemSIZE by \Leading
}


\def\JoinStack{%
 \ifnum\LengthOfStack=\MaxItems 
  \Warn{WARNING: Stack is full...some items will be lost!}
 \else
  \Point=\StatusStack
  \advance\Point by \LengthOfStack
  \global\count\Point=\ItemSTATUS
  \Point=\NumStack
  \advance\Point by \LengthOfStack
  \global\count\Point=\ItemNUMBER
  \Point=\TypeStack
  \advance\Point by \LengthOfStack
  \global\count\Point=\ItemTYPE
  \Point\SpanStack
  \advance\Point by \LengthOfStack
  \global\count\Point=\ItemSPAN
  \Point=\BoxStack
  \advance\Point by \LengthOfStack
  \global\setbox\Point=\vbox{\copy\ItemBOX}
  \global\advance\LengthOfStack \@ne
  \ifnum\ItemTYPE=\Figure 
   \global\MoreFigurestrue
  \else
   \global\MoreTablestrue
  \fi
 \fi
}


\def\LeaveStack#1{%
 {\Iteration=#1
 \loop
 \ifnum\Iteration<\LengthOfStack
  \advance\Iteration \@ne
  \GetItemSTATUS{\Iteration}
   \advance\Point by \m@ne
   \global\count\Point=\ItemSTATUS
  \GetItemNUMBER{\Iteration}
   \advance\Point by \m@ne
   \global\count\Point=\ItemNUMBER
  \GetItemTYPE{\Iteration}
   \advance\Point by \m@ne
   \global\count\Point=\ItemTYPE
  \GetItemSPAN{\Iteration}
   \advance\Point by \m@ne
   \global\count\Point=\ItemSPAN
  \GetItemBOX{\Iteration}
   \advance\Point by \m@ne
   \global\setbox\Point=\vbox{\copy\ItemBOX}
 \repeat}
 \global\advance\LengthOfStack by \m@ne
}


\newif\ifStackNotClean

\def\CleanStack{%
 \StackNotCleantrue
 {\Iteration=\z@
  \loop
   \ifStackNotClean
    \GetItemSTATUS{\Iteration}
    \ifnum\ItemSTATUS=\InStack
     \advance\Iteration \@ne
     \else
      \LeaveStack{\Iteration}
    \fi
   \ifnum\LengthOfStack<\Iteration
    \StackNotCleanfalse
   \fi
 \repeat}
}


\def\FindItem#1#2{%
 \global\StackPointer=\m@ne 
 {\Iteration=\z@
  \loop
  \ifnum\Iteration<\LengthOfStack
   \GetItemSTATUS{\Iteration}
   \ifnum\ItemSTATUS=\InStack
    \GetItemTYPE{\Iteration}
    \ifnum\ItemTYPE=#1
     \GetItemNUMBER{\Iteration}
     \ifnum\ItemNUMBER=#2
      \global\StackPointer=\Iteration
      \Iteration=\LengthOfStack 
     \fi
    \fi
   \fi
  \advance\Iteration \@ne
 \repeat}
}


\def\FindNext{%
 \global\StackPointer=\m@ne 
 {\Iteration=\z@
  \loop
  \ifnum\Iteration<\LengthOfStack
   \GetItemSTATUS{\Iteration}
   \ifnum\ItemSTATUS=\InStack
    \GetItemTYPE{\Iteration}
   \ifnum\ItemTYPE=\Figure
    \ifMoreFigures
      \global\NextItem=\Figure
      \global\StackPointer=\Iteration
      \Iteration=\LengthOfStack 
    \fi
   \fi
   \ifnum\ItemTYPE=\Table
    \ifMoreTables
      \global\NextItem=\Table
      \global\StackPointer=\Iteration
      \Iteration=\LengthOfStack 
    \fi
   \fi
  \fi
  \advance\Iteration \@ne
 \repeat}
}


\def\ChangeStatus#1#2{%
 \Point=\StatusStack
 \advance\Point by #1
 \global\count\Point=#2
}



\def\Zone{\InZoneA}

\ZoneBAdjust=\z@

\def\MakePage{
 \global\ZoneBSize=\PageHeight
 \global\TextSize=\ZoneBSize
 \global\ZoneAFullPagefalse
 \global\topskip=\TextLeading
 \MakePageInCompletetrue
 \MoreFigurestrue
 \MoreTablestrue
 \FigInZoneBfalse
 \FigInZoneCfalse
 \TabInZoneBfalse
 \TabInZoneCfalse
 \global\FirstSingleItemtrue
 \global\FirstZoneAtrue
 \global\setbox\ZoneABOX=\box\VOIDBOX
 \global\setbox\ZoneBBOX=\box\VOIDBOX
 \global\setbox\ZoneCBOX=\box\VOIDBOX
 \loop
  \ifMakePageInComplete
 \FindNext
 \ifnum\StackPointer=\m@ne
  \NextItem=\m@ne
  \MoreFiguresfalse
  \MoreTablesfalse
 \fi
 \ifnum\NextItem=\Figure
   \FindItem{\Figure}{\NextFigure}
   \ifnum\StackPointer=\m@ne \global\MoreFiguresfalse
   \else
    \GetItemSPAN{\StackPointer}
    \ifnum\ItemSPAN=\Single \def\Zone{\InZoneB}\relax
     \ifFigInZoneC \global\MoreFiguresfalse\fi
    \else
     \def\Zone{\InZoneA}
     \ifFigInZoneB \def\Zone{\InZoneC}\fi
    \fi
   \fi
   \ifMoreFigures\Print{}\FigureItems\fi
 \fi
\ifnum\NextItem=\Table
   \FindItem{\Table}{\NextTable}
   \ifnum\StackPointer=\m@ne \global\MoreTablesfalse
   \else
    \GetItemSPAN{\StackPointer}
    \ifnum\ItemSPAN=\Single\relax
     \ifTabInZoneC \global\MoreTablesfalse\fi
    \else
     \def\Zone{\InZoneA}
     \ifTabInZoneB \def\Zone{\InZoneC}\fi
    \fi
   \fi
   \ifMoreTables\Print{}\TableItems\fi
 \fi
   \MakePageInCompletefalse 
   \ifMoreFigures\MakePageInCompletetrue\fi
   \ifMoreTables\MakePageInCompletetrue\fi
 \repeat
 \ifZoneAFullPage
  \global\TextSize=\z@
  \global\ZoneBSize=\z@
  \global\vsize=\z@\relax
  \global\topskip=\z@\relax
  \vbox to \z@{\vss}
  \eject
 \else
 \global\advance\ZoneBSize by -\ZoneBAdjust
 \global\vsize=\ZoneBSize
 \global\hsize=\ColumnWidth
 \global\ZoneBAdjust=\z@
 \ifdim\TextSize<23pt
 \Warn{}
 \Warn{* Making column fall short: TextSize=\the\TextSize *}
 \vskip-\lastskip\eject\fi
 \fi
}

\def\MakeRightCol{
 \global\TextSize=\ZoneBSize
 \MakePageInCompletetrue
 \MoreFigurestrue
 \MoreTablestrue
 \global\FirstSingleItemtrue
 \global\setbox\ZoneBBOX=\box\VOIDBOX
 \def\Zone{\InZoneB}
 \loop
  \ifMakePageInComplete
 \FindNext
 \ifnum\StackPointer=\m@ne
  \NextItem=\m@ne
  \MoreFiguresfalse
  \MoreTablesfalse
 \fi
 \ifnum\NextItem=\Figure
   \FindItem{\Figure}{\NextFigure}
   \ifnum\StackPointer=\m@ne \MoreFiguresfalse
   \else
    \GetItemSPAN{\StackPointer}
    \ifnum\ItemSPAN=\Double\relax
     \MoreFiguresfalse\fi
   \fi
   \ifMoreFigures\Print{}\FigureItems\fi
 \fi
 \ifnum\NextItem=\Table
   \FindItem{\Table}{\NextTable}
   \ifnum\StackPointer=\m@ne \MoreTablesfalse
   \else
    \GetItemSPAN{\StackPointer}
    \ifnum\ItemSPAN=\Double\relax
     \MoreTablesfalse\fi
   \fi
   \ifMoreTables\Print{}\TableItems\fi
 \fi
   \MakePageInCompletefalse 
   \ifMoreFigures\MakePageInCompletetrue\fi
   \ifMoreTables\MakePageInCompletetrue\fi
 \repeat
 \ifZoneAFullPage
  \global\TextSize=\z@
  \global\ZoneBSize=\z@
  \global\vsize=\z@\relax
  \global\topskip=\z@\relax
  \vbox to \z@{\vss}
  \eject
 \else
 \global\vsize=\ZoneBSize
 \global\hsize=\ColumnWidth
 \ifdim\TextSize<23pt
 \Warn{}
 \Warn{* Making column fall short: TextSize=\the\TextSize *}
 \vskip-\lastskip\eject\fi
\fi
}

\def\FigureItems{
 \Print{Considering...}
 \ShowItem{\StackPointer}
 \GetItemBOX{\StackPointer} 
 \GetItemSPAN{\StackPointer}
  \CheckFitInZone 
  \ifnum\ItemFits=\Yes
   \ifnum\ItemSPAN=\Single
     \ChangeStatus{\StackPointer}{\InZoneB} 
     \global\FigInZoneBtrue
     \ifFirstSingleItem
      \hbox{}\vskip-\BodgeHeight
     \global\advance\ItemSIZE by \TextLeading
     \fi
     \unvbox\ItemBOX\ItemSep
     \global\FirstSingleItemfalse
     \global\advance\TextSize by -\ItemSIZE
     \global\advance\TextSize by -\TextLeading
   \else
    \ifFirstZoneA
     \global\advance\ItemSIZE by \TextLeading
     \global\FirstZoneAfalse\fi
    \global\advance\TextSize by -\ItemSIZE
    \global\advance\TextSize by -\TextLeading
    \global\advance\ZoneBSize by -\ItemSIZE
    \global\advance\ZoneBSize by -\TextLeading
    \ifFigInZoneB\relax
     \else
     \ifdim\TextSize<3\TextLeading
     \global\ZoneAFullPagetrue
     \fi
    \fi
    \ChangeStatus{\StackPointer}{\Zone}
    \ifnum\Zone=\InZoneC \global\FigInZoneCtrue\fi
  \fi
   \Print{TextSize=\the\TextSize}
   \Print{ZoneBSize=\the\ZoneBSize}
  \global\advance\NextFigure \@ne
   \Print{This figure has been placed.}
  \else
   \Print{No space available for this figure...holding over.}
   \Print{}
   \global\MoreFiguresfalse
  \fi
}

\def\TableItems{
 \Print{Considering...}
 \ShowItem{\StackPointer}
 \GetItemBOX{\StackPointer} 
 \GetItemSPAN{\StackPointer}
  \CheckFitInZone 
  \ifnum\ItemFits=\Yes
   \ifnum\ItemSPAN=\Single
    \ChangeStatus{\StackPointer}{\InZoneB}
     \global\TabInZoneBtrue
     \ifFirstSingleItem
      \hbox{}\vskip-\BodgeHeight
     \global\advance\ItemSIZE by \TextLeading
     \fi
     \unvbox\ItemBOX\ItemSep
     \global\FirstSingleItemfalse
     \global\advance\TextSize by -\ItemSIZE
     \global\advance\TextSize by -\TextLeading
   \else
    \ifFirstZoneA
    \global\advance\ItemSIZE by \TextLeading
    \global\FirstZoneAfalse\fi
    \global\advance\TextSize by -\ItemSIZE
    \global\advance\TextSize by -\TextLeading
    \global\advance\ZoneBSize by -\ItemSIZE
    \global\advance\ZoneBSize by -\TextLeading
    \ifFigInZoneB\relax
     \else
     \ifdim\TextSize<3\TextLeading
     \global\ZoneAFullPagetrue
     \fi
    \fi
    \ChangeStatus{\StackPointer}{\Zone}
    \ifnum\Zone=\InZoneC \global\TabInZoneCtrue\fi
   \fi
  \global\advance\NextTable \@ne
   \Print{This table has been placed.}
  \else
  \Print{No space available for this table...holding over.}
   \Print{}
   \global\MoreTablesfalse
  \fi
}


\def\CheckFitInZone{%
{\advance\TextSize by -\ItemSIZE
 \advance\TextSize by -\TextLeading
 \ifFirstSingleItem
  \advance\TextSize by \TextLeading
 \fi
 \ifnum\Zone=\InZoneA\relax
  \else \advance\TextSize by -\ZoneBAdjust
 \fi
 \ifdim\TextSize<3\TextLeading \global\ItemFits=\No
 \else \global\ItemFits=\Yes\fi}
}

\def\BeginOpening{%
  \ninepoint
  \thispagestyle{titlepage}%
  \global\setbox\ItemBOX=\vbox\bgroup%
    \hsize=\PageWidth%
    \hrule height \z@
    \ifsinglecol\vskip 6pt\fi 
}

\let\begintopmatter=\BeginOpening  

\def\EndOpening{%
  \One
  \egroup
  \ifsinglecol
    \box\ItemBOX%
    \vskip\TextLeading plus 2\TextLeading
    \@noafterindent
  \else
    \ItemNUMBER=\z@%
    \ItemTYPE=\Figure
    \ItemSPAN=\Double
    \ItemSTATUS=\InStack
    \JoinStack
  \fi
}


\newif\if@here  \@herefalse

\def\no@float{\global\@heretrue}
\let\nofloat=\relax 

\def\beginfigure{%
  \@ifstar{\global\@dfloattrue \@bfigure}{\global\@dfloatfalse \@bfigure}%
}

\def\@bfigure#1{%
  \par
  \if@dfloat
    \ItemSPAN=\Double
    \TEMPDIMEN=\PageWidth
  \else
    \ItemSPAN=\Single
    \TEMPDIMEN=\ColumnWidth
  \fi
  \ifsinglecol
    \TEMPDIMEN=\PageWidth
  \else
    \ItemSTATUS=\InStack
    \ItemNUMBER=#1%
    \ItemTYPE=\Figure
  \fi
  \bgroup
    \hsize=\TEMPDIMEN
    \global\setbox\ItemBOX=\vbox\bgroup
      \eightpoint\nostb@ls{10pt}%
      \let\caption=\fig@caption
      \ifsinglecol \let\nofloat=\no@float\fi
}

\def\fig@caption#1{%
  \vskip 5.5pt plus 6pt%
  \bgroup 
    \eightpoint\nostb@ls{10pt}%
    \setbox\TEMPBOX=\hbox{#1}%
    \ifdim\wd\TEMPBOX>\TEMPDIMEN
      \noindent \unhbox\TEMPBOX\par
    \else
      \hbox to \hsize{\hfil\unhbox\TEMPBOX\hfil}%
    \fi
  \egroup
}

\def\endfigure{%
  \par\egroup 
  \egroup
  \ifsinglecol
    \if@here \midinsert\global\@herefalse\else \topinsert\fi
      \unvbox\ItemBOX
    \endinsert
  \else
    \JoinStack
    \Print{Processing source for figure \the\ItemNUMBER}%
  \fi
}


\newbox\tab@cap@box
\def\tab@caption#1{\global\setbox\tab@cap@box=\hbox{#1\par}}

\newtoks\tab@txt@toks
\long\def\tab@txt#1{\global\tab@txt@toks={#1}\global\table@txttrue}

\newif\iftable@txt  \table@txtfalse
\newif\if@dfloat    \@dfloatfalse

\def\begintable{%
  \@ifstar{\global\@dfloattrue \@btable}{\global\@dfloatfalse \@btable}%
}

\def\@btable#1{%
  \par
  \if@dfloat
    \ItemSPAN=\Double
    \TEMPDIMEN=\PageWidth
  \else
    \ItemSPAN=\Single
    \TEMPDIMEN=\ColumnWidth
  \fi
  \ifsinglecol
    \TEMPDIMEN=\PageWidth
  \else
    \ItemSTATUS=\InStack
    \ItemNUMBER=#1%
    \ItemTYPE=\Table
  \fi
  \bgroup
    \eightpoint\nostb@ls{10pt}%
    \global\setbox\ItemBOX=\vbox\bgroup
      \let\caption=\tab@caption
      \let\tabletext=\tab@txt
      \ifsinglecol \let\nofloat=\no@float\fi
}

\def\endtable{%
  \par\egroup 
  \egroup
  \setbox\TEMPBOX=\hbox to \TEMPDIMEN{%
    \eightpoint\nostb@ls{10pt}%
    \hss
    \vbox{%
      \hsize=\wd\ItemBOX
      \ifvoid\tab@cap@box
      \else
        \noindent\unhbox\tab@cap@box
        \vskip 5.5pt plus 6pt%
      \fi
      \box\ItemBOX
      \iftable@txt
        \vskip 10pt%
        \noindent\the\tab@txt@toks
        \global\table@txtfalse
      \fi
    }%
    \hss
  }%
  \ifsinglecol
    \if@here \midinsert\global\@herefalse\else \topinsert\fi
      \box\TEMPBOX
    \endinsert
  \else
    \global\setbox\ItemBOX=\box\TEMPBOX
    \JoinStack
    \Print{Processing source for table \the\ItemNUMBER}%
  \fi
}

\def\UnloadZoneA{%
\FirstZoneAtrue
 \Iteration=\z@
  \loop
   \ifnum\Iteration<\LengthOfStack
    \GetItemSTATUS{\Iteration}
    \ifnum\ItemSTATUS=\InZoneA
     \GetItemBOX{\Iteration}
     \ifFirstZoneA \vbox to \BodgeHeight{\vfil}%
     \FirstZoneAfalse\fi
     \unvbox\ItemBOX\ItemSep
     \LeaveStack{\Iteration}
     \else
     \advance\Iteration \@ne
   \fi
 \repeat
}

\def\UnloadZoneC{%
\Iteration=\z@
  \loop
   \ifnum\Iteration<\LengthOfStack
    \GetItemSTATUS{\Iteration}
    \ifnum\ItemSTATUS=\InZoneC
     \GetItemBOX{\Iteration}
     \ItemSep\unvbox\ItemBOX
     \LeaveStack{\Iteration}
     \else
     \advance\Iteration \@ne
   \fi
 \repeat
}


\def\ShowItem#1{
  {\GetItemAll{#1}
  \Print{\the#1:
  {TYPE=\ifnum\ItemTYPE=\Figure Figure\else Table\fi}
  {NUMBER=\the\ItemNUMBER}
  {SPAN=\ifnum\ItemSPAN=\Single Single\else Double\fi}
  {SIZE=\the\ItemSIZE}}}
}

\def\ShowStack{%
 \Print{}
 \Print{LengthOfStack = \the\LengthOfStack}
 \ifnum\LengthOfStack=\z@ \Print{Stack is empty}\fi
 \Iteration=\z@
 \loop
 \ifnum\Iteration<\LengthOfStack
  \ShowItem{\Iteration}
  \advance\Iteration \@ne
 \repeat
}

\def\B#1#2{%
\hbox{\vrule\kern-0.4pt\vbox to #2{%
\hrule width #1\vfill\hrule}\kern-0.4pt\vrule}
}


\newif\ifsinglecol   \singlecolfalse

\def\onecolumn{%
  \global\output={\singlecoloutput}%
  \global\hsize=\PageWidth
  \global\vsize=\PageHeight
  \global\ColumnWidth=\hsize
  \global\TextLeading=12pt
  \global\Leading=12
  \global\singlecoltrue
  \global\let\onecolumn=\relax
  \global\let\footnote=\sing@footnote
  \global\let\vfootnote=\sing@vfootnote
  \ninepoint 
  \message{(Single column)}%
}

\def\singlecoloutput{%
  \shipout\vbox{\PageHead\vbox to \PageHeight{\pagebody\vss}\PageFoot}%
  \advancepageno
  \ifplate@page
    \shipout\vbox{%
      \sp@pagetrue
      \def\sp@type{plate}%
      \global\plate@pagefalse
      \PageHead\vbox to \PageHeight{\unvbox\plt@box\vfil}\PageFoot%
    }%
    \message{[plate]}%
    \advancepageno
  \fi
  \ifnum\outputpenalty>-\@MM \else\dosupereject\fi%
}

\def\ItemSep{\vskip\ItemSepamount\relax}

\def\ItemSepbreak{\par\ifdim\lastskip<\ItemSepamount
  \removelastskip\penalty-200\ItemSep\fi%
}


\let\@@endinsert=\endinsert 

\def\endinsert{\egroup 
  \if@mid \dimen@\ht\z@ \advance\dimen@\dp\z@ \advance\dimen@12\p@
    \advance\dimen@\pagetotal \advance\dimen@-\pageshrink
    \ifdim\dimen@>\pagegoal\@midfalse\p@gefalse\fi\fi
  \if@mid \ItemSep\box\z@\ItemSepbreak
  \else\insert\topins{\penalty100 
    \splittopskip\z@skip
    \splitmaxdepth\maxdimen \floatingpenalty\z@
    \ifp@ge \dimen@\dp\z@
    \vbox to\vsize{\unvbox\z@\kern-\dimen@}
    \else \box\z@\nobreak\ItemSep\fi}\fi\endgroup%
}


\def\gobbleone#1{}
\def\gobbletwo#1#2{}
\let\footnote=\gobbletwo 
\let\vfootnote=\gobbleone

\def\sing@footnote#1{\let\@sf\empty 
  \ifhmode\edef\@sf{\spacefactor\the\spacefactor}\/\fi
  \hbox{$^{\hbox{\eightpoint #1}}$}\@sf\sing@vfootnote{#1}%
}

\def\sing@vfootnote#1{\insert\footins\bgroup\eightpoint\b@ls{9pt}%
  \interlinepenalty\interfootnotelinepenalty
  \splittopskip\ht\strutbox 
  \splitmaxdepth\dp\strutbox \floatingpenalty\@MM
  \leftskip\z@skip \rightskip\z@skip \spaceskip\z@skip \xspaceskip\z@skip
  \noindent $^{\scriptstyle\hbox{#1}}$\hskip 4pt%
    \footstrut\futurelet\next\fo@t%
}

\def\footnoterule{\kern-3\p@ \hrule height \z@ \kern 3\p@}

\skip\footins=19.5pt plus 12pt minus 1pt
\count\footins=1000
\dimen\footins=\maxdimen

\def\note#1#2{%
  \let\@sf=\empty \ifhmode\edef\@sf{\spacefactor\the\spacefactor}\/\fi
  #1\insert\footins\bgroup
    \eightpoint\b@ls{10pt}\rm
    \interlinepenalty\interfootnotelinepenalty
    \splitmaxdepth\dp\strutbox \floatingpenalty\@MM
    \leftskip\z@skip \rightskip\z@skip \spaceskip\z@skip \xspaceskip\z@skip
    \noindent\footstrut #1$\,$#2\strut\par
  \egroup
  \@sf\relax}


\def\landscape{%
  \global\TEMPDIMEN=\PageWidth
  \global\PageWidth=\PageHeight
  \global\PageHeight=\TEMPDIMEN
  \global\let\landscape=\relax
  \onecolumn
  \message{(landscape)}%
  \raggedbottom
}


\output{%
  \ifLeftCOL
    \global\setbox\LeftBOX=\vbox to \ZoneBSize{\box255\unvbox\ZoneBBOX
      \ifvoid\footins\else
        \vskip\skip\footins\unvbox\footins\fi
    }%
    \global\LeftCOLfalse
    \MakeRightCol
  \else
    \setbox\RightBOX=\vbox to \ZoneBSize{\box255\unvbox\ZoneBBOX
      \ifvoid\footins\else
        \vskip\skip\footins\unvbox\footins\fi
    }%
    \setbox\MidBOX=\hbox{\box\LeftBOX\hskip\ColumnGap\box\RightBOX}%
    \setbox\PageBOX=\vbox to \PageHeight{%
      \UnloadZoneA\box\MidBOX\UnloadZoneC}%
    \shipout\vbox{\PageHead\vbox to \PageHeight{\box\PageBOX\vss}\PageFoot}%
    \advancepageno
    \ifplate@page
      \shipout\vbox{%
        \sp@pagetrue
        \def\sp@type{plate}%
        \global\plate@pagefalse
        \PageHead\vbox to \PageHeight{\unvbox\plt@box\vfil}\PageFoot%
      }%
      \message{[plate]}%
      \advancepageno
    \fi
    \global\LeftCOLtrue
    \CleanStack
    \MakePage
  \fi
}


\Warn{\start@mess}

\newif\ifCUPmtplainloaded 
\ifprod@font
  \global\CUPmtplainloadedtrue
\fi

\def\mnmacrosloaded{} 

\catcode `\@=12 



\fi
\input psfig.tex
%

\newif\ifAMStwofonts

\ifCUPmtplainloaded \else
  \NewTextAlphabet{textbfit} {cmbxti10} {}
  \NewTextAlphabet{textbfss} {cmssbx10} {}
  \NewMathAlphabet{mathbfit} {cmbxti10} {} 
  \NewMathAlphabet{mathbfss} {cmssbx10} {} 
  \ifAMStwofonts
    \NewSymbolFont{upmath} {eurm10}
    \NewSymbolFont{AMSa} {msam10}
    \NewMathSymbol{\upi}     {0}{upmath}{19}
    \NewMathSymbol{\umu}     {0}{upmath}{16}
    \NewMathSymbol{\upartial}{0}{upmath}{40}
    \NewMathSymbol{\leqslant}{3}{AMSa}{36}
    \NewMathSymbol{\geqslant}{3}{AMSa}{3E}

    \let\leq=\leqslant 
    \let\geq=\geqslant 
  \else
    \def\umu{\mu}
    \def\upi{\pi}
    \def\upartial{\partial}
  \fi
\fi


\pageoffset{-2.5pc}{0pc}

\loadboldmathnames



\pagerange{Ln--Ln}   
\pubyear{1998}

\begintopmatter  

\title{The terminal bulk Lorentz factor of relativistic electron-positron
 jets.}
\author{N. Renaud and G. Henri}
\affiliation{Laboratoire d'Astrophysique, Observatoire de Grenoble, 
BP 53X F-38041 Grenoble Cedex France}

\shortauthor{N. Renaud and G. Henri}
\shorttitle{The terminal bulk Lorentz factor of relativistic electron-positron
 jets.}


\acceptedline{Accepted ----. Received ----}

\abstract {We present numerical simulation of bulk Lorentz factor of 
relativistic
 electron-positron jet driven by Compton rocket effect from accretion disc 
radiation. The plasma is assumed to have a power-law distribution 
$n_{e}(\gamma)
\propto \gamma^{ -s}$ whith $1 < \gamma
< \gamma_{max}$ and is continuously reheated to compensate for
 radiation losses. We include full Klein-Nishina (hereafter KN) cross section,
 and
 study the role of energy upper cut-off $\gamma_{max}$, spectral index $s$, and
 source compactness.
We determine terminal bulk Lorentz factor in the case of
 supermassive black holes relevant to AGN and stellar black holes relevant to 
galactic microquasars. In the latter case, Klein-Nishina cross section effect
 are more important, and induce terminal bulk Lorentz factor smaller than
 in the former case. Our result are in good agreement with 
bulk Lorentz factors observed in 
galactic sources (GRS1915+105, GROJ1655-40) and extragalactic ones.
Differences in scattered radiation and acceleration mechanism efficiency
in AGN environment can be responsible for the variety of relativistic motion
 in those objects.
We also take into account the influence of the size of the accretion disc;
 if the external radius is small enough, the bulk Lorentz factor can be 
as high as 60.}

\keywords {galaxies: active - galaxies: jets - radiation mechanism: 
miscellaneous - stars: individual: GRS1915+105, GROJ1655-40 }

\maketitle  

\section{Introduction}

Superluminal motion observed in Active Galactic Nuclei (AGN), 
especially in the blazars class, 
seems to be closely linked with 
high-energy emission. Such motion was recently observed in the Galaxy 
(Mirabel \& Rodriguez 1994, Hjellming \& Rupen 1995, Tingay et al. 1995) 
in the so-called microquasars. 
Nevertheless differences
are noticeable in those two cases. The latter systems were observed with
small value of bulk Lorentz factor (around 2.5),
 while in the former ones values of
about 10-20 are frequent. It is well known that the radiation pressure
 acting on electron-positron plasma in the vicinity of a near Eddington 
accreting object is very efficient to accelerate the plasma outwards, since
 the gravitational force is around 1000 times weaker than for an 
electron-proton plasma. However, Phinney (1987) has shown that for a 
realistic accretion disc emission, only moderate values of bulk Lorentz 
factors can be reached. Li \& Liang (1996), have recently proposed
 that this mechanism could explain the relatively small $\gamma_b \sim 2.5$ 
observed in galactic objects. They considered plasma composed with
 both $e^+e^-$ and $e^-p$ and obtained the equation of motion using the
 Thomson cross section and including gravitation force. 
To explain higher values of $\gamma_b$, O'Dell (1981) proposed
 the so-called 'Compton rocket' effect, {\it i.e.} anisotropic
 Inverse Compton effect on a highly relativistic plasma.
It was then argued (Phinney 1982) that because Compton cooling 
is always much more rapid than bulk 
acceleration, only small value of $\gamma_{b\infty}$
 could be reached by this mechanism. 
However, taking into account
 that in the frame of the 'two-flow' model (Sol, Pelletier \& Ass\'eo 
1989), a pair plasma could be reheated by the turbulence triggered by a
 surrounding jet
(Henri \& Pelletier 1991), Marcowith, Henri \& Pelletier 
(1995) showed that the Compton rocket becomes much more efficient
 and accelerates pair plasma to Lorentz factor $\gamma_b \sim 10$. 
In another work Sikora et al. (1996) studied the radiation drag
 in AGN jets. They included relativistic electron-positron plasma accelerated 
{\it in situ} and relativistic protons which contribute to the inertia of 
the flow. They considered radiation emitted from an accretion disc, 
partially reprocessed by the outer part of the disc or by spherically
 distributed matter at a given distance from the central object. 
 All Compton interactions were computed using Thomson cross section. They showed
 that in most cases jets should undergo radiation drag, and that the efficiency
 of this mechanism becomes important for purely pair plasma dominating
 the jet luminosity.
The aim of this paper is to study how Compton rocket effect can accelerate a
 pure pair plasma in the vicinity of accretion disc, taking into account
 the full KN cross section.
Following Marcowith et al. (1995), we consider a blob of $e^+e^-$
 pair plasma with an isotropic energy distribution in the comoving frame
$n^\prime_e(\gamma^\prime) \propto \gamma^{\prime -s}$, 
 where s is the spectral index. We assume that the acceleration process
 is efficient enough in the jet to get a stationary energy distribution.
 We assume that the dynamics of the pair plasma is decoupled from the electron
 proton component and we do not include gravitation force.
 The radiation field is coming from an accretion disc. 
We include KN corrections in the description of the
 Inverse Compton interactions.
We study the influence of both compactness of the radiation source, spectral
 index and upper cut-off $\gamma_{max}$, and make then comparisons between 
AGN and galactic microquasars. We also consider
 the influence of scattered radiation by a Broad Line Region (BLR) and
 dusty torus around the central black hole. Finally we 
discuss the influence of 
accretion disc's size that could be relevant to the high value of bulk Lorentz
 factors.

\section{Compton rocket effect with Klein-Nishina corrections}
\subsection{Notations} 
All energies are measured in unit $m_ec^2$. We refer
 all quantities expressed in the blob rest frame by a prime $^\prime$,
 all quantities in the particle rest frame by a star $^\ast$ and quantities 
 in the disc frame are not labelled. Photon
 energies will be labelled by $\varepsilon$, and the unit direction vector
 by $\bmath{k}$. We use the KN differential cross section (Rybicki \& Lightman 
1979) given by:
$$
\displaylines{%
 \hbox to \hsize{}\cr
{{\rm d}\sigma \over {\rm d}\varepsilon _1^{\ast }{\rm d}\Omega _1^{\ast }}= 
{ 3 \sigma_T \over 16}\left ({\varepsilon _1^{\ast} \over \varepsilon ^{\ast }}
\right)
^2\left({\varepsilon _1^{\ast } \over \varepsilon ^{\ast }}+
{\varepsilon^{\ast } \over \varepsilon_1^{\ast }}-
\sin^2 \phi^{\ast}\right) \hfill\cr
\delta \left(\displaystyle
 \varepsilon_1^\ast-{\varepsilon^\ast \over 
1+\varepsilon^\ast(1-\cos \phi^\ast)} \right).\cr}
$$
$\sigma_T= \displaystyle{8 \pi \over 3} r_e^2$ 
is the Thomson cross-section  where $\displaystyle r_e={e^2 \over 
4 \pi \varepsilon_0 m_e c^2}$ is the electron classical radius.
This expression applies to the scattering
of a photon with energy $\varepsilon ^{\ast }$ and direction 
$\bmath{k^{\ast }}$ in a photon with energy $\varepsilon _1^{\ast
}$ and direction $\bmath{k_1^{\ast }}$, and 
$\cos \phi^\ast = \bmath{k^{\ast }} .
\bmath{k_1^{\ast }}$.  
\subsection{The general picture}
Figure 1 shows the general configuration of the model. 
\beginfigure{1}  
  \psfig{width=8cm,height=8cm,angle=-90,file=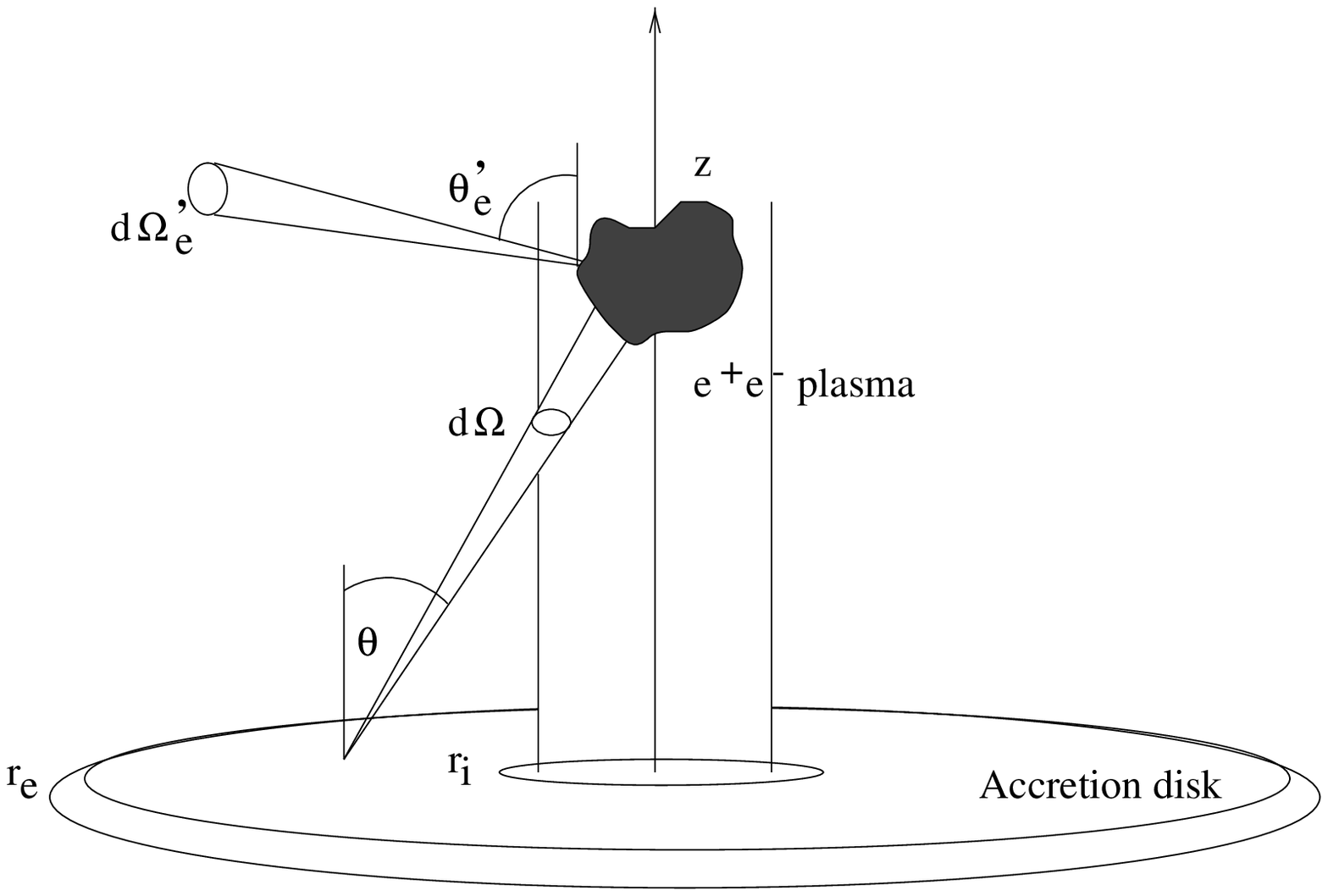}
  \caption{{\bf Figure 1.} The general configuration of the model.
 Quantities in the
 blob rest frame are denoted by a prime.}
\endfigure
The pair plasma 
is assumed to be described in the bulk rest frame by an energy distribution 
$n_e^\prime
(z,\gamma^\prime)
 \propto \gamma^{\prime -s}$ for $\gamma_{min} < \gamma^\prime
 < \gamma_{max}$, with
 $s$, $\gamma_{min}$ and $\gamma_{max}$ independent of $z$
 (see paragraph 3.2.1 for a further discussion of this assumption).
 The radiation force is due to soft photons coming from a standard 
accretion disc (Shakura \& Sunyaev 1973) around a Schwarzschild black
 hole. 
The inner radius of the accretion disc is $r_i=3 \,r_g$ 
(where $r_g$ is the Schwarzschild radius of the
 black hole). The outer radius $r_e$ is a free parameter. 
We use the black-body approximation for
 the disc emission so that the specific intensity at a radius $r$ is:
$$
I_{\nu}(r)=B_{\nu}(T_{eff}(r)), \eqno\stepeq 
$$
where $B_{\nu}$ is the Planck function and the effective temperature
 $T_{eff}$ is given by:
$$
T_{eff}(r)=\left ( {3 G M \dot{M}\over 8 \pi \sigma r^3}\right)^{1/4}
\left[1-b(r_i/r)^{1/2}\right]^{1/4}. \eqno\stepeq 
$$
$\dot{M}$ is the accretion rate, $M$ the mass of 
the central black hole and $b$ is a parameter describing the inner torque
 on the disc ($0\leq b \leq 1$). In section 4, we also consider 
the case of a two-temperatures 
disc model (Shapiro, Lightman \& Eardley 1976). This
 model applies to the hot inner part of the disc ($r<$ a few $r_g$). 
The specific intensity
 is then (Sunyaev \& Titarchuk 1980):
$$ 
\displaylines{%
\hbox to \hsize{} \cr
I_\nu(r)\propto{{\rm exp}(-x) \over x} \left(1+x+{1\over 2}x^2+
{1 \over 6} x^3+ {1 \over 24}x^4 \right) \,\, {\rm for } \,\, 
x>1 \hfill\cr
I_\nu(r)\propto x^4 \,\, {\rm for } \,\, x<1, \hfill\stepeq\cr}
$$
where $x=h\nu/kT_e$ and $T_e$ is the electronic temperature given by:
$$
\displaylines{%
\hbox to \hsize{} \cr
T_e(r)=7.10^8 {\rm K}\left( {M \over 3M_{\odot}} {10^{17}
 {\rm g.s^{-1}}\over \dot{M}}\right)^{1/6}
 \left[1-(r_i/r)^{1/2}\right]^{-1/6}\hfill\cr
\times (2r/r_g)^{1/4} .\hfill\stepeq\cr}
$$
In AGN, BLR surrounding the central black hole reprocesses a fraction of 
about 10 per cent of the disc radiation.
In paragraph 3.2.5, following Sikora et al. (1996), we chose 
two extreme cases 
to describe this emission. The first case is case 'a' in Sikora et al. (1996)
 where a fraction $\chi$ of the total disc luminosity $L_d$ is intercepted
 by the outer part of the disc and reprocessed in Ly$\alpha$ lines
 with a specific intensity given by:
$$
I_\nu(r)={\chi L_d r^{-\alpha} \over 4\pi^2 \int^{r_2}_{r_1}r^{1-\alpha} 
 {\rm d} r}g\left(\nu-\nu_{{\rm Ly}\alpha}\right), \eqno\stepeq 
$$
The radiation is re-emitted by a ring for $r_1 < r < r_2$. 
$g$ is a rectangular 
profile with a Doppler width of $v / c \sim 2.24\,10^{-2}$. 
In the second case spherically distributed matter at a distance $r_0$ from
the central engine reemits a fraction $\chi$ of the total disc luminosity.
The emissivity of this radiation source at a distance $z$ on the jet axis is:
$$
I_\nu(z,\theta)={\chi L_d \over 4 \pi r_0^2 \sqrt{1+\left({z \over r_0}\right)^2({\rm
cos}^2\theta-1)}}g\left(\nu-\nu_{{\rm Ly}\alpha}\right), \eqno\stepeq 
$$
For $z \leq r_0$, $-1 \leq {\rm cos}\theta \leq z/\sqrt{z^2+r_0^2}$ and
 for  $z > r_0$ , $\sqrt{z^2-r_0^2}/z \leq {\rm cos}\theta \leq 1$.

We also study the influence of possible scattered radiation from hot dust
which may be responsible for near infrared emission in AGN. The
emissivity of such a source modeled by a ring located
 between $r_3$ and $r_4$ can be expressed 
as follows :
$$
I_\nu(r)={\chi^{\prime} L_d \over 4 \pi \sigma \int^{r_4}_{r_3} T^4(r) r 
{\rm d} r} B_\nu((T(r)), \eqno\stepeq 
$$
$T$ is the dust temperature, $\sigma$ is the Stefan-Boltzmann 
constant, and $\chi^{\prime}$ is the fraction of the disc luminosity 
that is reprocessed.
\subsection{Computation of the radiative force}
Let $ \displaystyle{ {\rm d} \bmath{p}^\prime \over{\rm d} t^\prime}$ be 
the rates of momentum 
change for one electron due to Inverse Compton process 
and $\beta^\prime c=\sqrt{1-1/\gamma^{\prime 2}}c$ 
its velocity.
 The force exerced upon the pair plasma is then in the blob rest
 frame:
$$
F^{\prime z}=\int {\rm d}\Omega_e^\prime {\rm d}\gamma^\prime 
 n_e^\prime(\gamma^\prime,\Omega_e^\prime)
  {{\rm d} \bmath{p}^\prime \over{\rm d} t^\prime}.\bmath{u_z}, \eqno\stepeq 
$$ 
where $\bmath{u_z}$ defines the jet axis direction. 
This force is accompanied of course by an energy loss that tends to cool
 the plasma. However, as mentioned above, we assume that there exists a
 continuous acceleration mechanism that compensate for these losses,
 so that the particle distribution remains constant. We assume that this
 acceleration acts isotropically in the plasma frame, and so it does not
 contribute to the expression given in equation (8). We also neglect
 gravitation force, since for near-Eddington objects and cold pairs we have
 $\displaystyle{F_{rad} \over F_{grav}} \sim 
\displaystyle{m_p \over m_e} \sim 10^3$.

To calculate more easily $F^{\prime z}$, one can use two different approximations:
 for low energy particles, the KN cross-section tends to the Thomson
 limit and we can apply the result of Phinney (1982) for an isotropic
 pair plasma:
$$
   F^{\prime z}=\displaystyle {\sigma_T \over c} 4 \pi 
H^\prime \int (1+ {2 \over 3} 
\gamma^{\prime 2} \beta^{\prime 2}) n_e^\prime(\gamma^\prime) 
{\rm d}\gamma^\prime, \eqno\stepeq 
$$
where $H^\prime$
 is the second 
Eddington parameter of the radiation field
 (Marcowith et al. 1995, see also Appendix A) defined in the blob 
rest frame.  It is linked to the Eddington parameters defined in observer's
 rest frame by:
$$
H^\prime=\gamma_b^2[(1+\beta_b^2)H-\beta_b(J+K)], \eqno\stepeq 
$$
where $\beta_b c$ is the bulk velocity and $\gamma_b$ is the 
corresponding Lorentz factor.
 For high energy particles, KN corrections become important but we
 can assume the head-on approximation 
 (Blumenthal \& Gould 1970) to be valid. In this approximation, 
the velocity of all photons in the particle frame is assumed to be 
anti-parallel to the particle velocity in the observer frame. So
 the momentum loss rate is necessarily in the direction of the particle 
motion.
 Let $x$ denote this direction, 
the electron energy and momentum loss rate in its rest frame 
are in unit $m_e c^2$ ($\ast$ refers to this frame):
$$
{{\rm d}p_x^\ast \over {\rm d}t^\ast}=-\int {\rm d}n^\ast {
{\rm d}\sigma  \over{\rm d} \varepsilon _1^{\ast }{\rm d}\Omega _1^{\ast }}\left(
\varepsilon^{\ast}
-\varepsilon_1^{\ast}\cos\phi^{\ast}\right
){\rm d}\varepsilon _1^{\ast }{\rm d}\Omega _1^{\ast }, \eqno\stepeq 
$$
and
$$
{{\rm d}E^\ast \over{\rm d} t^\ast}=- c \int {\rm d}n^\ast {
{\rm d}\sigma \over {\rm d}\varepsilon _1^{\ast }{\rm d}\Omega _1^{\ast }}
\left(\varepsilon_1^{\ast}
-\varepsilon^{\ast}\right)
{\rm d}\varepsilon _1^{\ast }{\rm d}\Omega _1^{\ast }. \eqno\stepeq 
$$
${\rm d}n^\ast$ 
is the differential photon density in the electron rest frame.
The momentum change rate in the blob rest frame is then given by:
$$
  {{\rm d}p_x^\prime \over {\rm d}t^\prime}=\displaystyle 
 {{\rm d}p_x^\ast \over {\rm d}t^\ast} 
+ {\beta^\prime \over c} {{\rm d}E^\ast \over{\rm d} t^\ast}. \eqno\stepeq 
$$
Note that using the KN cross-section, the rate of momentum change
 in the electron rest frame is not zero, contrary to the case of Thomson
 limit or synchrotron emission. One must take it into account
 in computing the energy loss rate in the blob rest frame which 
is not a Lorentz invariant anymore (Blumenthal \& Gould 1970).
In the head-on approximation the integral (11) and (12) over $\varepsilon_1^\ast$ and
 $\Omega_1^\ast$ can be readily performed leading to:
$$
  {{\rm d}p_x^\prime \over {\rm d}t^\prime}=-\sigma_T \int {\rm d}n^\ast (f_p(
\varepsilon^\ast) +\beta^\prime f_E(\varepsilon^\ast)). \eqno\stepeq 
$$ 
The two functions $f_p$ and $f_E$ are given by:
$$
\displaylines{%
\hbox to \hsize{} \cr
f_p(\varepsilon^{\ast})=(1+\varepsilon^{\ast})\,f(\varepsilon^{\ast}),
\hfill\cr
f_E(\varepsilon^{\ast})=-\varepsilon^{\ast}\,f(\varepsilon^{\ast}), 
\hfill\cr}
$$
where the function $f(\varepsilon^{\ast})$ corresponds to the 
ultrarelativistic case ($\beta^\prime=1$) and is given by:
$$
\displaylines{%
\hbox to \hsize{}\cr
f(\varepsilon^{\ast})={\displaystyle {1 \over 8}} \,{\displaystyle 
 {18\,\varepsilon^{\ast} + 102\,\varepsilon^{\ast \ 2} + 186\,
\varepsilon^{\ast \ 3} + 102\,\varepsilon^{\ast \ 4} - 20\,
\varepsilon^{\ast \ 5} \over
\varepsilon^{\ast \ 3}\,(1 + 2\,\varepsilon^{\ast})^{3}}}\cr
+{\rm ln}(1 + 2\,\varepsilon^{\ast})  \times \cr
{\displaystyle  {1 \over 8}} \,
{\displaystyle  {(
 - 9 - 141\,\varepsilon^{\ast \ 2} - 60\,\varepsilon^{\ast} - 126\,
\varepsilon^{\ast \ 3} - 12\,\varepsilon^{\ast \ 4} + 24\,
\varepsilon^{\ast \ 5} ) \over \varepsilon^{\ast \ 3}\,(1 + 2\,
\varepsilon^{\ast})^{3}}}.\cr }
$$
We then use these expressions in equation (8).

We can estimate the errors in the two extreme regimes described above.
In the head-on approximation the first corrections are roughly $
\displaystyle{1 \over \gamma^2}$,
 while in the Thomson limit these corrections are $\sim 
\gamma \langle \varepsilon \rangle$. Here $\langle \varepsilon \rangle 
\propto \dot{M}^{1/4} M^{-1/2}$ is the average photon energy emitted
 from accretion disc. 
We connect the two regimes by defining a critical Lorentz factor
$\gamma^\prime_{crit}$ for which errors in the head-on approximation
 ($\gamma^\prime >
 \gamma_{crit}$) and in
 the Thomson limit ($\gamma^\prime < \gamma_{crit}$) are of same order. 
This gives $\gamma_{crit} \sim
 \langle \varepsilon \rangle^{-1/3}$. So we may have errors of order of
 $\displaystyle { 1 \over \gamma_{crit}^2} 
\sim \gamma_{crit}\langle \varepsilon \rangle
 \sim \langle \varepsilon \rangle^{2/3}$. 
For AGN $\langle \varepsilon \rangle \sim 10^{-4}$ and we find
 $\gamma_{crit}\sim 20-30$ with a maximum error $\sim 0.2 \%$ while
 for a microquasar $\langle \varepsilon \rangle \sim 10^{-2}$ and
 $\gamma_{crit}\sim 5$ with a maximum error $\sim 5 \%$ .

\subsection{Equation of motion}
Following Phinney (1982) we determine the acceleration of pair plasma by
considering the conservation of stress-energy tensor leading to
 Phinney's (7) and (8) equations in the bulk rest frame:
$$
\displaylines{%
 \hbox to \hsize{}\cr
    { \upartial \over  \upartial t^\prime} [(\rho^\prime+p^\prime)
\gamma_b-p/\gamma_b ] = 
F^{\prime 0}+ \beta_b F^{\prime z}, \hfill\stepeq\cr
     {\partial \over \partial t^\prime} [(\rho^\prime+p^\prime)
\gamma_b \beta_b ] = 
F^{\prime z}+\beta_b F^{\prime 0}. \hfill\stepeq\cr}
$$
Combining these two equations and 
for a reheated relativistic plasma 
(with $p^\prime=\rho^\prime / 3$) one finds the 
equation of motion (with $dz^\prime=\beta_b dt^\prime$
 and $z=z^\prime$):
$$  
{{\rm d} \gamma_b \over  {\rm d}z}={F^{\prime z} \over \rho^\prime} {1 \over
 {{1 \over 3\gamma_b^2}}+1}. \eqno\stepeq 
$$
To compute the radiative force  by equation (14)
we need the differential photon distribution
 in each electron rest frame. For this we use the Lorentz invariant 
$ \displaystyle {{\rm d}n \over \varepsilon}$ (see Blumenthal \& Gould 1970) 
 and the relation between energy in the accretion disc frame and the electron
 rest frame:
$$
\varepsilon^\ast=\gamma_b(1-\beta_b\mu_s)\gamma(1-\beta\mu^\prime)
\varepsilon. \eqno\stepeq 
$$
In the last equation $\mu_s$ is the cosinus of the angle between 
photon direction and jet axis in the accretion disc frame (see figure 1)
 and $\mu^\prime=\mu_e^\prime \mu_s^\prime+(1-\mu_e^\prime)^{1/2}
(1-\mu_s^\prime)^{1/2}\cos \phi_e^\prime$ is the cosinus between the
 electron and photon direction in the bulk rest frame. We use the averaged
 scattered photon energy over the azimutal angle $\phi_e$, and take  
equation (18)
 with $\mu^\prime=\mu^\prime_e \mu^\prime_s$.
Finally we integrate equation (17) between $z=10\,r_g$ and $z=10^4\,r_g$
(galactic case) or $z=10^5\,r_g$ (extragalactic case)  for
 different configurations to determine the final bulk Lorentz factor.

\section{Results}
\subsection{Equilibrium Lorentz factor}
In any axisymmetric radiation field different from a plane wave, there
 exists an equilibrium Lorentz factor $\gamma_{beq}$ for which
 the radiation force vanishes ($F^{\prime z}(\gamma_{beq})=0$, see figure 3). 
For $\gamma_b > \gamma_{beq}$ the pair plasma
 sees much more photons coming forward and then decelerates, while if 
$\gamma_b < \gamma_{beq}$ the plasma is pushed by radiation. The
 qualitative behaviour of a general solution of equation (17) is thus the following: $\gamma_b$
 is first set to the value $\gamma_{beq}$. This value increases gradually
 until an altitude $z=z_{crit}$ 
radiation force becomes
 too weak to accelerate the plasma. So it follows a ballistic
 motion at a constant Lorentz factor $\gamma_{b \infty}$.
 This mechanism ensures that
 in most cases the initial value of $\gamma_b$ does not influence the
 terminal value, if the blob is injected bellow $z_{crit}$. 
\beginfigure{2}
 \psfig{width=8cm,height=8cm,angle=-90,file=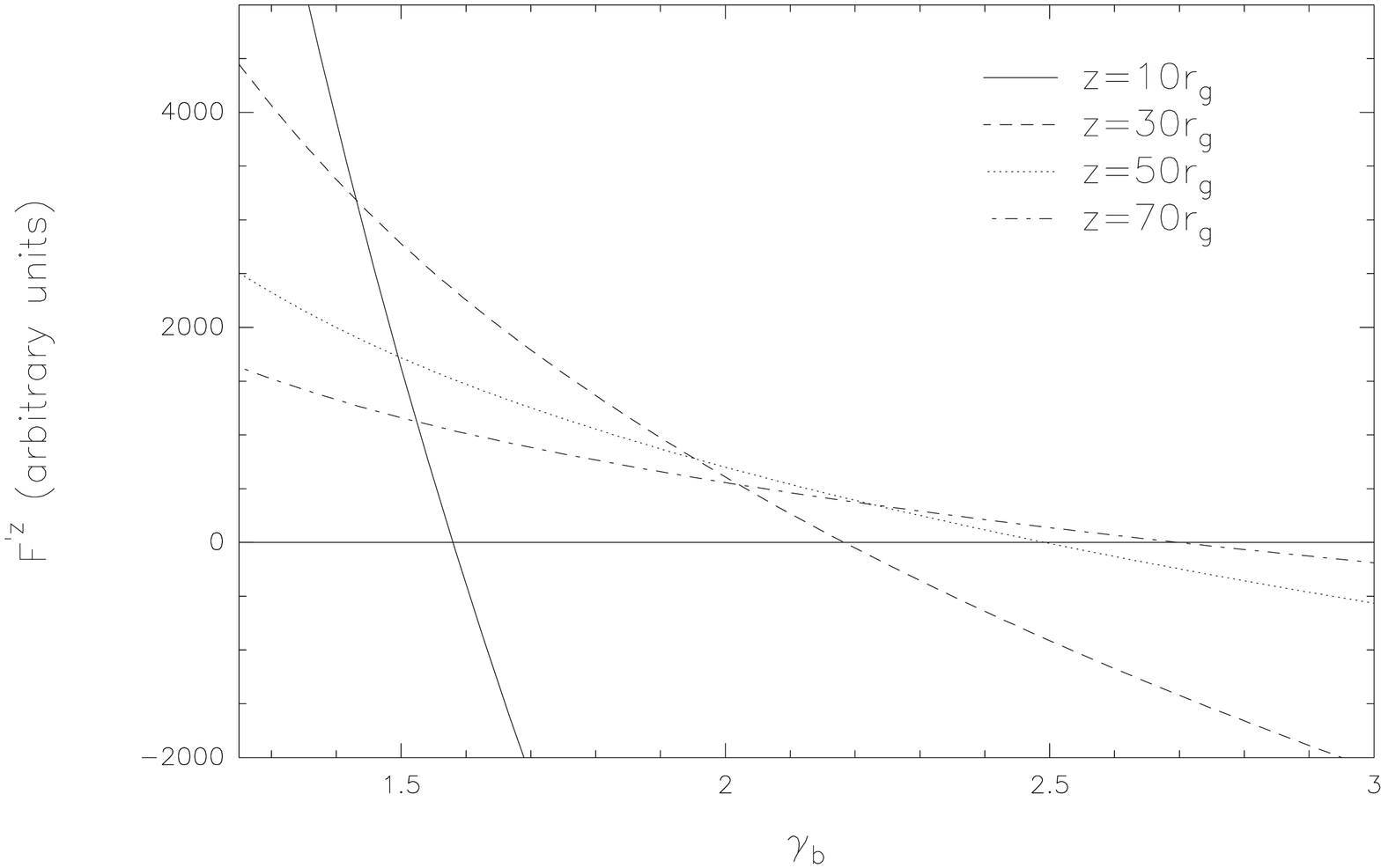}
  \caption{{\bf Figure 2.} The radiative force as a 
function of the bulk Lorentz
 factor $\gamma_b$ at different altitudes from the central black hole.}
\endfigure
The Thomson regime has been well studied by various authors 
(Phinney 1982, 1987, Marcowith et al. 1995, Sikora et al. 1996). In this
 case the equilibrium Lorentz factor is determined by the simple condition
 $H^\prime=0$, that is the radiation flux vanishes in the blob rest frame.
Analytical calculations in the Thomson regime give the behaviour of the 
equilibrium
 Lorentz factor and the solutions of the equation of motion for
 a standard accretion disc (Shakura \& Sunyaev 1973). This yields 
$\displaystyle \gamma_{beq} \propto z^{1/4}$ for $ z < r_e$ and 
$ \displaystyle \gamma_{beq} \propto z$ for $z > r_e$ 
(see Appendix A for this case). Figure 3
 shows the function $\gamma_{beq}$ obtained in the Thomson limit for
 different values of $r_e$ in comparison with these two asymptotic regimes.
 Note that for decreasing size of the accretion disc, the radiation field
 is more and more anisotropic and the function 
 $\gamma_{beq}$ increases more rapidily with $z$. 
 For $z_{crit} < r_e$ one  finally gets for a hot plasma 
$\displaystyle \gamma_{b\infty} \propto 
\left({\langle \gamma^{\prime 2} \rangle l \over \langle 
\gamma^\prime \rangle} \right)^{1/7}$, 
 while for $ z_{crit} > r_e$, $ \displaystyle \gamma_{b\infty} 
\propto \left(\left({r_i \over r_e}\right)^{3/4}
{\langle \gamma^{\prime 2} \rangle 
l \over \langle \gamma^\prime \rangle} \right)^{1/4}$ (see
 Appendix A). $\displaystyle l= 
 {\sigma_T L \over 4 \pi m_e c^3 r_i}$ is the compactness of the source,
 and $\langle f \rangle=\displaystyle
\left (\int n_e^\prime(\gamma^\prime) {\rm d}\gamma^\prime \right)^{-1}
\times \int f\, n_e^\prime(\gamma^\prime) {\rm d}\gamma^\prime $.
\beginfigure{3}
 \psfig{width=8cm,height=8cm,angle=-90,file=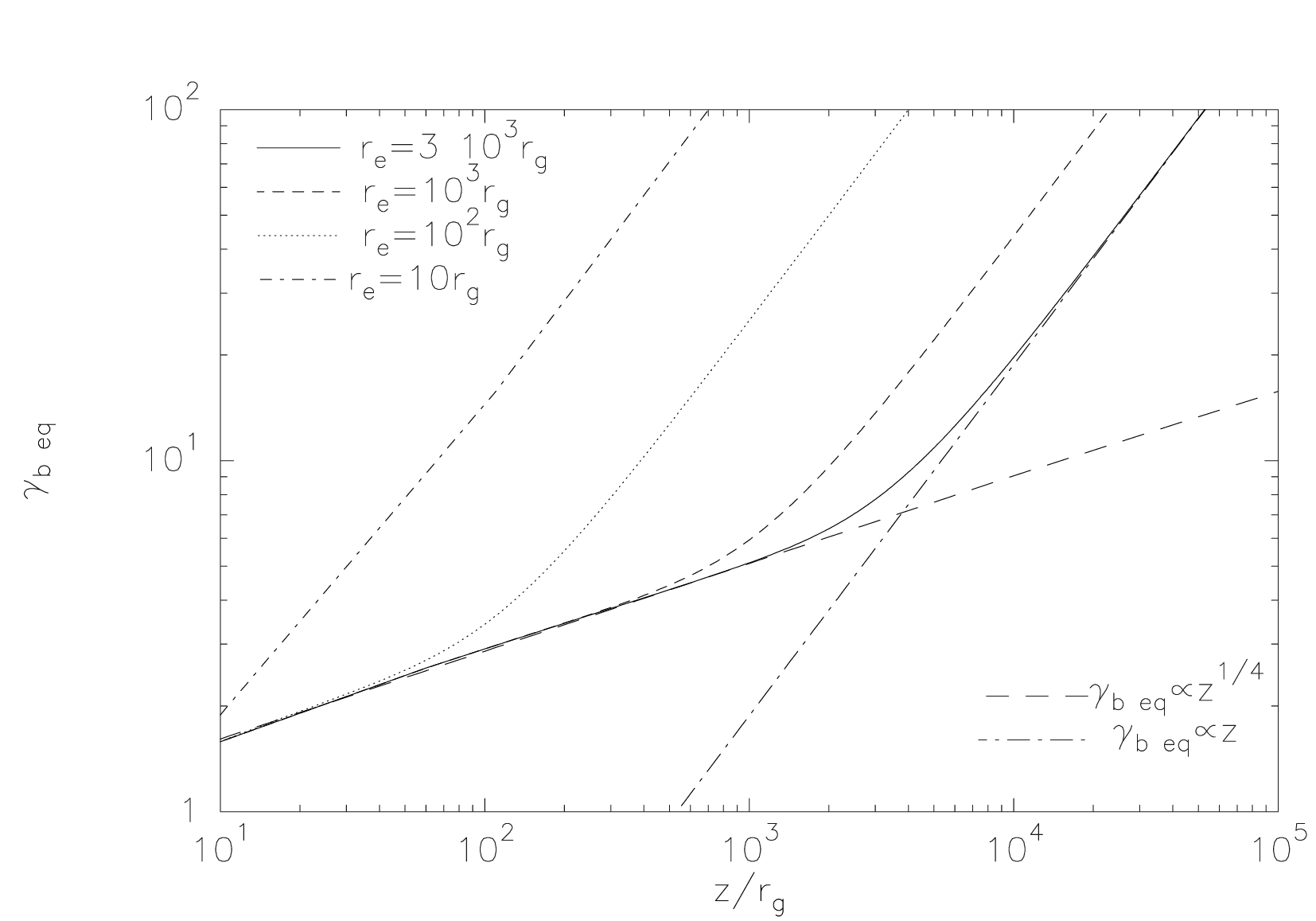}
  \caption{{\bf Figure 3.} Equilibrium Lorentz factor $\gamma_{beq}$ 
as a function of $z$ in the Thomson limit. $\gamma_{beq}$ is given for
 $r_e=10 \,r_g$, $r_e=10^2 \,r_g$, $r_e=10^3 \,r_g$ and $r_e=3.10^3 \,r_g$.
 We also represent the two asymptotic regimes $\gamma_{beq} \propto
 z^{1/4}$ for $z<r_e$ and $\gamma_{beq} \propto z$ for $z>r_e$, with
 $r_e=3.10^3 \,r_g $.}
\endfigure
Figure 4 shows the equilibrium Lorentz factor calculated including
 KN corrections for different configurations. We compare the results
 with the correspondant Thomson solution.
The effect of the KN corrections
 is to reduce the contribution of high energetic collisions 
($\varepsilon^\ast \geq 1$) in
 the net radiation force seen by the plasma. Since the more energetic photons
 come from the inner part of the accretion disc, in the blob rest frame they 
contribute to accelerate the plasma. On the other hand the outer and colder
part contribute to decelerate it. We then expect that the equilibrium 
Lorentz factor is reduce when including KN corrections as shown in figure 4. 
The importance of the difference depends
 on the particle and photon distributions. It is predominant 
for microquasars in
which accretion disc radiates more energetic photons than in the 
extragalactic case. 
\beginfigure{4}
 \psfig{width=8cm,height=8cm,angle=-90,file=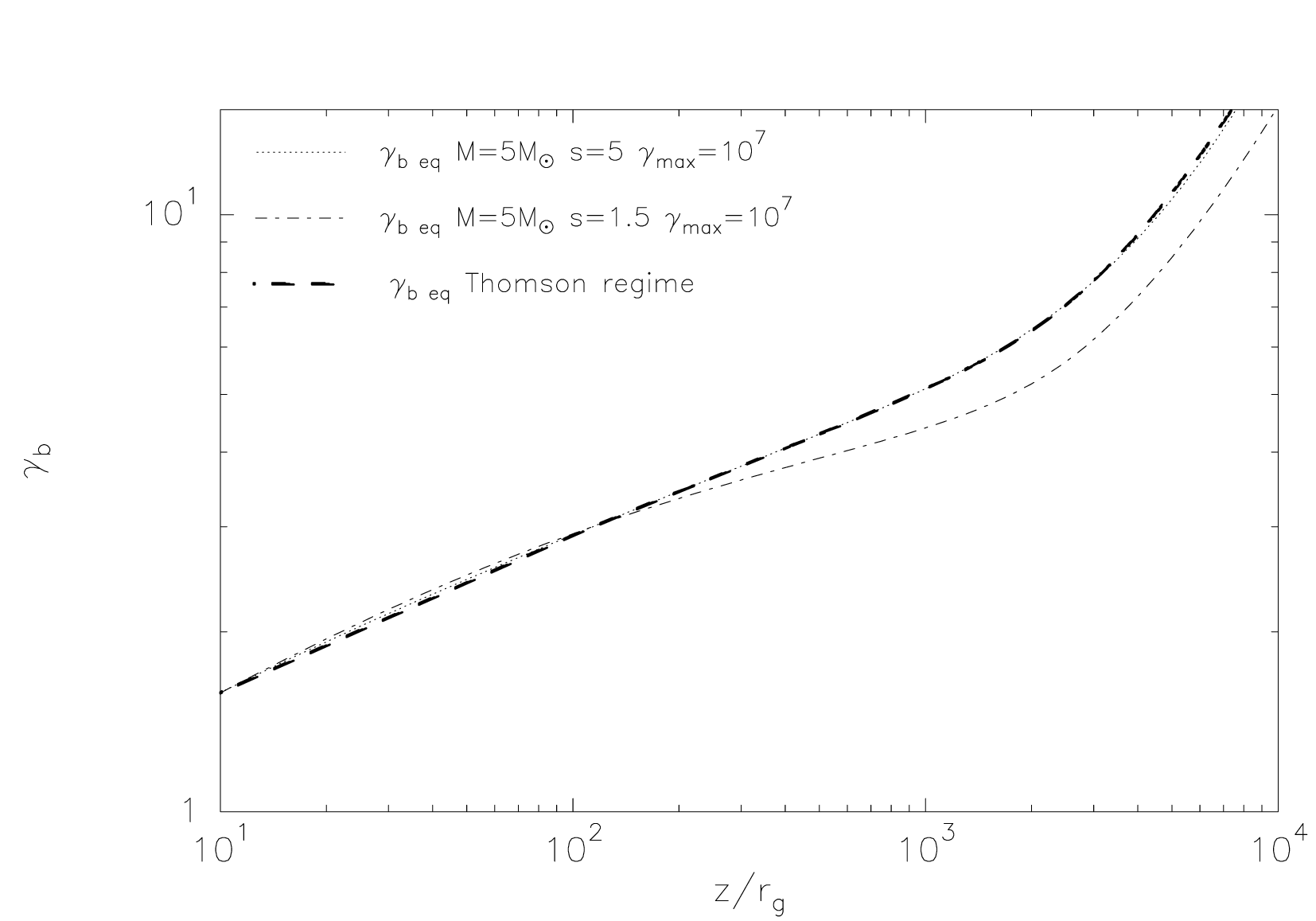}
  \caption{{\bf Figure 4.} The Equilibrium Lorentz factor calculated including
 KN corrections compared to Thomson solution. The figure corresponds
 to a stellar black hole $M=5M_\odot$ with $\gamma_{max}=10^7$ for 
spectral index $s=1.5$ and $s=5$. The external radius
 is $r_e=3\,10^3\,r_g$ and $L=L_{Edd}$.}
\endfigure

\subsection{Influence of the different parameters}
\subsubsection{General considerations}
To determine the terminal Lorentz factor for different configurations,
 we solve numerically the equation of motion for various radiation fields
 and particle distributions. We show in figure 5 typical solutions in the
 case of a stellar black hole. The terminal value of the
 bulk Lorentz factor does not depend on the initial value of 
$\gamma_b\, {\it init}$  as discussed in paragraph 3.1. The general
 behaviour of the solution discussed in this paragraph still holds  
 even including KN corrections. The critical point $z_{crit}$ is reached
 rather close to the central engine (before $10^4$ Schwarzchild radii).
 For $z> z_{crit}$ the motion is nearly balistic and so independent of
 the radiation force which has become too weak. It is so independent of
 any variation of the pair distribution unless these variations strengthen
 the radiation force. This scenario would require a
 more efficient acceleration mechanism when moving away from the central
 source, which is very unlikely. Therefore our assumption of a stationary 
pair energy distribution on a large range of $z$ does not influence strongly
 the terminal value of $\gamma_b$, or in other words this value is essentially
 determined by the local parameters at the critical distance.
 
\beginfigure{5}
 \psfig{width=8cm,height=8cm,angle=-90,file=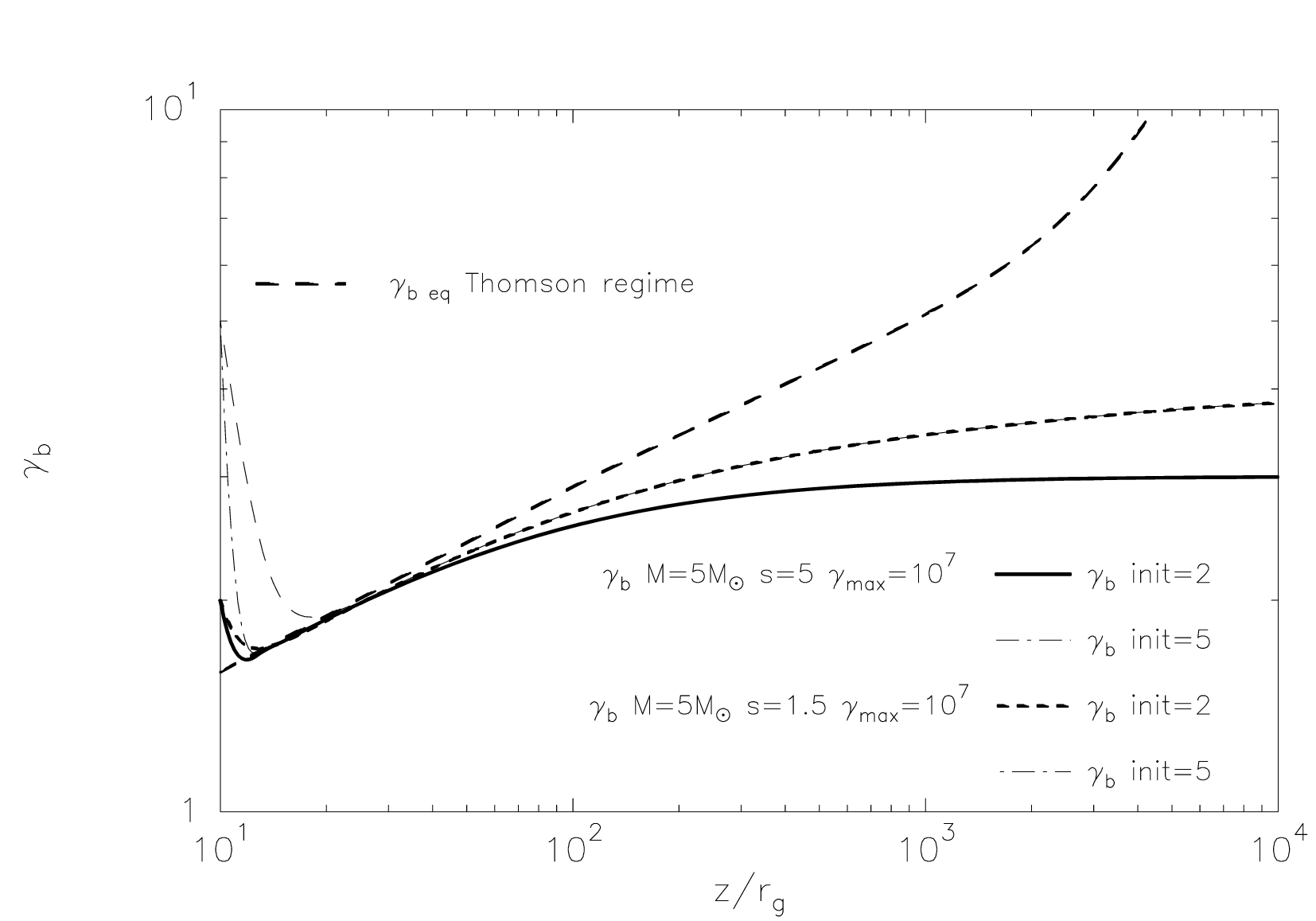}
  \caption{{\bf Figure 5.} Solutions of the equation of motion
 in the case of a stellar black hole ($M=5M_\odot$). We chose
 two different initial conditions $\gamma_{b} \, {\it init}=2$ and 
$\gamma_{b} \, {\it init}=5$. $r_{e} =3\, 10^3 \,r_g$ and $L=L_{Edd}$.}
\endfigure
\subsubsection{Influence of the energy upper cut-off and spectral index} 
Figure 6 illustrates
 the influence of the spectral index (for $1.5 \leq s \leq 5$) 
and the energy cut-off (for $ 10^3 \leq \gamma_{max} \leq 10^7$) on
 the terminal Lorentz factor. We chose $M=5M_\odot$ representative for 
stellar black holes and $M=10^9M_\odot$ for supermassive black holes.
 The calculations were carried out for $L=L_{Edd}$ and $L=0.1L_{Edd}$, where
 $L$ is the luminosity of the accretion disc.

\beginfigure*{6}
  \psfig{width=17cm,rheight=12cm,angle=-90,file=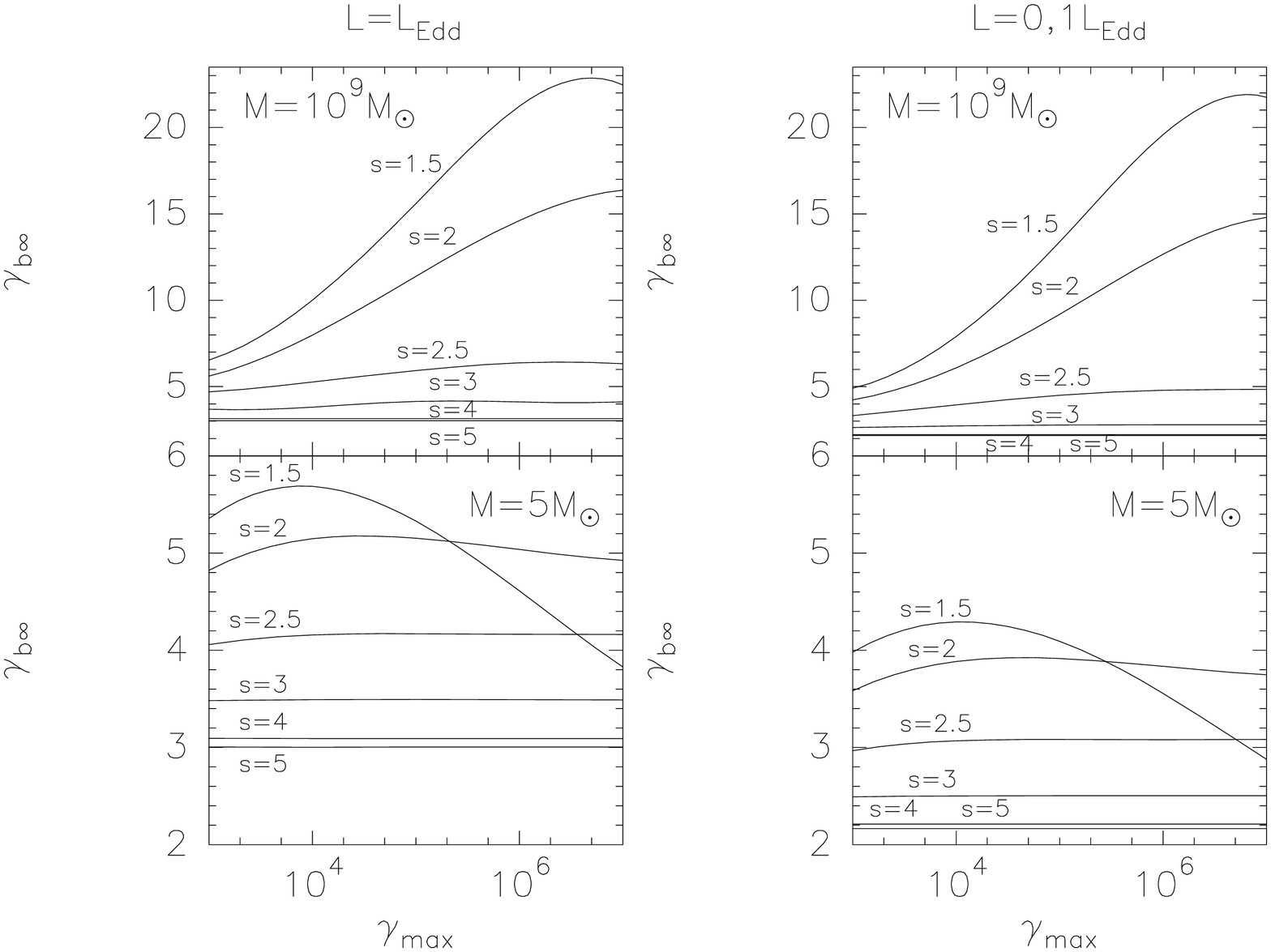}
  \caption{{\bf Figure 6.} Terminal Lorentz factor $\gamma_{b\infty}$ as a function
of $\gamma_{max}$ for different value of spectral index $s$. The two top
 pannels correspond to $M=10^9 M_{\odot}$ black hole with $r_e=3.10^3 \,r_g$,
 $L=L_{Edd}$ (left) and $L=0.1L_{Edd}$(right). The two bottom
 pannels correspond to $M=5 M_{\odot}$ black hole with $r_e=3.10^3 \,r_g$,
 $L=L_{Edd}$ (left) and $L=0.1L_{Edd}$(right).} 
\endfigure
Results are very sensitive to the spectral index value. There are 3 different
 behaviours according to the value of $s$.
\beginlist
\item (i) for $s < 2$ there exists a maximum terminal Lorentz factor
 as a function of $\gamma_{max}$ and the curve intercepts the other ones.
\item (ii) for $2 < s < 3$ there still exists a maximum, but less pronounced.
 The variation as a function of $\gamma_{max}$ are smoother.
\item (iii) for $s >3$ no variation with $\gamma_{max}$.
\endlist
We find that for low value of $\gamma_{\max}$, our solutions agree with
 the Thomson regime solutions. Nevertheless KN corrections reduce the 
efficiency of the Compton rocket effect. As a matter of fact, in the Thomson
 regime, an increase of $\gamma_{\max}$ leads to an increase of 
$\langle \gamma^{\prime 2} \rangle$ (for $s<3$) and so of $\gamma_{b\infty}$.
This mechanism is valid until KN corrections begin to dominate, roughly
 when $\gamma_{max}\langle\varepsilon \rangle \sim 1$. So
 when $\gamma_{max}$ is greater than $\langle\varepsilon \rangle^{-1}$
 the radiation force does not increase anymore whereas the plasma inertia
 $\rho^{\prime}$ is much more important. This leads to a less efficient
 acceleration mechanism. This effect is larger for small indexes explaining
 the inversion of the curve for high $\gamma_{\max}$. Indeed we find
 that acceleration is much more efficient for $s=2$ and $\gamma_{max}=10^7$ 
than
 for $s=1.5$ and the same value of $\gamma_{max}$, in the case of stellar 
black hole.
 When steepening the pair distribution, the radiation force is 
dominated by the low energy part of the distribution. This fact
 explains why no variation is apparent with $\gamma_{max}$ for $s>3$.
 The plasma behaves dynamically as a cold one, and we find small value of 
terminal Lorentz factor.
 Finally including KN corrections in the calculations gives rise to 
an absolute upper limit to maximal Lorentz factor for a given luminosity.

\subsubsection{Influence of the black hole mass}
As discussed above the influence of the mass of the central black hole
 is predominant. Stellar black holes with soft X-ray emission 
($\langle\varepsilon \rangle \sim 10^{-2}$) are less efficient
 in accelerating blob of pair plasma than supermassive black hole with 
 softer radiated emission ($\langle\varepsilon \rangle \sim 10^{-4}$), 
because KN saturation effects occur at much lower energy. A more realistic
 description of the accretion disc around stellar black holes reinforces
 this discrepency. As shown in figure 7 the radiation emitted from
 a two-temperature disc (Shapiro et al. 1976) 
 leads to smaller $\gamma_{b\infty}$ than in the case of standard accretion
 disc radiation. Because it is well established that accretion disc
 around stellar black holes should radiate up to a few keV 
(as in a two-temperatures disc), KN corrections play an 
important role in this case.

 \beginfigure*{7}
 \psfig{width=17cm,rheight=7cm,angle=-90,file=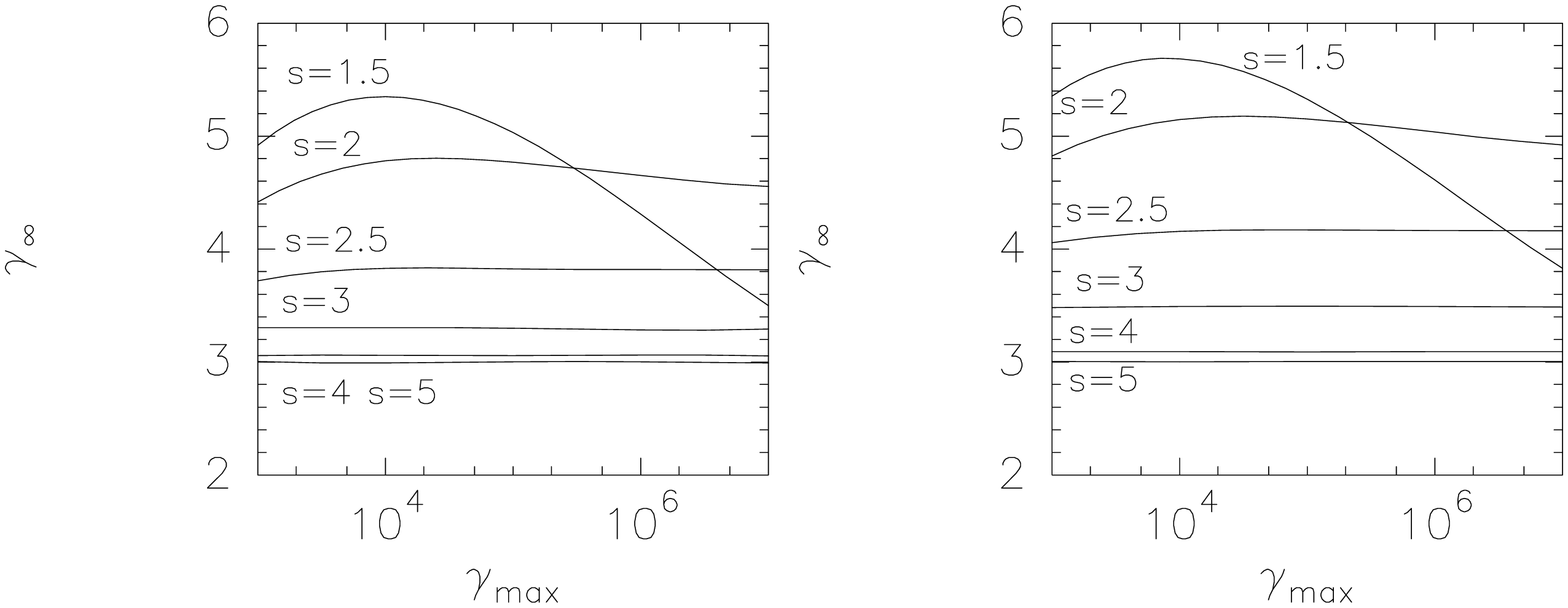}
  \caption{{\bf Figure 7.} Comparisons of terminal
 Lorentz factor for a standard accretion disc (right) and
 a two-temperature disc (Shapiro et al. 1976) extended
 up to $15 \,r_g$ (left).
 In both case $r_e=3 10^3 \,r_g$, $M=5M_{\odot}$ and the accretion rate 
is the Eddington one.}
\endfigure
\subsubsection{Influence of the luminosity}
As $\langle \varepsilon \rangle 
\propto \dot{M}^{1/4} M^{-1/2}$, the luminosity of the disc also directly 
influences the maximum of the function $\gamma_{b \infty}$ as a function of
 $\gamma_{max}$. As shown in figure 6, the maximum takes place at higher 
$\gamma_{max}$ when the luminosity decreases. Besides less luminous systems
 contribute to a lower radiation force and so to a less efficient acceleration.
In the Thomson regime one has a dependance
 $\gamma_{b \infty}\propto L^{1/7}$.

\subsubsection{Effect of scattered radiation}
All the results described above are obtained when studying the disc radiation
 alone. We also include in our calculation BLR radiation fields 
corresponding to two cases:

\beginlist
\item a) re-emission from a ring located between $r_1$ and $r_2$ and with an
emissivity given by equation (5)
\item b) re-emission from spherically distributed 
matter at a distance $r_0$ from the central black hole with an emissivity
 given by equation (6). 
\endlist
Figure 8 displays the equilibrium Lorentz factor in the presence of a BLR
 located between $r_1=10^4 \,r_g$ and $r_2=10^5\,r_g$ (case a), figure 8) 
in the Thomson regime. We also plotted the equlibrium Lorentz factor including
 KN corrections as well as the solution of the equation of motion for a plasma
 with $s=2$ and $\gamma_{max}=10^5$. As one can see, the effect of BLR on
 $\gamma_{beq}$ is very weakened by KN corrections. This can be understood 
because the photons coming from the BLR are blueshifted by the 
relativistic motion in the blob rest frame whereas the photons coming from 
the disc are redshifted. 
So the dragging force from the BLR is much more reduced by KN 
corrections than the accelerating one due to disc photons. Due to the
 weakness of the radiation field, the dynamical solution $\gamma_b(z)$ is
 still less affected than the equilibrium value. 
The case of a spherical shell located at $r_0=10^4\,r_g$ (case b)) is
 illustrated in figure 9. One can see that the diffused radiation field
 strongly affects the motion that is almost stopped at the crossing of
 the shell, where radiation density is dominated by the isotropically 
scattered photons. However, the plasma is quickly reaccelerated after the 
crossing, and high Lorentz factor can be reached again. 

\beginfigure{8}
 \psfig{width=8cm,height=8cm,angle=-90,file=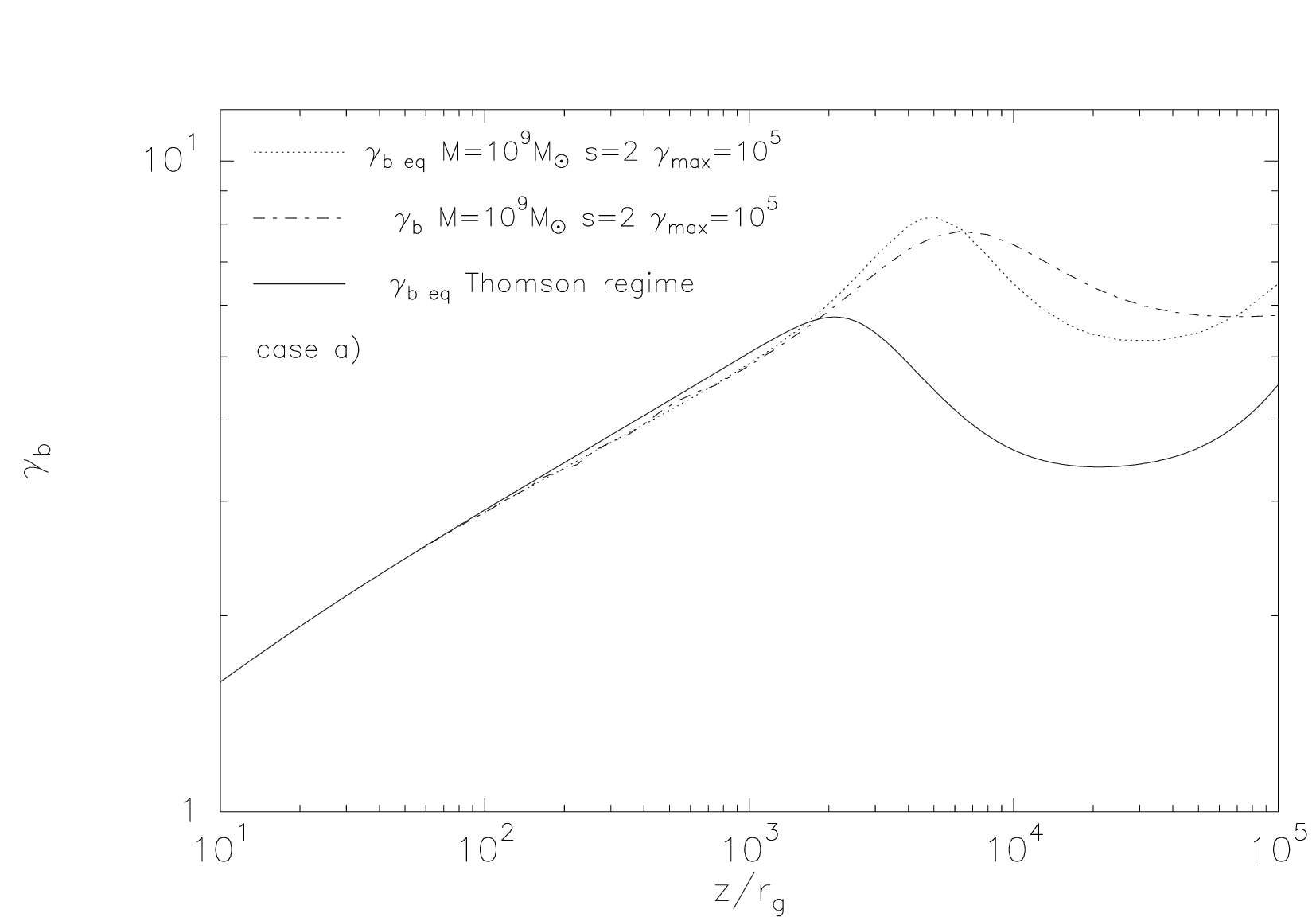}
  \caption{{\bf Figure 8.} Equilibrium Lorentz factor as a function of $z$ for
radiation coming from an accretion disc and BLR radiation from a ring
 located between $r_1=10^4r_g$ and $r_2=10^5r_g$. The emissivity
 is described by equation (5) with $\chi=0.1$ and $\alpha=2$. We also show the solution of the equation of motion for this case.
The solid curve shows $\gamma_{beq}$ in Thomson regime. 
The plasma parameters are $s=2$ and $\gamma_{max}=10^5$. The
 black hole mass is $M=10^9M_\odot$ and $L=L_{Edd}$.}
\endfigure

\beginfigure{9}
 \psfig{width=8cm,height=8cm,angle=-90,file=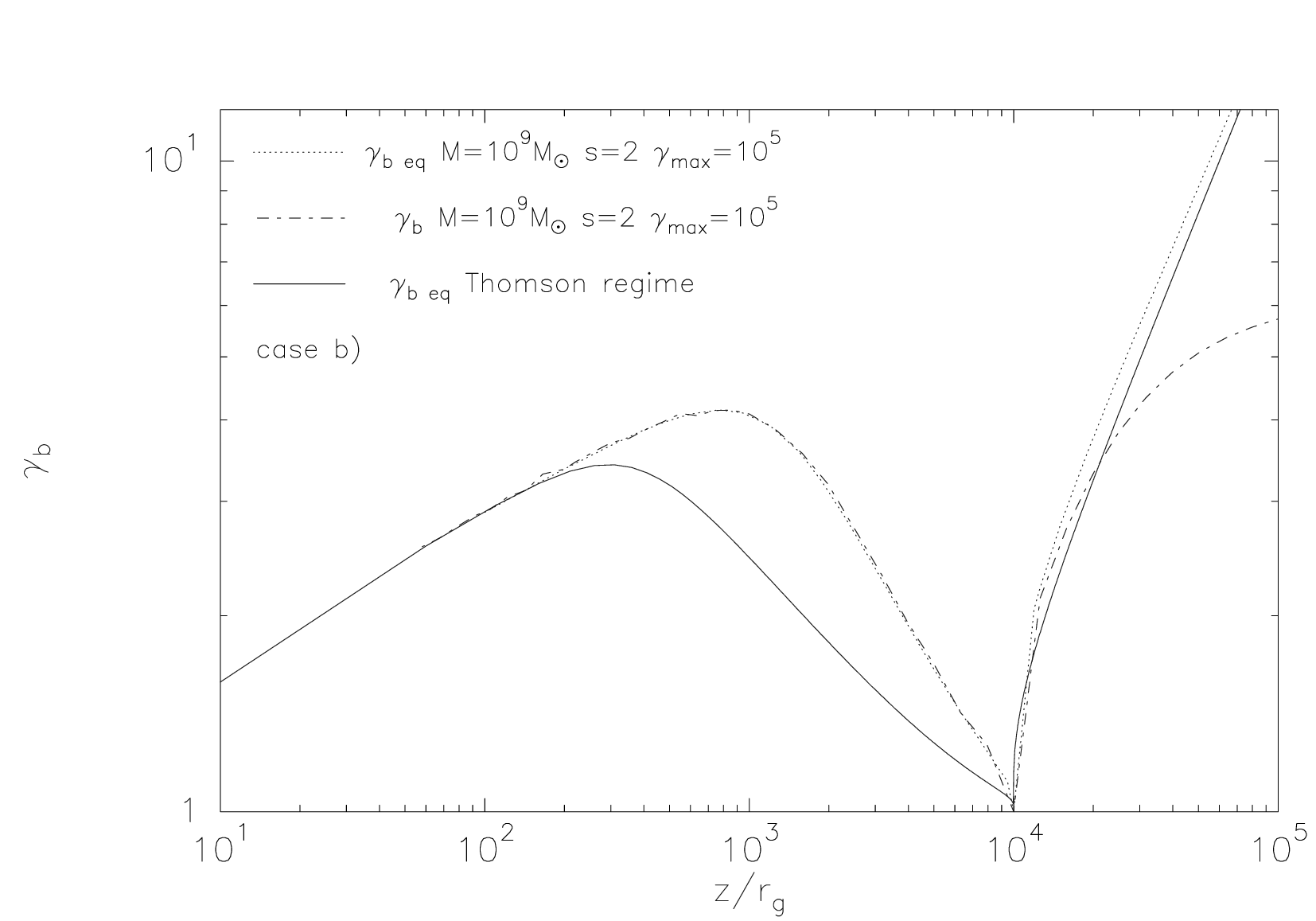}
  \caption{{\bf Figure 9.} Same as figure 8 for a BLR radiation from
 a shell located at $r_0=10^4r_g$. The emissivity is described by equation (6)
 with $\chi=0.1$.}
\endfigure
 Sikora et al. (1996) 
argued that scattered radiation should lead
 to efficient radiation drag. However, the influence of scattered
radiation is strongly governed by the position of scattering clouds with
respect to the critical distance $z_{crit}$ : if they lie below this
critical distance, the plasma will be temporarily braked during the
crossing of scattering region, but will be quickly reaccelerated after
it. If the clouds lie above $z_{crit}$, the terminal
Lorentz factor can indeed be strongly affected. Figure 10 and 11 display the
 terminal Lorentz factor as a function of $\gamma_{max}$ for different
 distances of the BLR. It can be seen that if it is close enough, the BLR
 can give still higher Lorentz factor than for the disc alone for the
 highest value of $\gamma_{max}$. This is because the solid angle substended
 by it at $z_{crit}$ is so small that its radiation field has an accelerating
 rather than decelerating effect. But even for a BLR between $10^4$ and
 $10^5 \, r_g$, terminal Lorentz factor around 10 are clearly reachable. We
conclude that the presence of broad lines can affect $\gamma_{b\infty}$,
but does not prevent in general highly relativistic motions. 
\beginfigure{10}
 \psfig{width=8cm,height=8cm,angle=-90,file=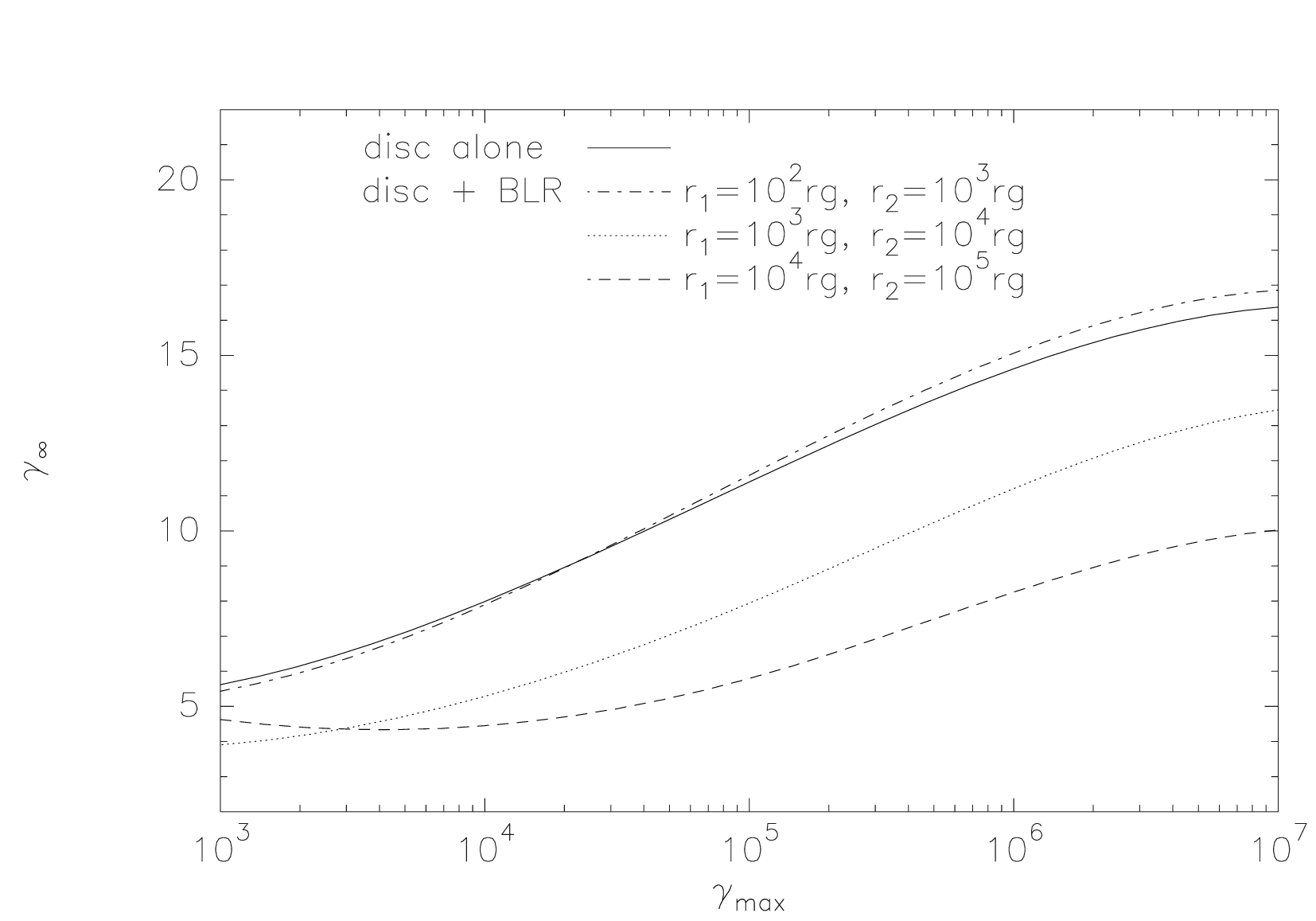}
\caption{{\bf Figure 10.}
 Terminal Lorentz factor $\gamma_{b\infty}$ as a function
of $\gamma_{max}$ for the spectral index $s=2$. The solid line shows
 the solution obtained including KN corrections for a standard accretion
 disc. The scattered radiation is described by equation (5) with $\chi=0.1$,
 $\alpha=2$ and different locations of the BLR ring.}
\endfigure

\beginfigure{11}
\psfig{width=8cm,height=8cm,angle=-90,file=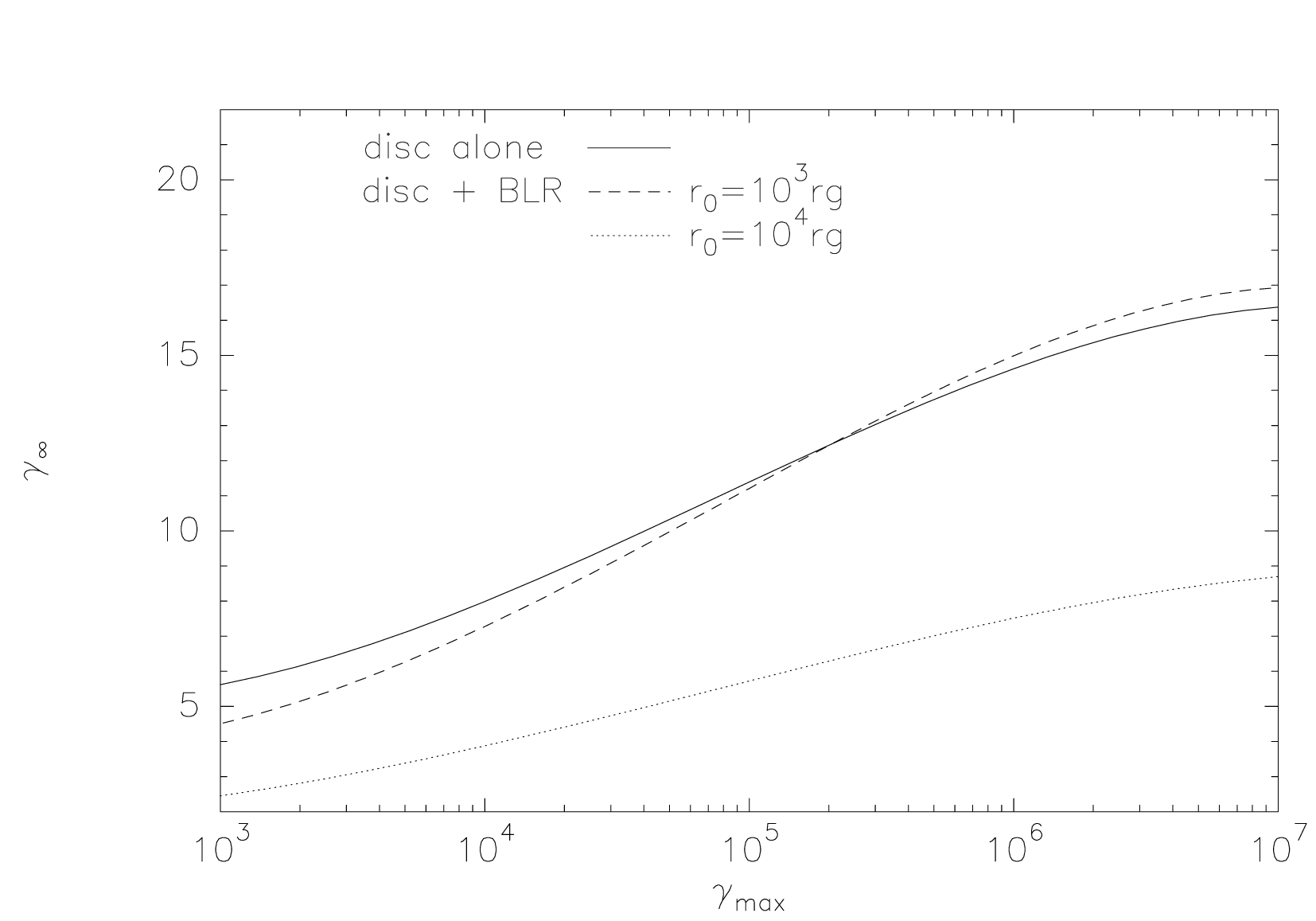}
\caption{{\bf Figure 11.}
 Same as figure 10 for scattered radiation described by equation (6)
 with $\chi=0.1$
 and different locations of the BLR shell.}
\endfigure
 
 
We also discuss the effect of a possible emission from a hot dusty torus
surrounding the central black hole outside the BLR region. The effect of such 
a scattered radiation field is much more important because KN corrections
 are here almost negligible. Figure 12 shows the terminal Lorentz factor
 in the presence of an IR emitting ring heated at $T=1500K$, located between
 $r_3$ and $r_4$. It is obvious that the terminal Lorentz factor is strongly
 reduced to values around 3 for the less favourable case ($r_3=10^4\, r_g$, 
$r_4=10^5\, r_g$). The situation is a little less dramatic for more distant
 sources ($r_3=10^5\, r_g$, $r_4=10^6\, r_g$) because the isotropic 
radiation density is lowered, and $\gamma_{b \infty}$ of 5 can be reached.
 
\beginfigure{12}
\psfig{width=8cm,height=8cm,angle=-90,file=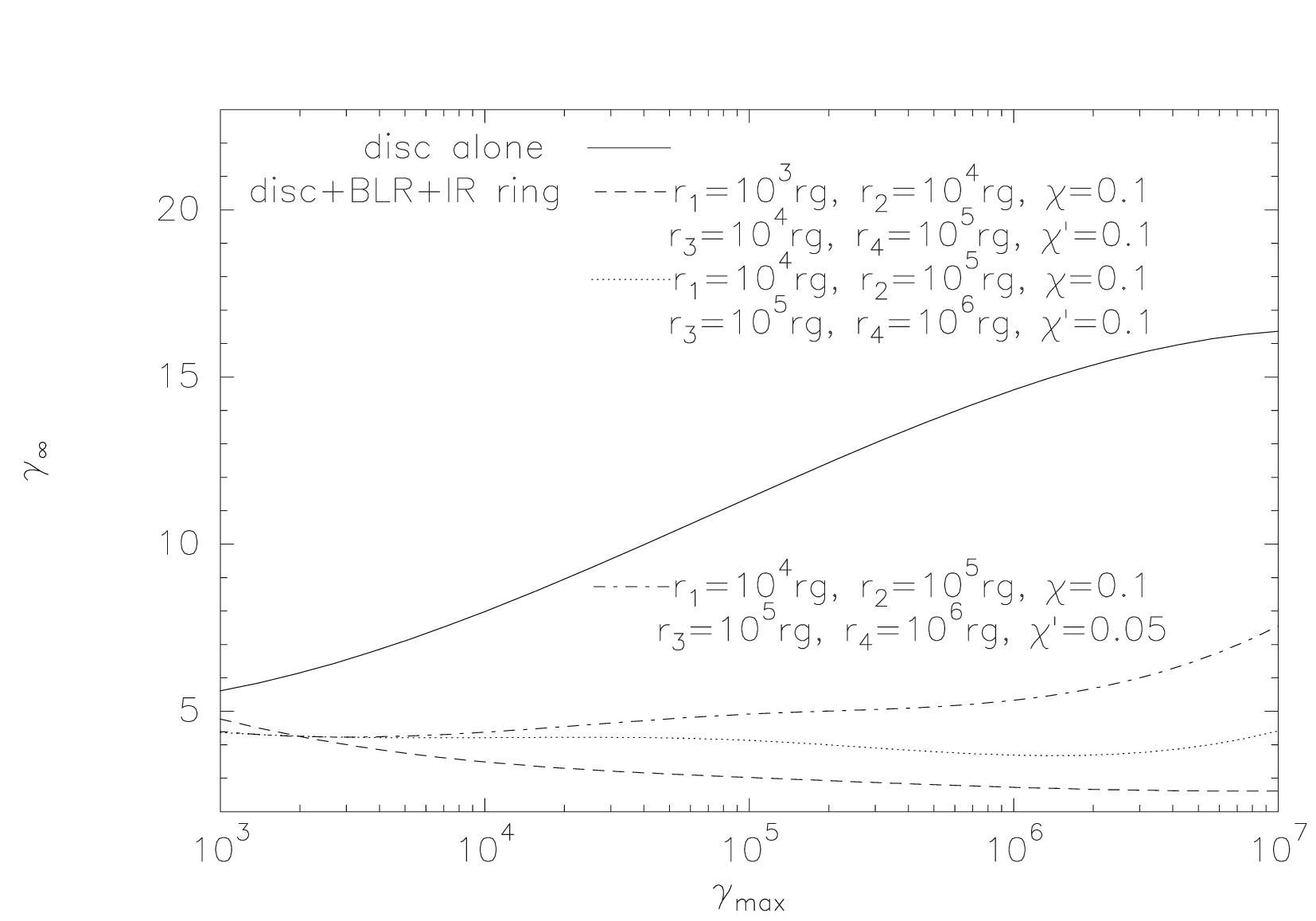}
\caption{{\bf Figure 12.}
 The influence of a dusty ring combined with a BLR on the terminal Lorentz 
factor $\gamma_{b\infty}$ as a function
of $\gamma_{max}$, for the spectral index $s=2$. The solid line shows
 the solution obtained including KN corrections for a standard accretion
 disc. The BLR ring emissivity 
is described by equation (5) with $\chi=0.1$, 
$\alpha=2$. The dusty ring emissivity is given by equation (7) with $\chi^{\prime}=0.1$ or $\chi^{\prime}=0.05$. We chose different locations for these
two components.}
\endfigure

\subsubsection{Influence of the accretion disc size}
\beginfigure{13}
 \psfig{width=8cm,height=8cm,angle=-90,file=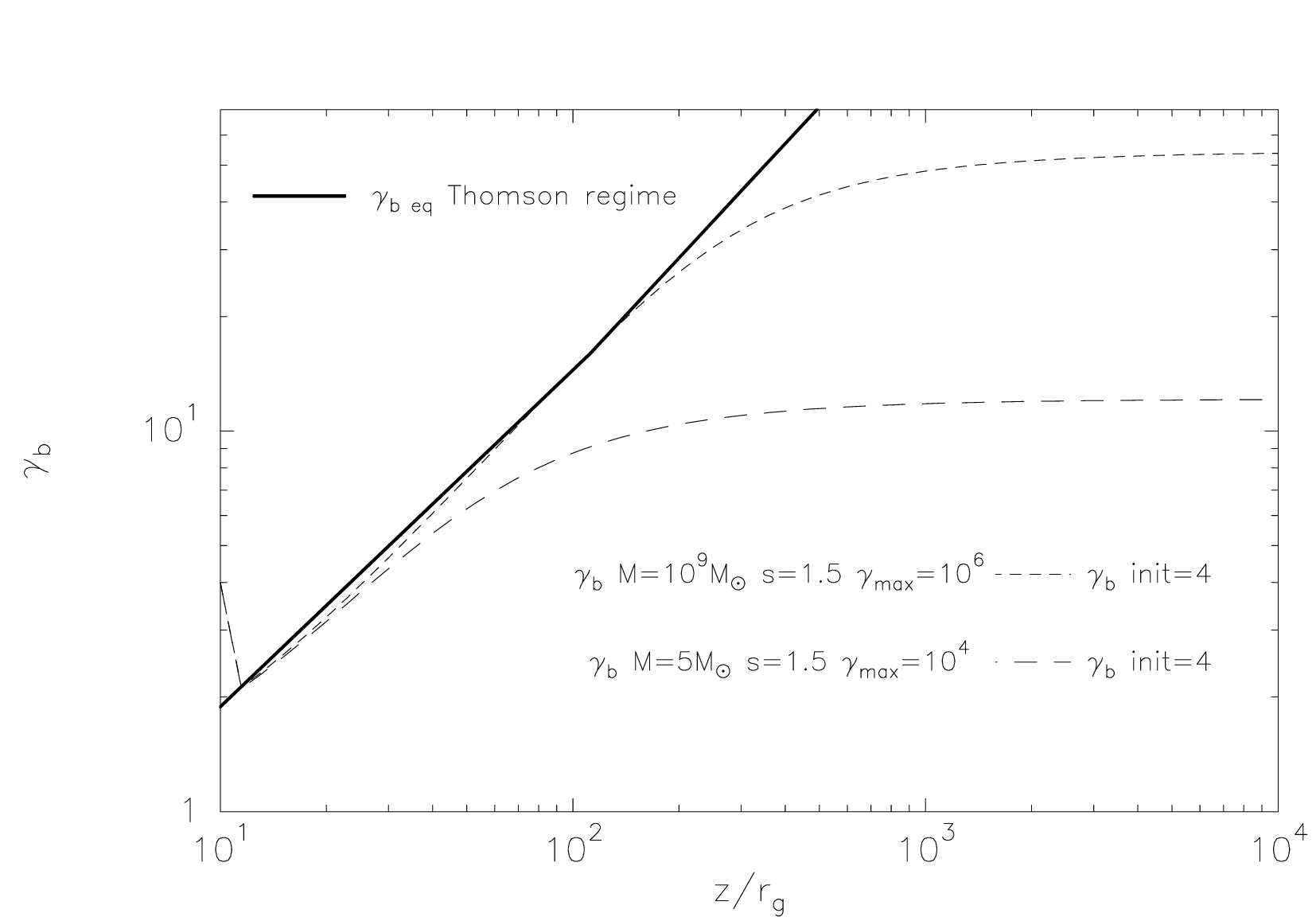}
  \caption{{\bf Figure 13.} Solutions of the equation of motion
 for a supermassive black hole ($M=10^9 M_\odot$) and a stellar black hole
 ($M=5M_\odot$) with a compact accretion disc. $r_e=10\,r_g$ and $L=L_{Edd}$. $\gamma_b\,\, init$ is the initial condition of the solution.
 We also represent the equilibrium Lorentz factor
 in Thomson regime .}
\endfigure
For small value of $r_e$ the radiation is more anisotropic and so more 
efficient
 to accelerate the pair plasma (see figure 13). In this figure $r_e=10 \,r_g$ and
 we show the solutions for which we obtain the largest $\gamma_{b \infty}$
 in extragalactic and galactic cases. In this configuration the radiation force is more efficient and 
 $\gamma_{b\infty}$ can be as high as 60 in the extragalactic case. 
The dependance on $s$ and $\gamma_{max}$ is shown in figure 14, 
where we extend the previous 
calculation to the case $r_e=10 \,r_g$. We find the same global 
behaviour of $\gamma_{b \infty}$ with spectral index and $\gamma_{max}$.

\beginfigure*{14}
\psfig{width=17cm,rheight=7cm,angle=-90,file=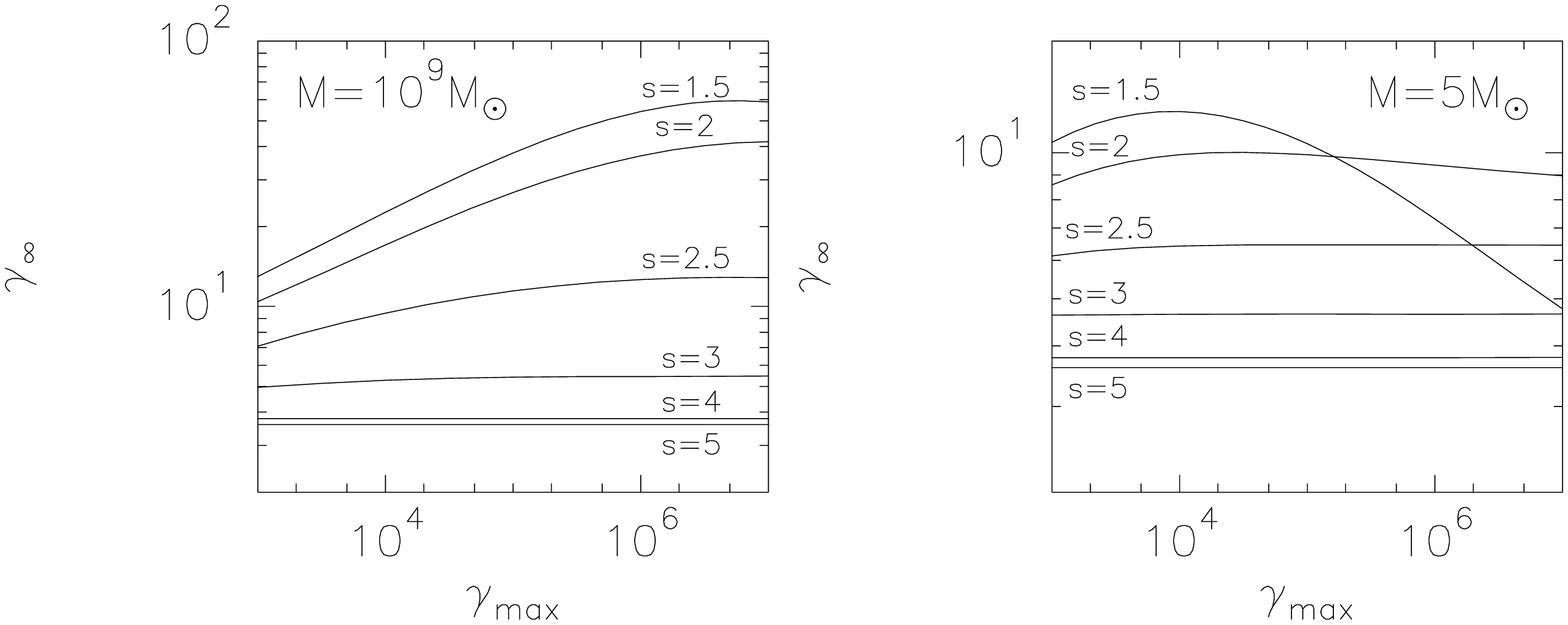}
\caption{{\bf Figure 14.}
 Terminal Lorentz factor $\gamma_{b\infty}$ as a function
of $\gamma_{max}$ for different value of spectral index $s$. The two
 pannels correspond to $r_e=10 \,r_g$, $L=L_{Edd}$, 
$M=10^9 M_{\odot}$(left) and $M=5 M_{\odot}$ (right).}
\endfigure

\section{Comparisons with observations}
Our model gives good agreement with observations of relativistic
 blob ejection in AGN and microquasars for the 'disc alone' solution. 
Figure 6 shows that in
 the most favourable configuration ($s=1.5$, $\gamma_{max}\sim 10^6$ and
 $L=L_{Edd}$) Compton 
rocket effect is able to accelerate pair plasma up to Lorentz factor
 $\gamma_{b\infty} \geq 20$. This value is only gradually reached, the jet 
being much slower at small distances. One can obtain higher value 
for super-Eddington systems. One can notice that observations
 of the faster extragalactic superluminal motion correspond to such value
 of bulk Lorentz factor (Vermeulen \& Cohen 1994).

This value is however strongly dependent on the spectal index $s$ and the
 high energy particle distribution cut-off $\gamma_{max}$. 
The precise value of $s$ is not obvious to
derive from observations. High energy spectra show typical X-ray spectral
indexes around $0.5$, which correspond to $s \simeq 2$. There is very
often a spectral break in the MeV range: this could be attributed to a
break in the particle distribution, that would correspond to
 $\gamma_{max}=10^2$. However, Marcowith et al. (1995) have
shown that this break could be very well reproduced by $\gamma-\gamma$
absorption with an actual particle distribution giving the primary
 photon spectrum that can extend to much higher energy.
 So high energy spectra may not be good
indicators of the upper cut-off $\gamma_{max}$. Moreover, the high energy
emission probably takes place at relatively small distances (around $10^2
r_g$) from the center, well below $z_{crit}$: the final bulk Lorentz
factor is not yet reached at this distance. On the other hand, the
detection of photons with at least 30 GeV, and even above TeV for some
BL Lacs, implies an upper cut-off $\gamma_{max} \geq 10^5$.

Radio spectral indexes are also difficult to assess because of
synchrotron self-absorption, especially for radio-flat quasars where the
observed spectrum results probably from the superposition of many
self-absorbed spectra. The optically thin index ranges mostly from 0.5 to
1, which corresponds to $2\leq s \leq 3$. With reasonable parameters 
($2 \leq s \leq 3$ and $\gamma_{max} \leq 10^5$), our model predicts 
typical values
$4 \leq \gamma_{b\infty} \leq  10$ which are the most frequently observed
 (Vermeulen \& Cohen 1994). Moreover, as shown in Sect. 3.2.5, KN effects 
prevents from strong Compton drag
 induced by BLR radiation in the vicinity of the central 
black hole. The fastest superluminal motions may be attributed to those
objects for which the BLR is weak and/or closest to the central object. 

On the other hand, the slowest motions ($\gamma_{b \infty} \leq 4$)
can be obtained in the presence
of a rich and extended environment of scattering material, such as
 BLR clouds and dusty torus. The presence of dust is inferred from an 
enhancement in the near infrared in some quasars spectra (Bairvainis 1987). 
Nevertheless a nonthermal infrared component, attributed to synchrotron radiation from the relativistic jet, is also usually observed in 
radio-loud AGN. This component is generally predominant in flat 
spectrum radio quasars (Neugebauer et al. 1986) 
and is necessary to explain rapid variations
observed in infrared flaring objects. We can speculate that these objects with
 the largest superluminal motion 
are those where the scattered thermal component is particularly weak.
Such 'nonthermal' objects should
not suffer strong radiation drag, while 'thermal' ones should have the
lowest terminal Lorentz factors. 
In conclusion the diversity among observed superluminal motions can be
 reproduced by our model 
by considering both influence of the plasma acceleration mechanism and
AGN environment.

\beginfigure{15}
 \psfig{width=8cm,height=8cm,angle=-90,file=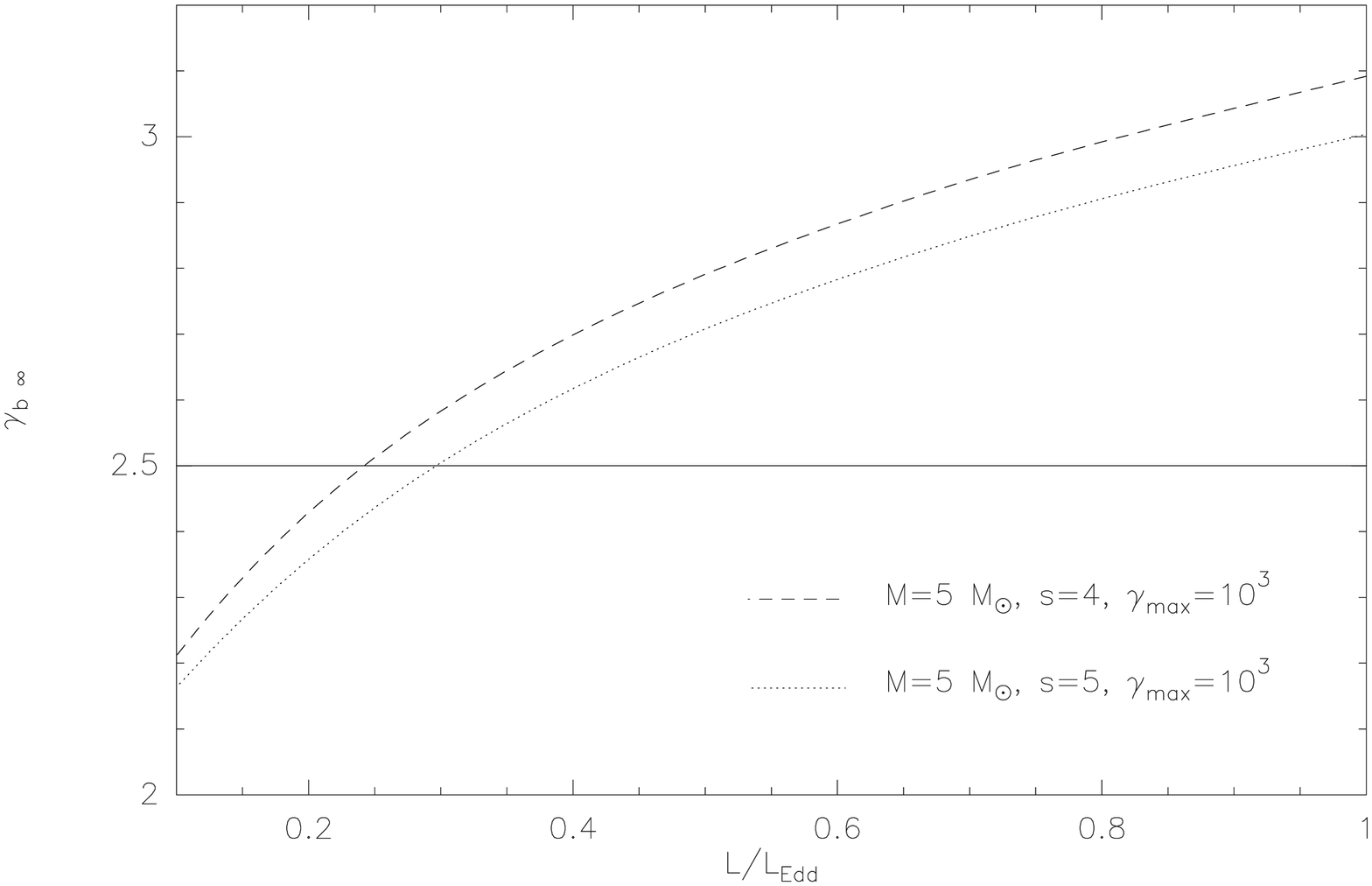}
  \caption{{\bf Figure 15.} The terminal Lorentz factor as a function
 of $L/L_{Edd}$ and for
 two spectral indexes ($s=4$ and $s=5$). $M=5M_{\odot}$, $\gamma_{max}=10^3$.}
\endfigure

The observed value of Lorentz factor of 
about 2.5 for the two
 microquasars (GRS1915+105, Mirabel \& Rodriguez 1994 and GROJ1655-40
, Hjemlling \& Rupen 1995) with large spectral indexes (respectively 
$s \sim 4$, Finoguenov et al. 1994 and $s \sim 4.6$, 
Harmon et al. 1995) are consistent with our results.
We show figure 15 the dependance of $\gamma_{b \infty}$ on compactness of 
the source for $s=4$ and $s=5$. We find that observations require
 $L\sim 0.2-0.3 L_{Edd}$, which is very close to the result
 by Li \& Liang (1996). This result is not strongly 
affected by the mass of the central black hole if the compactness of 
the source remains the same.
 The steep spectrums observed in these two objects ensure us that the
 terminal Lorentz factor is just dependant on the low energy cut-off
 of the electron distribution. In this case a more realistic 
description of the 
 accretion disc emission does not change our results for such spectral indexes
  (figure 7). Yet the maximal value of $\gamma_{b \infty}$ is model dependant.
The part of high energy emission contributes to decrease this value
 from the one obtained for a standard accretion disc. We find that value 
of about 6 can be reached in the most favourable case.

Finally, $\gamma_{b\infty}$ is much higher for a small sized accretion disc.
 We find value of about 60 in the case of an external radius of $10 \,r_g$
 in the case of supermassive black hole. 

\section{Conclusion}
We have considered the bulk acceleration of an electron-positron pair
 plasma in the vicinity of a central black hole. The acceleration is due
to Compton rocket effect on the plasma and the radiation force originates
 from a standard accretion disc emission. The pair plasma 
is continuously reheated
 by an efficient turbulent mechanism wich takes place in the frame of the
'two-flow' model. Thus we assumed in our calculations a stationary power-law
 energy distribution for the pair. We included KN corrections in the computation
 of the radiation force, and solved numerically the equation of motion. We studied configurations relevant to 
relativistic motion in AGN and galactic microquasars.
The main results of our calculations  
can be summarized as follows:
\beginlist
\item (i) For a given luminosity, the terminal Lorentz factor $\gamma_{b\infty}$
 admits an absolute maximum due to KN corrections. 
Values of about 20 can be reached
 in the extragalactic case, for sufficiently flat spectrum ($s\sim 1.5$)
 and accretion at the Eddington rate which may correspond
 to the highest observed relativistic motion. 
For more reasonable plasma parameters
 ($s\leq 2$ and $\gamma_{max}\leq 10^6$) the Compton rocket effect can 
account for the typical value of terminal Lorentz factor inferred from 
observations ($\gamma_{b\infty}$).
\item (ii) Scattered radiation from extended BLR or dusty torus
 can brake efficiently relativistic motion, even for a high energy 
plasma. This Compton drag and a weak plasma heating 
can be responsible for the lowest terminal Lorentz factor observed.
 The highest superluminal motion could be attributed to objects with
 particularly weak diffuse component and very efficient heating.
\item (iii) For stellar black holes KN corrections are important 
leading to smaller values than for a supermassive 
black hole. Recent observations of relativistic ejection
 in galactic microquasar are consistent with our results.


\endlist

\section*{Acknowledgments}
We would like to thank the anonymous referee for his helpful remarks,
 especially for the description of scattered radiation.

\appendix
\section{Equilibrium Lorentz factor and terminal Lorentz factor
 in Thomson regime}
Eddington parameters are defined as follows:
$$
\displaylines{%
 \hbox to \hsize{}\cr
 J={1 \over 4 \pi}\int I_{\nu}(\Omega){\rm d}\Omega{\rm d} \nu, \hfill\cr 
 H={1 \over 4 \pi}\int \mu I_{\nu}(\Omega){\rm d}\Omega {\rm d}\nu, 
\hfill \stepeq \cr
 K={1 \over 4 \pi}\int \mu^2 I_{\nu}(\Omega){\rm d}\Omega {\rm d}\nu. 
\hfill  \cr}
$$
 Marcowith et al. (1995) show that these parameters can be expressed 
in the case of standard accretion disc (Shakura \& Sunyaev 1973) as 
a function of the integral:
$$
I(\alpha,\beta,\gamma)=\int^{u_e}_{1}[1/u^\alpha(1+\varepsilon^2 u^2)^\beta]
[1-(1/u)^{1/2}]^\gamma {\rm d}u, \hfill\stepeq
$$
with $u=r/r_i$, $u_e=r_e/r_i$ and $\varepsilon=r_i/z$. They obtain:
$$
\displaylines{%
 \hbox to \hsize{}\cr
    J=(L/8\pi^2r_i^2)(1-2b/3)^{-1} \varepsilon^2 I(2,3/2,1),\hfill \cr 
    H=(L/8\pi^2r_i^2)(1-2b/3)^{-1} \varepsilon^2 I(2,2,1), \hfill\stepeq\cr
    K=(L/8\pi^2r_i^2)(1-2b/3)^{-1} \varepsilon^2 I(2,5/2,1).\hfill  \cr}
$$
They study the case $r_i<z<r_e$ for which $\varepsilon<1$ and $u \geq 1$.
They find that the equilibrium Lorentz factor is given by:
$$
\gamma_{beq}\sim \varepsilon^{1/4}={z^{1/4} \over r_i^{1/4}}\hfill\stepeq
$$
In the case $z>r_e$ we have $\varepsilon u = r/z < 1$ for $1<u<u_e$.
So we can expand the term $(1+\varepsilon^2 u^2)^{-\beta}$ into series:
$$
(1+\varepsilon^2 u^2)^{-\beta}=\sum_{k}(-1)^k {\beta (\beta+1)...(\beta+k-1)\over k!}
\varepsilon^{2 k} u^{2k}.\hfill\stepeq 
$$
Hence in the case $\alpha=2$, $\gamma=1$ and integrating over $u$
,we obtain:
$$
\displaylines{%
\hbox to \hsize{}\cr
I(2,\beta,1)=\sum_{k}\varepsilon^{2k} (-1)^k {\beta (\beta+1)
...(\beta+k-1)\over k!}\hfill\cr
\times
 \left({u_e^{2k-1}-1 \over 2k-1}-{u_e^{2k-3/2}-1 \over 2k-3/2}\right)
.\hfill\stepeq\cr}
$$
The first coefficients in $\varepsilon$ are then:
$$
I(2,\beta,1)=A_0-\beta A_1 \varepsilon^2 +{\beta (\beta+1) \over2} 
A_2 \varepsilon^4 ,\hfill\stepeq
$$
with
$$
\displaylines{%
 \hbox to \hsize{}\cr
 A_0={1 \over 3}-u_e^{-1}+{2 \over 3}u_e^{-3/2}\sim {1 \over 3},\hfill \cr 
 A_1=u_e-2u_e^{1/2}+1 \sim u_e, \hfill\stepeq\cr
 A_2={u_e^3\over 3}-{2u_e^{5/2} \over 5}+{1\over 15} \sim {u_e^3\over 3}. 
\hfill \cr}
$$
The equivalents are given for $u_e\gg 1$. This gives for the 
Eddington parameters:
$$
\displaylines{%
 \hbox to \hsize{}\cr
    J=(L/8\pi^2r_i^2)(1-2b/3)^{-1} \varepsilon^2 (A_0-3/2A_1\varepsilon^2
+15/8A_2\varepsilon^4) , \hfill\cr 
    H=(L/8\pi^2r_i^2)(1-2b/3)^{-1} \varepsilon^2 (A_0-2A_1\varepsilon^2
+3A_2\varepsilon^4), \hfill\stepeq\cr
    K=(L/8\pi^2r_i^2)(1-2b/3)^{-1} \varepsilon^2 (A_0-5/2A_1\varepsilon^2
+35/8A_2\varepsilon^4). \hfill \cr}
$$
The equilibrium Lorentz factor is $\gamma_{beq}=(1-\beta_{beq}^2)^{-1/2}$ 
where 
$$
\beta_{beq}=x-(x^2-1)^{1/2},\eqno\stepeq
$$
with 
$$
x={J+K\over 2 H}. \eqno\stepeq
$$
Using equations (A9) we find with $u_e\gg 1$:
$$
x=1+{1 \over 8}u_e^3 \varepsilon^4, \eqno\stepeq
$$
$$
\beta_{beq}=1-{u_e^{3/2} \over 2}\varepsilon^2, \eqno\stepeq
$$
and finally
$$
\gamma_{beq}=u_e^{-3/4}\varepsilon={z \over r_e^{3/4} r_i^{1/4}}. 
\hfill\stepeq
$$
The asymptotic solution of equation (17) is approximatively $\gamma_{b\infty}
=\gamma_{beq}(z_{crit})$, where $z_{crit}$ is the location where the radiative
 force becomes to weak to maintain $\gamma_b(z)=\gamma_{beq}(z)$.
 This occurs when the evolution of $\gamma_{beq}$, {\it i.e.}
 $\displaystyle \Delta z_0= \gamma_{beq} /({\rm d}\gamma_{beq}/{\rm d}z)
= {\rm dlog}(z)/(z {\rm dlog}(\gamma_{beq}))$, is larger than
 the evolution of $\gamma_b$ towards $\gamma_{beq}$, {\it i.e.} 
$ \displaystyle \Delta z_1=({\rm d}\gamma_b / {\rm d}z)/(\gamma_b-\gamma_{beq})$.
 Using equation (17) one finds:
$$
z_{crit}\sim \left({1\over \rho^\prime}\left .{{\rm d} F^{\prime z} 
\over {\rm d} \gamma_b}\right |_{
\gamma_{beq}} \right)^{-1}. \hfill\stepeq
$$
In the Thomson regime we can use equation (9) and (10) which give in the 
relativistic case:
$$
{ {\rm d} F^{\prime z} \over {\rm d} \gamma_b}=-
{\sigma_T \over c} {8 \pi \over 3} \langle \gamma^{\prime 2} \rangle
{ {\rm d}H^\prime 
\over  {\rm d}\gamma_b}=
{2H\over \gamma_b}(1-x/\beta_b). \hfill\stepeq
$$
With (A9), (A12), (A13), (A14) and ignoring some terms of
 order unity one finally gets:
$$
z_{crit} \sim \left( r_e^{9/4}r_i^{3/4}{L \sigma_T \over m_e c^3}
{\langle \gamma^{\prime 2} \rangle \over \langle \gamma^{\prime} \rangle }
\right) ^{1/4} , \hfill\stepeq
$$
and,
$$
\gamma_{b\infty} \sim \left( l \left( {r_i \over r_e} \right) ^{3/4}
{\langle \gamma^{\prime 2} \rangle \over \langle \gamma^{\prime} \rangle }
\right) ^{1/4}. \hfill\stepeq
$$
In the case $r_i < z < r_e$, using equation (A4) one finds:
$$
z_{crit} \sim r_i\left({L \sigma_T \over m_e c^3 r_i}
{\langle \gamma^{\prime 2} \rangle \over \langle \gamma^{\prime} \rangle }
\right) ^{4/7} , \hfill\stepeq
$$
and finally,
$$
\gamma_{b\infty} \sim \left( l {\langle \gamma^{\prime 2} \rangle 
\over \langle \gamma^{\prime} \rangle }
\right) ^{1/7}. \hfill\stepeq
$$

\section*{References}

\beginrefs
\bibitem Barvainis, R. , 1987, ApJ, 320, 537
\bibitem Blumenthal, G. R., Gould, R. J., 1970, Rev. of mod. physics, 42,
 2237, Bremsstrahlung, Synchrotron radiation, and Compton scattering 
of high-energy electrons traversing dilute gases
\bibitem Finoguenov, A. et al., 1994, ApJ, 424, 940
\bibitem Harmon, B. A. et al., 1995, Nature, 374, 703
\bibitem Henri, G., Pelletier, G., 1991, ApJ Letters, 383, L7
\bibitem Henri, G., Pelletier, G., Roland, J., 1993, ApJ Letters, 404, L41
\bibitem Hjellming, R. M., Rupen, M. P., 1995, Nature, 375, 464
\bibitem Li, H., Liang, E. P., 1996, ApJ, 458, 514
\bibitem Marcowith, A., Henri, G., Pelletier, G., 1995, MNRAS, 277, 681
\bibitem Mirabel, I. F., Rodriguez, L. F., 1994, Nature, 371, 46
\bibitem Neugebauer, G. et al., 1986, ApJ, 308, 815
\bibitem O'Dell, S. L., 1981, ApJ Letters, 243, L147
\bibitem Phinney, E. S., 1982, MNRAS, 198, 1109
\bibitem Phinney, E. S., 1987, in Superluminal Radio Sources, ed.
 Zensus, J. A. \& Pearson, T. J., 301 (Cambridge, UK)
\bibitem Rybicki, G. B., Lightman, A. P., 1979, Radiative Processes
 in Astrophysics (New York: John Wiley)
\bibitem Shakura, N. I., Sunyaev, R. A. , 1973, A\&A, 24, 337
\bibitem Shapiro, S. L., Lightman, A. L., Eardley, D. M., 1976, ApJ, 204, 1
87
\bibitem Sikora, M., Sol, H., Begelman M. C., Madejski, G. M., 1996, MNRAS, 280, 781
\bibitem Sol, H., Pelletier, G., Ass\'eo, E., 1989, MNRAS, 327, 411
\bibitem Sunyaev, R. A., Titarchuk, L. G., 1980, A\&A, 86, 121
\bibitem Tingay, S. J. et al., 1995, Nature, 374, 141
\bibitem Vermeulen, R. C., Cohen, M. H., 1994, ApJ, 430, 467
\endrefs
\bye